\documentclass[pdflatex, sn-basic, iicol]{sn-jnl}

\usepackage{bm}
\usepackage[utf8]{inputenc}
\usepackage[T1]{fontenc}
\usepackage{dsfont}
\usepackage{parskip}
\usepackage{graphicx}
\usepackage{xcolor}
\usepackage{svg} 
\usepackage{hyperref}

\usepackage{array} 
\usepackage{booktabs}
\usepackage{multirow}
\usepackage{dcolumn}

\newcolumntype{L}{>{\raggedright\arraybackslash}p{3cm}}
\usepackage{arydshln} 

\usepackage{float}
\usepackage{microtype}

\usepackage[super]{nth}
\usepackage{amsfonts}
\usepackage{amsmath, amsthm, amssymb}
\usepackage{nicefrac}
\usepackage{bm}
\usepackage{physics}
\usepackage{braket}
\usepackage[version=4]{mhchem}
\usepackage{siunitx}

\newcommand{\xdata}{\mathbf{x}}

\usepackage{lipsum}
\usepackage{comment}

\newcommand{\red}{\color{\red}}

\newcommand{\revision}[1]{\textcolor{black}{#1}}

\begin{document}

\title[Quantum Kernel Methods under Scrutiny: A Benchmarking Study]{Quantum Kernel Methods under Scrutiny: A Benchmarking Study}

\author*[1]{\fnm{Jan} \sur{Schnabel}}\email{jan.schnabel@ipa.fraunhofer.de}
\author*[1]{\fnm{Marco} \sur{Roth}}\email{marco.roth@ipa.fraunhofer.de}
\affil[1]{\orgdiv{Artificial Intelligence and Machine Vision}, \orgname{Fraunhofer Institute for Manufacturing Engineering and Automation IPA}, \orgaddress{\street{Nobelstraße 12}, \city{Stuttgart}, \postcode{D-70569}, \country{Germany}}}

\date{\today}

\abstract{
    Since the entry of kernel theory in the field of quantum machine learning, quantum kernel methods (QKMs) have gained increasing attention with regard to both probing promising applications and delivering intriguing research insights. Benchmarking these methods is crucial to gain robust insights and to understand their practical utility. \revision{In this work, we present a comprehensive large-scale study examining QKMs based on fidelity quantum kernels (FQKs) and projected quantum kernels (PQKs) across a manifold of design choices.} Our investigation encompasses both classification and regression tasks for five dataset families and 64 datasets, systematically comparing the use of FQKs and PQKs quantum support vector machines and kernel ridge regression. This resulted in over 20,000 models that were trained and optimized using a state-of-the-art hyperparameter search to ensure robust and comprehensive insights. We delve into the importance of hyperparameters on model performance scores and support our findings through rigorous correlation analyses. \revision{Additionally}, we provide an in-depth analysis addressing the design freedom of PQKs and explore the underlying principles responsible for learning. Our goal is not to identify the best-performing model for a specific task but to uncover the mechanisms that lead to effective QKMs and reveal universal patterns.
}

\keywords{quantum computing, quantum machine learning, quantum kernel methods, fidelity quantum kernels, projected quantum kernels, benchmarking, hyperparameter optimization}

\maketitle

\section{Introduction}
Within the rapidly evolving field of quantum machine learning (QML)~\citep{Biamonte2017,Cerezo2022}, quantum kernel methods (QKMs)~\citep{Schuld2019,Havlicek2019,Peters2021,Tomono2022, Hubregtsen2022,Jerbi2023,gan2023traceinducedquantum} have emerged as a particularly interesting and promising branch of research. For example, Liu \emph{et al.}~\citep{Liu2021} have proven a rigorous quantum advantage for a classification task engineered from the discrete logarithm problem, although it requires a fault-tolerant quantum computer due to its reliance on a Shor-type data encoding. Beyond that, Refs.~\citep{Schuld2019, schuld2021supervised} formally reveal that supervised QML models are kernel methods. In particular this implies that QKMs can be embedded into the rich mathematical framework of conventional kernel theory~\citep{schölkopf2002learning}. 

The key idea of kernel methods are kernel functions that implicitly map input data into a higher-dimensional space where the learning problem \revision{can often be formulated as a linear model (albeit there is no theoretical guarantee).} Quantum kernel methods leverage the principles of quantum mechanics to perform these mappings into the exponentially large Hilbert space of quantum states. In practice, this is realized by parameterized quantum circuits (PQCs)~\citep{Schuld2019, lloyd2020,Schuld2021,Jerbi2023}, which in the context of quantum kernels are referred to as data encoding circuits. Two common approaches to evaluate the corresponding quantum kernel functions have become established: fidelity quantum kernels (FQKs)~\citep{Schuld2019, Havlicek2019, Blank2020, Hubregtsen2022} and projected quantum kernels (PQKs)~\citep{Huang2021,gan2023traceinducedquantum, suzuki2023}. A unified framework for generalized trace-induced quantum kernels that includes FQKs and linear PQKs has been proposed recently in Ref.~\citep{gan2023traceinducedquantum}.

Wide-ranging application-oriented research in diverse domains emerged from the promising potentials of QKMs. \revision{This ranges from financial classification tasks~\citep{miyabe2023}, quantum support vector machines (QSVMs) for modelling tranisition probabilities in health and disability insurance~\citep{Djehiche2021}, and a quantum kernel classifier for real high-dimensional data from cosmology~\citep{Peters2021}, and QKMs for solving differential equations~\citep{Paine2023}.} Moreover, there are recent efforts to automize the encoding circuit architecture search, cf., e.g., Ref.~\citep{Rapp2025} and references therein.

Equally, QKMs have been studied from a theoretical perspective. Kübler \emph{et al.}~\citep{Kuebler2021} provide insights into the inductive bias of quantum kernels and conclude that quantum speed-ups may only occur if one manages to encode knowledge about the problem at hand into underlying quantum circuits, while encoding the same bias into a classical model would be hard. In this regard, in Refs.~\citep{Shaydulin2022,canatar2023}, the authors show that tuning the kernel's bandwidth (which can be seen as a data preprocessing step) controls the model's inductive bias. Consequently, these works identify the quantum kernel bandwidth as the key hyperparameter controlling the expressiveness of the model and give a theory, which shows that varying the bandwidth enables generalization. However, the exponential size of the quantum feature space can hinder generalization and cause exponentially concentrated quantum kernel values~\citep{suzuki2023, thanasilp2024exponential}. To circumvent this setback, Huang \emph{et al.}~\citep{Huang2021} introduced the family of PQKs.\footnote{We note that Ref.~\citep{thanasilp2024exponential} also derives exponential concentration bounds for PQKs. However, in practice PQKs usually suffer less from exponentially large Hilbert spaces, provided certain assumptions on corresponding data encoding circuits are taken into account~\citep{suzuki2023}.} Their key findings include proving that classical learners can approximate quantum outputs with enough data and introducing methods for assessing potential quantum advantages. An illustrative application of these concepts is given in Ref.~\citep{Slattery2023}, where the authors provide numerical evidence against quantum advantage for FQKs on classical data. They show that tuning the kernel's bandwidth improve the model performance and thus enables generalization, but also results in classically tractable kernels due to unfavorable geometric difference values.

Despite the variety of previous work, a comprehensive understanding of the wealth of design choices and their underlying mechanisms for various datasets is still incomplete. Our work attempts to make a contribution to closing this gap further by systematically analyzing the diversity of design criteria on a large scale and systematically examining both FQK and PQK approaches. We address this through state-of-the-art hyperparameter search and correlation analysis. 

We conduct an extensive benchmarking study for classification and regression tasks using various QKMs such as quantum support vector classification (QSVC), quantum kernel ridge regression (QKRR), and quantum support vector regression (QSVR). Additionally, we generate a broad and general database by exploring nine popular data encoding circuits from the literature, offering a diverse analysis on the impact of the choice of encoding circuit. Beyond that, we propose two different encoding strategies to introduce feature redundancies and systematically compare them. Furthermore, the present work delves deep into the different design aspects of PQKs by examining the effects of measurement operators and outer kernels on model performance. We additionally address the question of which components of the PQK definition are responsible for learning, i.e., the projected quantum circuits versus the outer kernel. Finally, we numerically support all findings by rigorous correlation analysis. In total, this study resulted in over $20,000$ quantum kernel models that were trained to provide an extensive database and to ensure robust and comprehensive insights.
To facilitate this, we developed a software tool. 

Previous studies have focused mostly on classification tasks. For example, Bowles \emph{et al.}~\citep{bowles2024} conducted a large-scale evaluation of popular QML models, including Quantum Neural Networks (QNNs), FQKs and PQKs across various datasets. While QKMs where part of the study, the influence of the different design choices was not a particular focus. Their work was preceded by first attempts to systematically explore certain aspects of quantum model design, cf., e.g., Refs.~\citep{Moussa2024, kashif2023}. Egginger \emph{et al.}~\citep{egginger2024hyperparameter} have conducted a hyperparameter study for QKMs which extends the findings of Ref.~\citep{Slattery2023} for PQKs. However, unlike this work, they exclusively focus on a single feature map (Hamiltonian evolution). In contrast to these previous findings, the present study offers an in-depth analysis of QKMs that significantly broadens the scope of investigation in multiple aspects and aims to answer aspects that have been previously untouched. This includes addressing regression problems, encoding mechanisms and a detailed analysis of PQKs. By extending previous findings we aim to contribute to a holistic understanding of QKMs.

This work is organized as follows. Section~\ref{sec:background} introduces the theoretical basics of QKMs. Hereafter, we give detailed insights about the study design in Sec.~\ref{sec:research-design}, including an overview on the multitude of different quantum kernel models considered in this work, an introduction of the datasets as well as implementation details for the corresponding hyperparameter optimization pipeline and the final experimental setup. In Sec.~\ref{sec:results} we first thoroughly investigate model performances and the influence of hyperparameters for classification and regression tasks across all datasets, encoding circuits and quantum kernel models and support corresponding findings with correlation analyses. Secondly, we 
give an in-depth analysis on different design options within the PQK approach, i.e., choice of outer kernel function and measurement operator. We discuss universal findings and patterns across all experiments and comment on the necessity of entanglement in data encoding circuits in Sec.~\ref{sec:discussion}. \revision{Finally, we summarize our study in Sec.~\ref{sec:conclusion}.}

\section{\label{sec:background}Theoretical Background}
One of the most interesting aspects of QKMs is that they can be formally embedded into the rich and powerful mathematical framework of classical kernel theory~\citep{Schuld2019, schuld2021supervised}. 

\subsection{\revision{Conventional Kernel Theory}}

The key idea behind the conventional kernelized approach to (supervised) machine learning is to find and analyze patterns by transforming the respective learning problem from the original input data domain $\mathcal{X}$ to a higher-dimensional (potentially infinite-dimensional) \emph{feature space} $\mathcal{F}$, where the learning tasks \revision{can often be expressed in a linear form.} This mapping is accomplished by a \emph{feature map} $\phi: \mathcal{X}\to\mathcal{F}; \xdata\mapsto\phi(\xdata)$. Kernels, are real- or complex-valued symmetric and positive semi-definite functions of two input data points, $k: \mathcal{X}\times\mathcal{X}\to\mathbb{C}$\revision{, induced by the inner product in the feature space, i.e.,}
\begin{align}
    \label{eq:kernel}
    k(\xdata, \xdata') = \langle\phi(\xdata), \phi(\xdata')\rangle_{\mathcal{F}}\,.
\end{align}
\revision{Hence, less formally, one can think of kernels as similarity measures between two data points $\xdata$ and $\xdata^\prime$.}

\revision{A more detailed introduction to conventional kernel theory and respective methods is given in appendices~\ref{appendix:conventional-kernel-theory} and~\ref{appendix:KRR-vs-SVR}.}

\subsection{\revision{Quantum Kernel Methods}}

In QML we \revision{commonly} process input data by encoding (embedding) them into quantum states of the form 
\begin{align}
    \label{eq:data-encoding}
    \ket{\psi_{\boldsymbol{\theta}}(\xdata)} &= U(\xdata, \boldsymbol{\theta})\ket{0}^{\otimes n}\,,
\end{align}
where $\boldsymbol{\theta}$ are variationally trainable parameters to refine the embedding into the quantum Hilbert space $\mathcal{H}^Q$. In practice, the unitary encoding operator $U(\xdata, \boldsymbol{\theta})$ can be implemented by a data encoding quantum circuit which manipulates an initial $n$-qubit quantum state $\ket{0}^{\otimes n}$. This reveals the striking similarity to kernel methods: Both utilize mathematical frameworks that map information into high-dimensional spaces for processing. In particular, as shown in Refs.~\citep{Schuld2019,schuld2021supervised}, the central concept of QKMs is that they can be formulated as a classical kernel method (e.g. SVM or KRR) whose kernel is computed using a quantum computer. Since quantum computations inherently feature the quantum mechanical principles of superposition and entanglement, the resulting quantum kernels hold the prospect of designing machine learning models that are able to learn complex problems that are out of reach for conventional machine learning methods~\citep{Liu2021}. 

\begin{figure}
    \includegraphics[width=\columnwidth]{"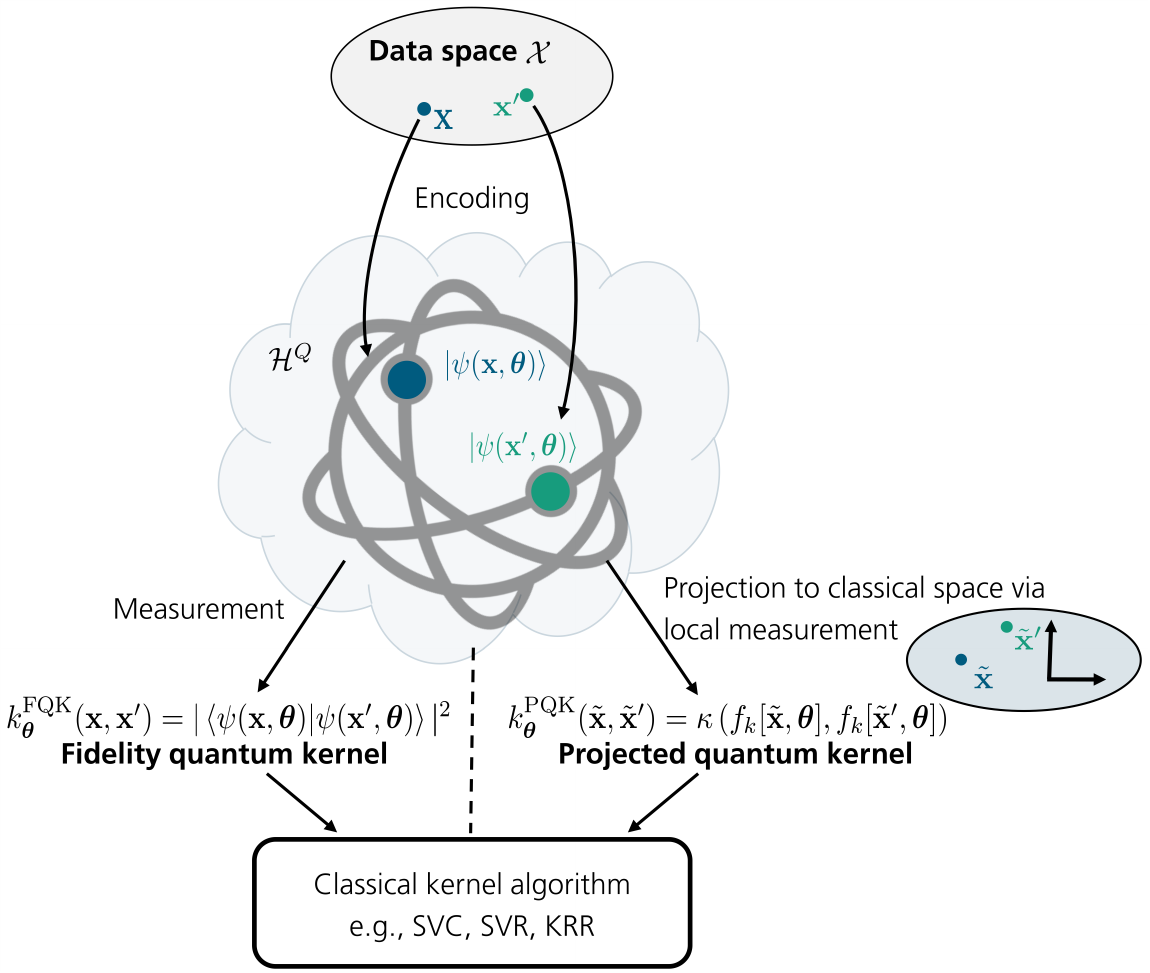"}
    \caption{\label{fig:qkm_scheme}Schematic illustration of the basic working principle of QKMs and its two most common approaches to compute respective quantum kernel Gram matrices. Data points are mapped from the input space $\mathcal{X}$ to the quantum Hilbert space $\mathcal{H}^Q$ by encoding them into quantum states $\ket{\psi(\mathbf{x},\boldsymbol{\theta})}$. Access to $\mathcal{H}^Q$ is provided by measurements, which can be expressed by inner products of quantum states in full analogy to classical kernel theory. \textbf{Left:} By using the Hilbert-Schmidt inner product and leveraging this fidelity-type metric to define quantum kernels leads to FQKs, cf. Eq.~\eqref{eq:Q-Kernel}. \textbf{Right:} Instead of directly processing quantum states within the quantum Hilbert space it has been shown that it can be beneficial to first project them to an approximate classical representation using, e.g., reduced physical observables. This concept gives rise to the family of PQKs. One of the simplest forms of defining PQKs is given in Eq.~\eqref{eq:PQK-general} and corresponds to measuring $k$-particle reduced density matrices and process the result with a classical kernel function $\kappa$. In both cases (FQK and PQK) the resulting kernel Gram matrices are subsequently passed to a classical kernel algorithm.} 
\end{figure}

In quantum computing, access to the Hilbert space of quantum states $\mathcal{H}^Q$ is given by measurements, which, in analogy to conventional kernel theory, can be expressed by inner products of quantum states. This is schematically shown in Fig.~\ref{fig:qkm_scheme}.
To formalize this, we define the density matrix $\rho_{\boldsymbol{\theta}}(\xdata)=\dyad{\psi(\xdata,\boldsymbol{\theta})}$ as the corresponding data-encoding feature map~\citep{schuld2021supervised,Havlicek2019}. 
With this, we can leverage the native geometry of the quantum state space and naturally define a quantum kernel using the Hilbert-Schmidt inner product, i.e.,
\begin{align}
    \label{eq:Q-Kernel}
    k^{\text{FQK}}_{\boldsymbol{\theta}}(\xdata, \xdata^\prime) &= \mathrm{tr}\left[\rho_{\boldsymbol{\theta}}(\xdata)\rho_{\boldsymbol{\theta}}(\xdata^\prime)\right]\,.
\end{align}
For pure states this definition reduces to 
\begin{align}
    \label{eq:FQK}
    k^{\text{FQK}}_{\boldsymbol{\theta}}(\xdata, \xdata^\prime) &= \left|\Braket{\psi_{\boldsymbol{\theta}}(\xdata)|\psi_{\boldsymbol{\theta}}(\xdata^\prime)}\right|^2\,,
\end{align}
which, since it represents a fidelity-type metric, is referred to as the fidelity quantum kernel (FQK)~\citep{Huang2021}. 

Recent work~\citep{Huang2021, thanasilp2024exponential, Slattery2023} demonstrates that with increasing problem size, FQKs can potentially suffer from \emph{exponential concentration} leading to quantum models that may become untrainable. To alleviate this problem, Ref.~\citep{Huang2021} introduced the family of PQKs which project the quantum states to an approximate classical representation by using, e.g., reduced physical observables. As such, PQKs can be thought of defining features in a classical vector space by taking a detour through a quantum Hilbert space. The result is typically hard to compute due to the quantum detour but still retains desirable properties of the classical feature space. They thus have several appealing properties such as a linear scaling in terms of the needed quantum computing resources compared to FQKs.

A simple way of defining PQKs is based on measuring $k$-particle reduced density matrices ($k$-RDMs),
\revision{
\begin{align}
    \label{eq:kRDM}
    \rho_{\boldsymbol{\theta}}(\xdata)_K &= \mathrm{tr}_{j\notin K}\left[\rho_{\boldsymbol{\theta}}(\xdata)\right]\,,
\end{align}
}
where $K$ is the subset of $k$ qubits from $n$, with $k\leq n$, and $\mathrm{tr}_{j\notin K}$ is the partial trace over all qubits not in subset $K$. The projected quantum circuit results for $\xdata$ and $\xdata^\prime$ are then used as features in some conventional outer kernel $\kappa$ (e.g., RBF, Mat\'ern, etc.). Measuring $k$-RDMs with respect to some observable $O$, corresponds to evaluating
\revision{
\begin{align}
    \label{QNN}
    h_{\boldsymbol{\theta}}(\xdata)_k &= \mathrm{tr}(\rho_{\boldsymbol{\theta}}(\xdata)_K O)\,,\notag \\
    &=\mathrm{tr}(\rho_{\boldsymbol{\theta}}(\xdata)O_{k})\,,
\end{align}
}
where $O_{k}$ represents a $k$-local measurement operator acting on $k\leq n$ qubits, e.g. $P^{\otimes k}\otimes\mathds{1}^{\otimes (n-k)}$, with $P$ a Pauli operator. As such, PQKs can be generally defined as
\revision{
\begin{align}
    \label{eq:PQK-general}
    k^{\text{PQK}}_{\boldsymbol{\theta}}(\xdata,\xdata^\prime) &= \kappa\left[h_{\boldsymbol{\theta}}(\xdata)_k, h_{\boldsymbol{\theta}}(\xdata^\prime)_k\right]\,.
\end{align}
}

The most common PQK definition is based on measuring the 1-RDM on all qubits with respect to all Pauli operators $P\in\lbrace X, Y, Z\rbrace$, i.e.~\citep{Huang2021}
\revision{
\begin{align}
    &k^{\text{PQK}}_{\boldsymbol{\theta}}(\xdata,\xdata^\prime) \nonumber \\ &=\exp(-\gamma\sum_{k,P}\left[\mathrm{tr}(\rho_{\boldsymbol{\theta}}(\xdata)P_{k}) \nonumber - \mathrm{tr}(\rho_{\boldsymbol{\theta}}(\xdata^\prime)P_{k})\right]^2)\notag\\
    &= \exp\left[-\gamma F_{\boldsymbol{\theta}}(\xdata, \xdata^\prime)\right]\,,\label{eq:PQK-std}
\end{align}
}
where $\gamma\in\mathbb{R_+}$ is a hyperparameter. Here, we introduced the latter notation for later use in Sec.~\ref{subsec:Indepth-PQK}. Note that unless otherwise stated, we are referring to the form as given in Eq.~\eqref{eq:PQK-std} when considering PQKs.

\section{\label{sec:research-design}Study Design}
\begin{figure}[tb]
    \includegraphics[width=\columnwidth]{"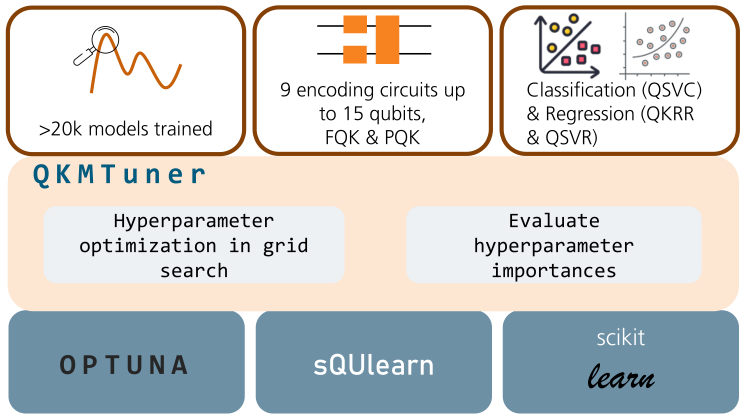"}
    \caption{\label{fig:sketch-implementation}Schematic illustration of the scope of this work and the basic functional principle of our software tool \texttt{QKMTuner}~\citep{schnabel24gitlab} used for the hyperparameter search of QKMs. We thoroughly investigate classification and regression tasks of five different dataset families and $64$ datasets using QSVC as well as QSVR and QKRR, respectively leveraging both FQK and PQK approaches for evaluating the corresponding quantum kernel matrix. Corresponding data are embedded using nine data encoding circuits from the literature with up to 15 qubits. The code is based on the QML library sQUlearn~\citep{kreplin2025squlearn}, the (classical) hyperparameter optimization framework Optuna~\citep{optuna2019} and the (classical) machine learning library scikit-learn~\citep{scikit-learn}.
    }
\end{figure}

In this study, we systematically investigate the interplay of various hyperparameters in QKMs with the aim to identify patterns and mechanisms that enhance model performance. \revision{A schematic view on the scope of this work is shown in Fig.~\ref{fig:sketch-implementation}, sketching the models and the basic functionality of the implementation to realize this comprehensive analysis.}
The following section details the methodological aspects of our study design.

\subsection{\label{subsec:research-design-models}Models}
The amount and range of hyperparameters that are inherently present in quantum kernel models generate a plethora of design choices. The most important hyperparameters are:
\begin{itemize}
    \item \emph{Number of qubits} $n_{\mathrm{qubits}}$  of the underlying encoding circuit used to encode the features of the dataset
    \item \emph{Number of layers} $n_{\mathrm{layers}}$ of the corresponding encoding circuit
    \item The feature range $[f_{\text{min}},f_{\text{max}}]$ used for scaling the dataset's features to respect the gate periodicity of the embedding and thus to prevent information loss. To obtain a single hyperparameter, we define the \emph{width of embedding} in later analyses: 
    \begin{align}
        \label{eq:embedding-width}
        w_e &= f_{\mathrm{max}} - f_{\mathrm{min}}
    \end{align}
    \item $k$-local \emph{measurement operator} $O_{k}$ used for measuring $k$-RDMs as given in Eq.~\eqref{QNN}, which are subsequently used to define PQKs according to Eq.~\eqref{eq:PQK-general}
    \item \emph{Regularization parameters}, these are the Tikhonov regularization strengths $\lambda$ and $C$ for QKRR and QSVR/QSVC, respectively. Additionally, for QSVR there is also the hyperparameter $\varepsilon$, which specifies the range within which no penalty is associated in the training loss function with a point predicted within a distance $\varepsilon$ from the actual value
    \item The functional form of the selected \emph{outer kernel} $\kappa$ in an PQK approach and therein the corresponding \emph{length scale parameter(s)}, e.g., $\gamma$ in case of a Gaussian (RBF) kernel.
\end{itemize}
We point out that data encoding unitaries (cf. Eq.~\eqref{eq:data-encoding}) additionally have multiple degrees of freedom. Data encoding circuits are at the heart of each QML method and significantly influence core properties of the resulting model. Generating problem-specific encoding circuits with, e.g., proper gate sets and corresponding structure marks a distinct research branch, cf., e.g., Refs.~\citep{Altares2021,Rapp2025}. This is beyond the scope of this work, wherefore we restrict ourselves to nine data encoding circuits from the QML literature~\citep{Haug2023,Peters2021,Hubregtsen2022,Kreplin2024reductionoffinite,kreplin2025squlearn,canatar2023,thanasilp2024exponential,qiskit2024}. In this context, we randomly initialize the variationally trainable parameters of those encoding circuits that involve $\boldsymbol{\theta}$ (cf. Eq.~\eqref{eq:data-encoding}) using a fixed seed. For details on these encoding circuits as well as corresponding illustrations, we refer to Appendix~\ref{sec:appendix-encoding-circuits}.

\subsection{\label{subsec:datasets}Datasets}
Selecting meaningful datasets for the sake of a conclusive study constitutes a highly nontrivial task. Therefore, our choice is mainly driven by aiming for datasets that are not too easy and whose complexity is ideally adjustable. Additionally, the size and dimensionality needs to be suited for QML applications with reasonable simulation time. Unless stated otherwise, we use $M=240$ training- and $M^\prime = 60$ test data points within each dataset.

For studying classification, we use two binary classification datasets\footnote{The resulting classification problems are balanced.} introduced in Refs.~\citep{buchanan2021dTwoCurvesDiff, Goldt2020HiddenManifold}. These datasets have been also used in a previous benchmarking study~\citep{bowles2024}. 

{\setlength{\parindent}{0pt}
\bmhead{two curves diff} This dataset describes the curvature and distance of two one-dimensional curves embedded into a $d$-dimensional space. We follow the data generation procedure of Ref.~\citep{bowles2024} and use low-degree ($D$) Fourier series to embed two sets of data sampled from a 1-d interval as curves into $d$ dimensions, while adding Gaussian noise $\sigma = 0.01$. The respective complexity can be controlled by fixing $d=4$ and vary $D=2,\ldots, 20$, while adapting the offset $\Delta = \nicefrac{1}{2D}$.

\bmhead{hidden manifold diff} This dataset is created by generating inputs on a low-dimensional manifold and label them by a simple neural network initialized at random. The inputs are then projected to a final $d$-dimensional space. We use the dataset generation procedure as described in Ref.~\citep{bowles2024} and vary the dimensionality $m$ of the manifold between $m=2,\ldots, 20$ but keep the feature dimension constant at $d=4$.
}
\begin{figure}
    \includegraphics[width=\columnwidth]{"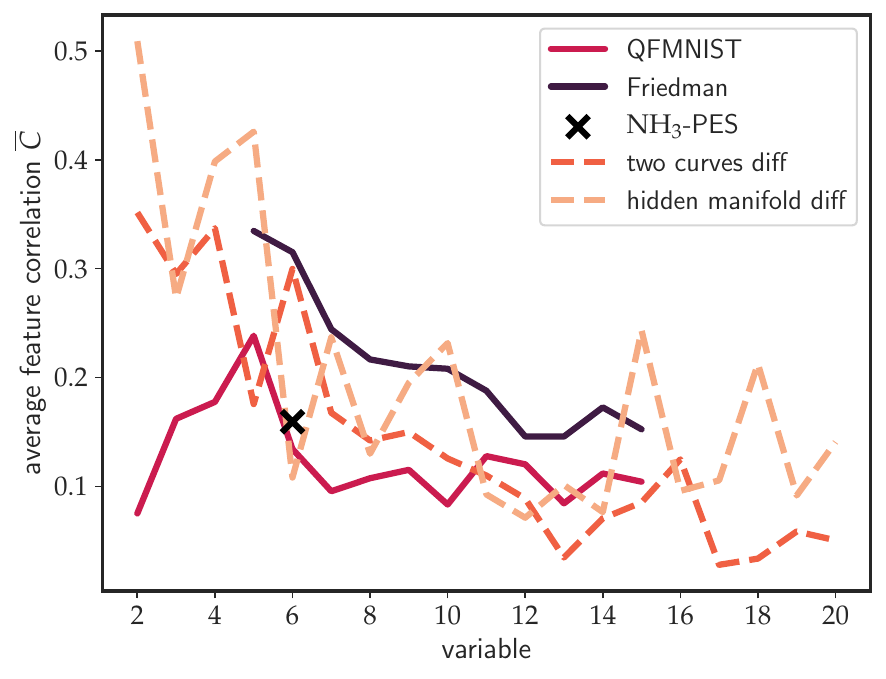"}
    \caption{\label{fig:dataset-complexities} Average \revision{Spearman} correlations $\overline{C}$ of \revision{all} $[0,1]$-normalized features to the outputs to assess the dataset complexity for the datasets considered in this study. The classification datasets depend on variables between $2$ and $20$ that can be seen as controlling the difficulty, while for regression datasets this variable corresponds to the number of features. Note that higher values of this measure indicate simpler problems.}
\end{figure}

The degree $D$ and the dimensionality $m$ of two curves diff and hidden manifold diff, respectively can be seen as control parameters adjusting the dataset complexity. To quantify this complexity, we use the average \revision{of the (absolute value of the) Spearman} correlations $\overline{C}$ of \revision{all} respective $[0,1]$-normalized features to the output as suggested in Ref.~\citep{Lorena2018} and as implemented in the ECoL package~\citep{Lorena2018,Lorena2019}.

The results are displayed in Fig.~\ref{fig:dataset-complexities}. 
Note that the dataset difficulty is inverse to the $\overline{C}$ values, i.e., the lower means more difficult.\revision{For these two classification tasks, we observe increasing complexity with decreasing qubit count.}

For regression, we select the following datasets:
{\setlength{\parindent}{0pt}
\bmhead{Friedman} The Friedman \#1 regression problem is described in Refs.~\citep{Friedman1991,Breiman1996}. The dataset consists of $d \geq 5$ independent features distributed uniformly on the interval $[0,1]$\revision{, with labels generated by a nonlinear function; cf. Appendix~\ref{sec:appendix-details-dataset} for detailed information.} Due to its non-linearity as well as its interaction between features and noise characteristics, this dataset mimics real-world regression scenarios. For this study, we generate datasets for $d=5,\ldots 15$ features.

\bmhead{QFMNIST} This dataset is based on the fashion-MNIST dataset~\citep{xiao2017fashionmnistnovelimagedataset}, which is send through a quantum circuit to create a quantum-based dataset. We use the procedure as described in Ref.~\citep{Jerbi2023}. For this, first the fashion-MNIST feature dimension is reduced by principle component analysis. The resulting $d$ features are then encoded into a quantum state using an $d$-qubit encoding circuit as proposed by Havlicek \emph{et al.}~\citep{Havlicek2019}. The subsequent arbitrary single qubit rotations as applied in Ref.~\citep{Jerbi2023} are omitted. Finally, the labels are generated by computing expectation values w.r.t. measuring the first qubit in the Pauli-$Z$ basis. We generate datasets with $d=2,\ldots 15$ principal components.

\bmhead{\ce{NH3} Potential energy surface (PES)} This real world dataset is taken from \revision{the bechmark database described in} Ref.~\citep{Schmitz2019PESData}. Here, we use the ammonia data labeled with ``STATIC-g32n1-1M'' and ``DZERO'' \revision{and transformed the $xyz$-data to internal coordinates (i.e. bond lengths and angles).} Since nonlinear molecules with $N$ atoms show $f=3N-6$ degrees of freedom, this dataset consists of $d=6$ features with corresponding ground state energies (in Hartee). The dataset consists of 193 samples from which $M=155$ are used for training and $M^\prime=38$ for testing. 
}

For the Friedman\#1 and QFMNIST regression problems the number of features $d$ can be viewed as control parameter to adjust the dataset complexity (cf. Fig.~\ref{fig:dataset-complexities}). \revision{The \ce{NH3}-PES is completely defined by its six degrees of freedom. Making its complexity adjustable would require using dimension reduction techniques, which we consider not meaningful for a four-atomic molecule. Therefore, the \ce{NH3} dataset is only represented as a single point in Fig.~\ref{fig:dataset-complexities}.}

\revision{Within the hyperparameter optimization, we generally found that the computation time increases as the dataset complexity increases for both quantum and classical methods in classification and regression tasks. This is especially noticeable for (quantum) support vector machines for datasets with $\overline{C}<0.1$.}

\subsection{\label{subsec:setup-and-implementation}Experimental setup and Implementation}
In the following, we summarize the experimental setup for the hyperparameter search and provide some insights into the respective implementation.

All simulations in this study are based on sQUlearn~\citep{kreplin2025squlearn} with the PennyLane~\citep{bergholm2022pennylane} statevector simulator device.

{\setlength{\parindent}{0pt}
\bmhead{Design choices of quantum kernels} We use sQUlearn for evaluating quantum kernel Gram matrices, which provides FQKs according to Eq.~\eqref{eq:FQK} and allows for defining PQKs as generally given in Eq.~\eqref{eq:PQK-general}. For PQKs, we investigate the impact of different outer kernel functions $\kappa$. Specifically, we consider the Gaussian (RBF) kernel (cf., Eq.~\eqref{eq:PQK-std}, which is the default in sQUlearn), the Mat\'{e}rn kernel~\citep{GPRBook} with $\nu=\nicefrac{3}{2}$, \revision{cf. Eq.~\eqref{eq:Matern},} and the RationalQuadratic kernel~\citep{KernelCookbook}\revision{, cf. Eq.~\eqref{eq:rationalq}}

Moreover, we study the effect of using different $k$-local measurement operators \revision{$O_{k=w(O)}$}\footnote{\revision{Here we denote $w(O)$ the \emph{weight} of the observable $O$, i.e., the number of qubits on which it acts nontrivially.}} for defining PQKs (cf., Eqs.~\eqref{QNN} and~\eqref{eq:PQK-general}). We use the 1-RDM on all qubits w.r.t. different (combinations of) Pauli operators
\revision{
\begin{equation}
    \begin{aligned}
        \label{eq:1RDM-Ok}
        O_{k=1} &\in \lbrace X_{k=1}, Z_{k=1}, (X_{k=1}+Z_{k=1}), \sum_{P\in\mathcal{P}}P_{k=1}\rbrace\,,
    \end{aligned}
\end{equation}
}
where $\mathcal{P} = \lbrace X,Y,Z\rbrace$ is the set of Pauli operators (cf. Eq.~\eqref{eq:PQK-std}) and $P_{k=1}$ denotes all possible $1$-qubit operators with Pauli operator $P$ of a $n$-qubit system, i.e., 
\revision{
\begin{align}
    P_{k=1} &= \sum_{i=1}^n P_i\,,
\end{align}
where $P_i=\mathds{1}_1\otimes\dots\otimes P_i\otimes \mathds{1}_{i+1}\otimes\dots\otimes\mathds{1}_n$ is the Pauli operator $P$ acting on the $i$-th qubit}. Additionally, we consider measuring the 2-RDM on all qubit combinations w.r.t. different Pauli operator configurations
\revision{
\begin{equation}
    \begin{aligned}
        \label{eq:2RDM-Ok}
        O_{k=2} &\in\lbrace X_{k=2}, Z_{k=2}, (X_{k=2} + Z_{k=2}), \sum_{P\in\mathcal{P}}P_{k=2}\rbrace\,,
    \end{aligned}
\end{equation}
}
where $P_{k=2}$ generally represents all possible permutations of $2$-qubit Pauli measurements from $n$ qubits, i.e.
\revision{
\begin{align}
    P_{k=2} = \sum_{i<j} P_iP_j\,.
\end{align}
}
Finally, we also check for one PQK definition with a combination of 1-RDM and 2-RDM measurements
\begin{align}
    \label{eq:1plus2RDM-Ok}
    P^{1+2} &= \sum_{P\in\mathcal{P}}(P_{k=1} + P_{k=2})\,.
\end{align}

\bmhead{Hyperparameters of quantum kernel methods} The QKMs as implemented in sQUlearn work analogously to their classical counterparts in \emph{scikit-learn}. In this work, we use QKRR and QSVR for regression tasks and QSVC for solving classification problems. The corresponding regularization hyperparameters are $\lambda$, $C$ and $\varepsilon$ as well as $C$, respectively as introduced above. \revision{While QKRR and QSVR leverage the same quantum kernel, they differ in the formulation of their classical kernel method; cf. Appendix~\ref{appendix:KRR-vs-SVR}. Considering both is essential to choose the best approach for a given dataset characteristic.}

\bmhead{\texttt{QKMTuner} for hyperparameter search} We develop the tool \texttt{QKMTuner} to facilitate the extensive hyperparameter search of this study. The code is based on sQUlearn~\citep{kreplin2025squlearn}, Optuna~\citep{optuna2019} and scikit-learn~\citep{scikit-learn}. 
\texttt{QKMTuner} consists of two main routines: a hyperparameter optimization within a grid search and a method for studying hyperparameter importances; cf. Fig.~\ref{fig:sketch-implementation}. \revision{Details on the implementation and how to use \texttt{QKMTuner} are outlined in Appendix~\ref{sec:appendix-qkmtuner}. The code is available via GitLab~\citep{schnabel24gitlab}, where we provide additional documentation.}

\bmhead{Data preprocessing} We scale the dataset features to the range $[f_{\text{min}}, f_{\text{max}}]$ with $f_{\text{min}}\in [-\frac{\pi}{2},0)$ and $f_{\text{max}}\in(0,\frac{\pi}{2}]$ as in Ref.~\citep{bowles2024} and consider $f_{\text{max}}$ ($f_{\text{min}}$), or equivalently $w_e =f_{\text{max}} - f_{\text{min}}$, cf. Eq.~\eqref{eq:embedding-width}, as additional hyperparameters which are optimized within \texttt{QKMTuner}\footnote{This holds for all data encoding unitaries, except for the \texttt{ChebyshevPQC}, which encodes features non-linearly as $\arccos(x)$, wherefore we adapt $\tilde{f}_{\text{min}}\in [-1.0,0)$ and $\tilde{f}_{\text{max}}\in(0,1.0]$}. For all regression datasets we additionally scale the corresponding target values to $[0,1]$. By using pipelines, we ensure that data scaling is part of the cross-validation step.
}

\section{\label{sec:results}Results}

In this section, we report all findings from the different types of experiments that have been performed, before we discuss these results in Sec.~\ref{sec:discussion}.

\subsection{\label{subsec:model-performance}Model Performance}

\begin{figure*}[tb]
    \centering
    \begin{minipage}[t]{.49\textwidth}
        \includegraphics[width=\linewidth]{"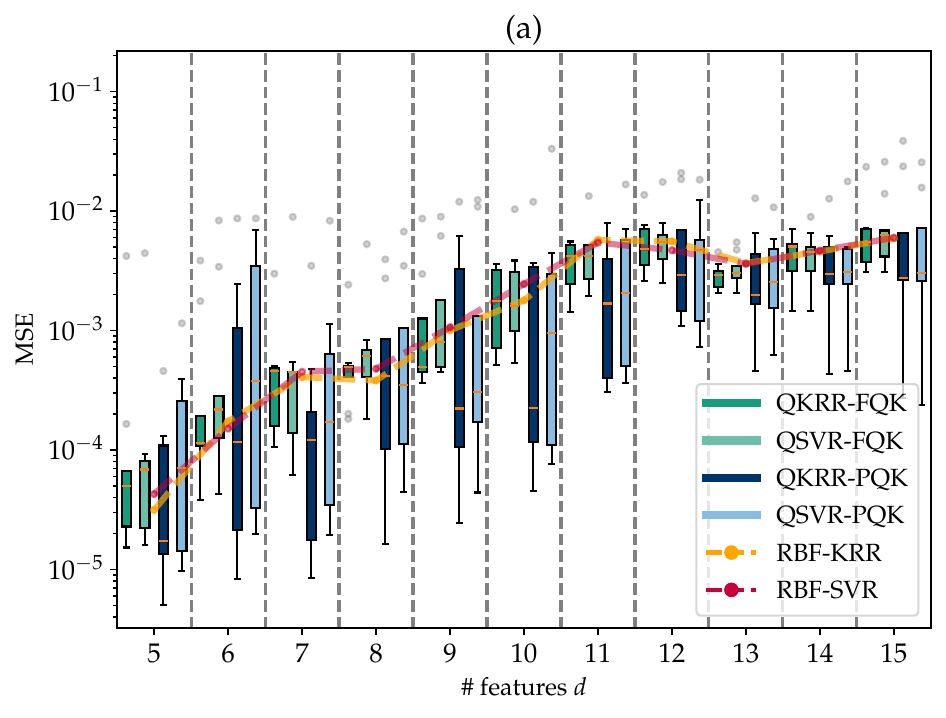"}
    \end{minipage}
    \begin{minipage}[t]{.49\textwidth}
        \includegraphics[width=\linewidth]{"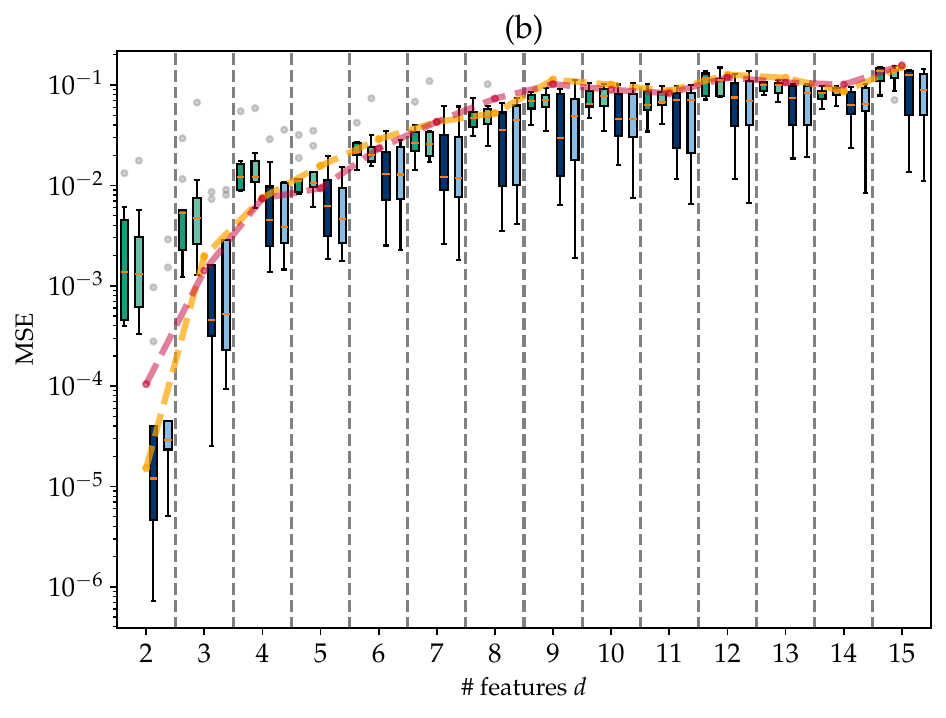"}
    \end{minipage}
    \begin{minipage}[t]{.49\textwidth}
        \includegraphics[width=\linewidth]{"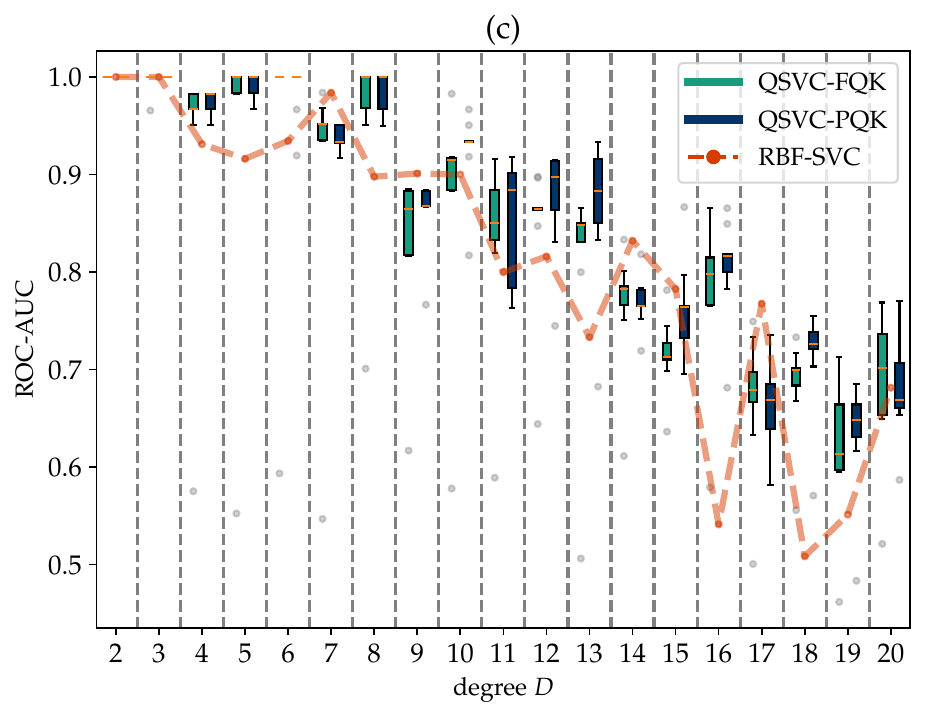"}
    \end{minipage}
    \begin{minipage}[t]{.49\textwidth}
        \includegraphics[width=\linewidth]{"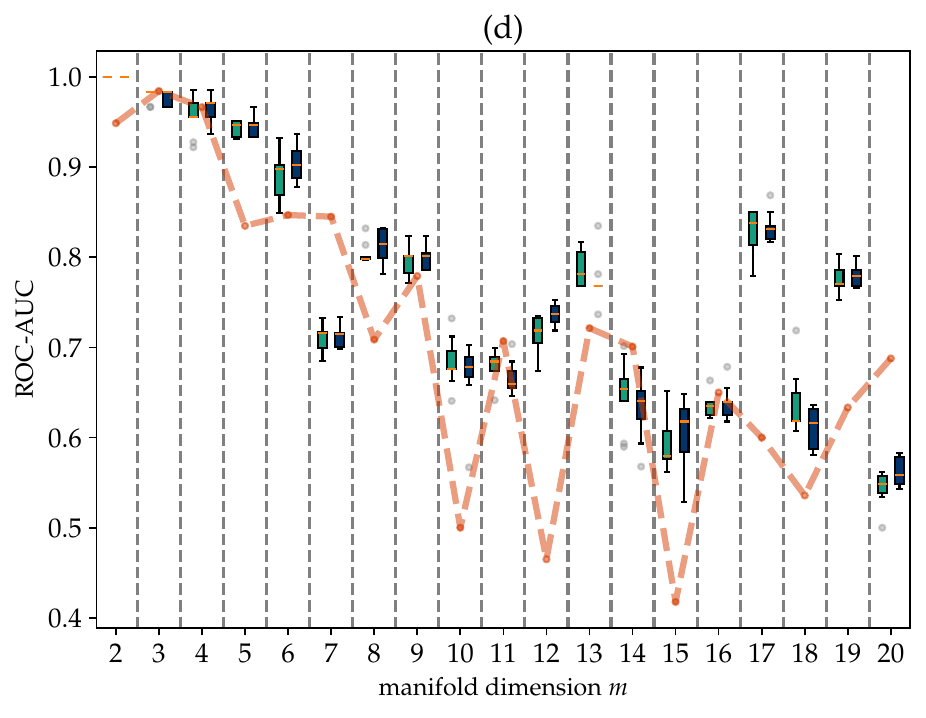"}
    \end{minipage}
    \caption{\label{fig:default-clf-reg-model-performance}Overview of test performance scores of respective QKMs as a function of increasing dataset complexity. Results within each dataset are aggregated across all data encoding circuits with corresponding optimal $n_{\mathrm{layers}}^*$ yielding minimum/maximum test performance scores for regression/classification, respectively. \revision{For comparison, we provide classical KRR/SVR and SVC results each based on a RBF kernel.} The \textbf{upper panel} displays two \textbf{regression} tasks, where the MSE is used to measure the prediction accuracy and the number of features controls the dataset complexity. The \textbf{Friedman} dataset family is shown in~\textbf{(a)}.
    The \textbf{QFMNIST} dataset family is shown in~\textbf{(b)}. The \textbf{lower panel} illustrates the two \textbf{classification} tasks of this study, where we use the ROC-AUC score to assess classification accuracy. In~\textbf{(c)} we show the \textbf{two curves diff} dataset family with the degree $D$ controlling the complexity. The \textbf{hidden manifold diff} family is given in~\textbf{(d)} with the manifold dimension $m$ as respective control parameter.}
\end{figure*}

To start with, we delve into the relationships concerning the generalizability of QKMs. For studying this, we fix $n_{\mathrm{qubits}}$ for every dataset to the respective number of features for all upcoming investigations and perform hyperparameter optimizations for every data encoding circuit with $n_{\mathrm{layers}}\in [1,8]$\,. \revision{To gauge the performance of QKMs, we also provide classical results with KRR/SVR and SVC based on RBF kernels. The corresponding hyperameters were optimized with Optuna using the same parameter ranges as introduced in Sec.~\ref{subsec:setup-and-implementation} for QKMs.}

This analysis of test performances as a function of increasing dataset complexity is given in Fig.~\ref{fig:default-clf-reg-model-performance}, which illustrates the results for the Friedman, QFMNIST, two curves diff, and hidden manifold diff datasets. In these cases, results for each QKM and dataset within the respective dataset family are aggregated across the data encoding circuits with corresponding obtained optimal $n_{\mathrm{layers}}^*$, yielding the best MSE or ROC-AUC score, respectively. 

A comparison with Fig.~\ref{fig:dataset-complexities} indicates that the behavior of test performance scores 
qualitatively aligns with the increasing complexity of the respective datasets. This confirms that $\overline{C}$ is a useful measure of complexity. The comparison between FQK and PQK approaches in Figs.~\ref{fig:default-clf-reg-model-performance}~(a) and~(b), particularly for larger problem instances with increasing $d$, which transfers to increasing $n_{\mathrm{qubits}}$, reveals no significant performance differences. While PQKs demonstrate slightly superior performance, the difference is not as substantial as one might anticipate given the challenges posed in connection with exponential concentration~\citep{thanasilp2024exponential}, which might already be apparent as we approach 15 qubits. The absence of this observation may be explained by the presence of the bandwidth tuning parameter $w_{\mathrm{e}}$, which shows significant correlations with performance scores as discussed below. The comparison between QKRR and QSVR, each equipped with either a FQK or a PQK, does not reveal a clear advantage for either method. Notably, the QFMNIST ($d=2$) dataset constitutes the only case for which we observe a clear performance difference between FQK and PQK. Interestingly, this marks the most complex regression task, cf., Fig.~\ref{fig:dataset-complexities}. Whether this is merely a coincidence or actually due to a more fundamental mechanism might be interesting to investigate in the future. Regarding the classification tasks, we do not observe any significant difference between both QKMs.

We generally observe some outliers exhibiting particularly poor test performance scores. The comparison with associated training performance scores in Fig.~\ref{fig:default-clf-reg-train-performance} shows that this is probably a result of overfitting for certain model combinations. Since we investigate a large variety of combinations, it is unavoidable that some of the combinations result in models that are too expressive for some tasks. As shown in Appendix~\ref{subsec:model-performance-correlation-appendix} this is most dominant for QKRR-FQK/PQK approaches with increasing dataset complexity, while QSVR appears to be more robust. 

\revision{The classical baselines follow the general trend of QKMs for the regression tasks. In the classification examples, the RBF-SVC results align with the overall difficulty trends of both problems and face challenges with the same datasets as the quantum models. In some instances, classical models outperform QKMs, but they perform poorly in the hidden manifold diff family, with ROC-AUC values below 0.5. This is due to significant overfitting in SVC models, as noted in Ref. \citep{bowles2024}, with overfitting sometimes more severe in classical baselines. This may be due to dataset size limitations or using default classical models. Future studies should explore classical simulability \citep{Slattery2023} and the impact of providing more data \citep{Huang2021}, to assess requirements for quantum advantage and associated geometric differences.}

\subsection{\label{subsec:influence-hyperparameters}Influence of Hyperparameters}

\begin{figure}[h!]
    \centering
    \includegraphics[width=\linewidth]{"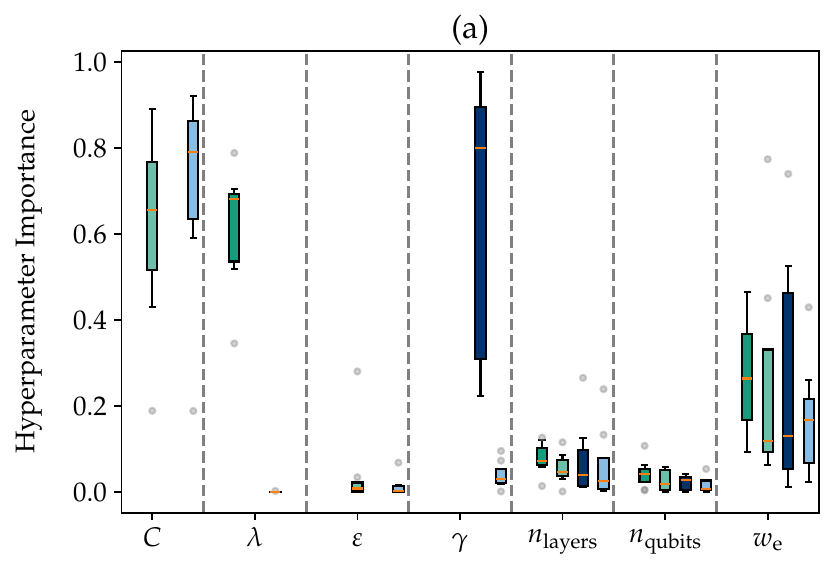"}
    \includegraphics[width=\linewidth]{"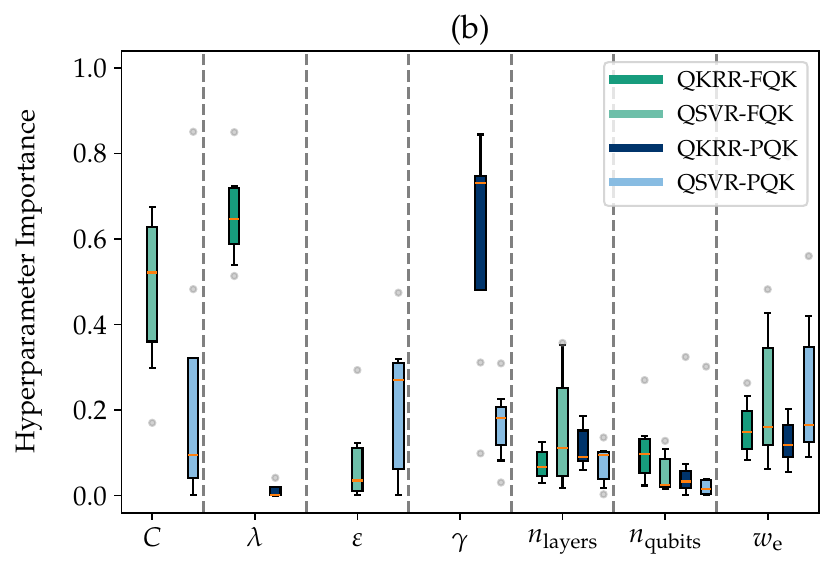"}
    \includegraphics[width=\linewidth]{"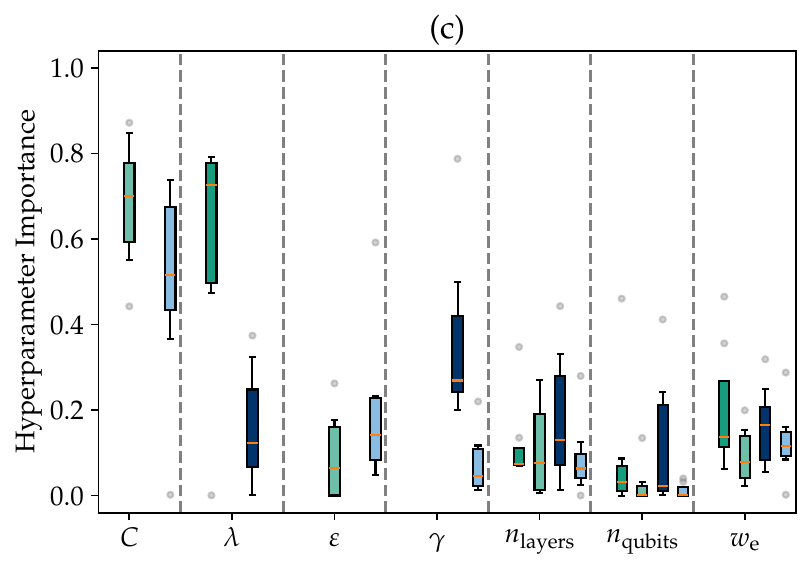"}
    \caption{\label{fig:hyperparam-importance-reg}Comparison of hyperparameter importances for optimizing the five-fold cross-validation score in the corresponding hyperparameter searches for the regression tasks of this study. \revision{We chose half the interquartile range to define the whisker  length.} The \textbf{Friedman} dataset with $d=5$ features is shown in~\textbf{(a)}, the \textbf{QFMNIST} results with $d=5$ components are illustrated in~\textbf{(b)}, and the \textbf{\ce{NH3}-PES} data are given in~\textbf{(c)}. The results for each model and dataset are aggregated over different encoding circuits in each case. Here, we always impose that $n_{\text{qubits}}$ can only be integer multiples of the number of features present in the respective dataset, with a maximum of $n_{\text{qubits}}^{\text{max}}=15$.}
\end{figure}

\begin{figure*}[tb]
    \centering
    \begin{minipage}[t]{.49\textwidth}
        \includegraphics[width=\linewidth]{"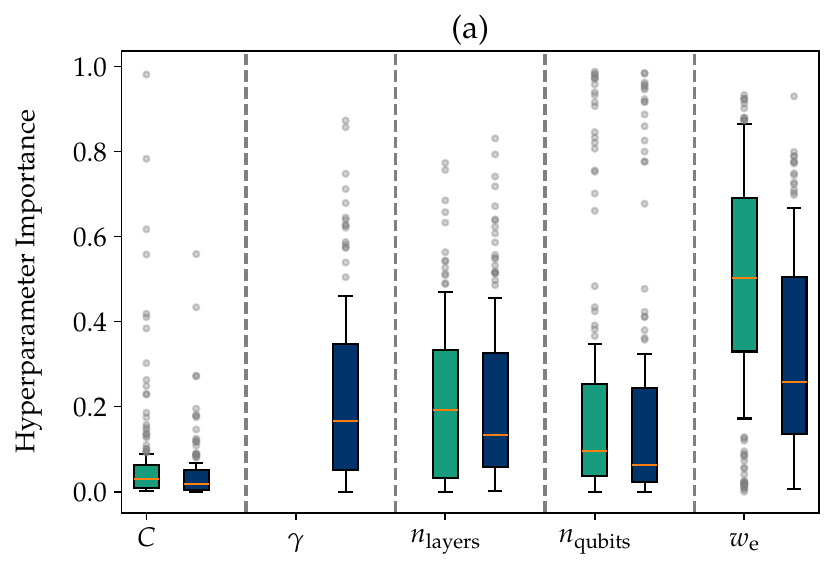"}
    \end{minipage}
    \begin{minipage}[t]{.49\textwidth}
        \includegraphics[width=\linewidth]{"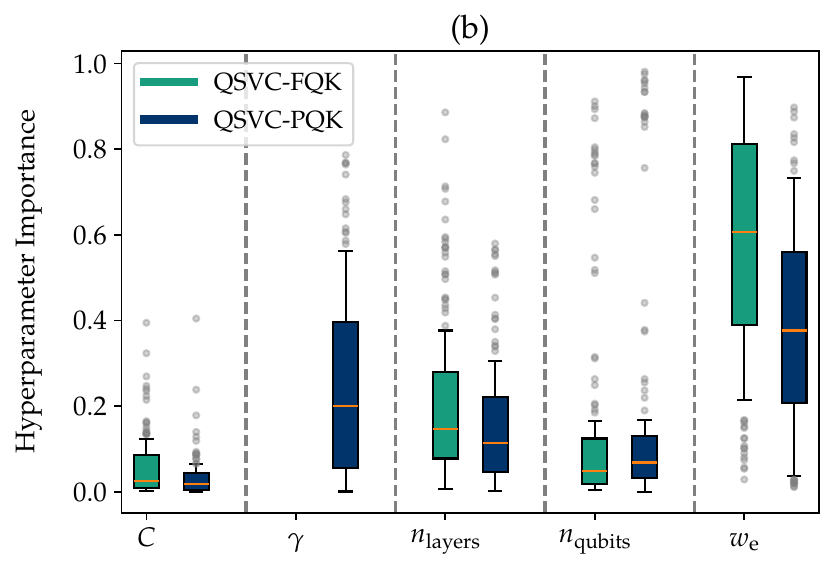"}
    \end{minipage}
    \caption{\label{fig:hyperparam-importance}Comparison of hyperparameter importances for optimizing the five-fold cross-validation score in the corresponding hyperparameter searches for the classification tasks of this study. The results for each model are aggregated across all datasets within the correlating dataset family and all encoding circuits. \revision{We chose half the interquartile range to define the whisker  length.} The results corresponding to the \textbf{two curves diff} family are shown in~\textbf{(a)}. The \textbf{hidden manifold diff} dataset family is depicted in~\textbf{(b)}. In both cases we impose that $n_{\mathrm{qubits}}$ can only be integer multiples of the number of features ($d=4$), with a maximum of $n_{\text{qubits}}^{\text{max}}=15$.}
\end{figure*}

To gain a deeper understanding of model performance, it is essential to study the influence of respective hyperparameters. In this regard, the analysis of the importance of individual hyperparameters for optimizing 
the five-fold cross-validation scores helps with effective model tuning. Moreover, knowing which hyperparameters are more influential provides insights into how the model behaves and can reveal how robust it is to small changes. We compute these hyperparameter importances as implemented in Optuna~\citep{hutter2014}. Here, the sum of the individual importance values is normalized to one and higher values imply that the associated parameters are more important. For the regression tasks, we analyze the \ce{NH3}-PES dataset as well as the Friedman and QFMNIST datasets with five features, respectively. For classification, we consider all datasets within the two curves diff and hidden manifold diff families, i.e., we consider $D, m=2,\ldots,20$. In both cases we investigate various model configurations in terms of encoding circuits, kernel type and QKM. Here, we always impose that $n_{\mathrm{qubits}}$ can only take values that are integer multiples of the number of features of the respective dataset with a maximum of $n_{\text{qubits}}^{\text{max}}=15$.

The hyperparameter importances for the regression datasets, aggregated for each model across the different encoding circuits, are shown in Fig.~\ref{fig:hyperparam-importance-reg}. By comparing the results from~(a) to~(c), we observe some similar trends. In all but one case, the regularization parameters $\lambda$ (QKRR) and $C$ (QSVR), respectively appear to be most important. Only for QKRR-PQK, the length scale parameter $\gamma$ of the Gaussian outer kernel, cf. Eq.~\eqref{eq:PQK-std}, shows largest importance. The feature scaling $w_{\mathrm{e}}$ is similarly important across all datasets and models, which can be directly related to the concept of bandwidth tuning~\citep{Shaydulin2022, canatar2023}. Beyond that it is worth noting that $n_{\mathrm{layers}}$ and $n_{\mathrm{qubits}}$ of the underlying data encoding circuits mostly represent the least important parameters across all QKMs. This might be due to the comparatively low dataset complexity of the respective datasets ($d=5$ for QFMNIST and Friedman and \ce{NH3} with $d=6$), cf. Fig.~\ref{fig:dataset-complexities}. Thus, large model expressivity as controlled by the number of qubits and layers might not be required. 

For classifications tasks, Fig.~\ref{fig:hyperparam-importance} displays the hyperparameter importances. Here, for both FQK and PQK QKMs, we aggregate results across all corresponding data subsets and encoding circuits. The results clearly reveal that $w_{\mathrm{e}}$ is most important in order to obtain well trained quantum classifiers, which is in contrast to the regression results. Again, \revision{this} can be directly related to the concept of kernel bandwidth tuning and is in accordance with the results presented therein. Moreover, it perfectly agrees with the recent outcomes of Ref.~\citep{egginger2024hyperparameter}. In addition to that, we provide some evidence that the importance of feature prescaling is slightly more pronounced for FQKs than for PQKs. This, however, is not surprising, given the fact that the $\gamma$-parameter in PQKs of the form of Eq.~\eqref{eq:PQK-std} acts as an additional (classical) bandwidth tuning parameter. Here, $\gamma$ is the second most important parameter for PQK models. The regularization parameter $C$ of QSVC appears to be comparatively unimportant, except for some outliers. This is even more pronounced for PQKs than for FQKs. In comparison to Fig.~\ref{fig:hyperparam-importance-reg}, the results in Fig.~\ref{fig:hyperparam-importance} clearly indicate that $n_{\text{qubits}}$ and $n_{\text{layers}}$ appear to be more important, with tuning the number of layers being slightly more important than potentially encoding features redundantly on the number of qubits. This may be due to the increased dataset complexity for $D,m\geq 6$, cf. Fig.~\ref{fig:dataset-complexities}, which might require more expressive models. This can be achieved by introducing feature redundancies and adjusting the number of layers accordingly~\citep{schuld2021supervised,Schuld2021}.

Nevertheless, given the relatively low median values for the hyperparameter importance of $n_{\mathrm{qubits}}$ in both Figs.~\ref{fig:hyperparam-importance-reg} and~\ref{fig:hyperparam-importance}, we \emph{a posteriori} justify our previous choice of fixing this parameter in the respective investigations of model performance.

\begin{figure*}[tb]
    \centering
    \begin{minipage}[b]{0.23\linewidth}
        \includegraphics[width=\linewidth]{"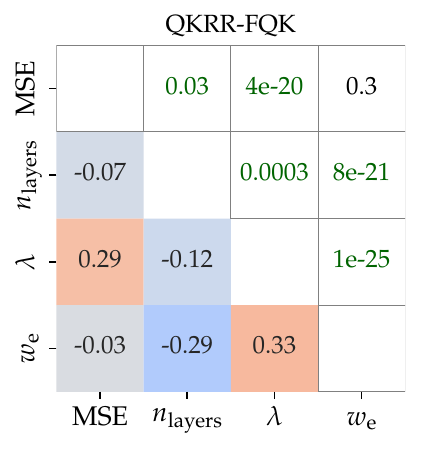"}
    \end{minipage}
    \hfill
    \begin{minipage}[b]{0.23\linewidth}
        \includegraphics[width=\linewidth]{"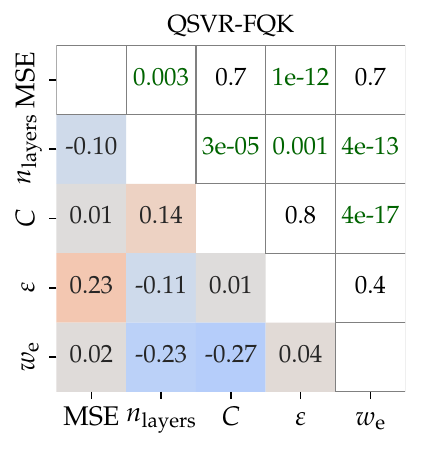"}
    \end{minipage}
    \hfill
    \begin{minipage}[b]{0.23\linewidth}
        \includegraphics[width=\linewidth]{"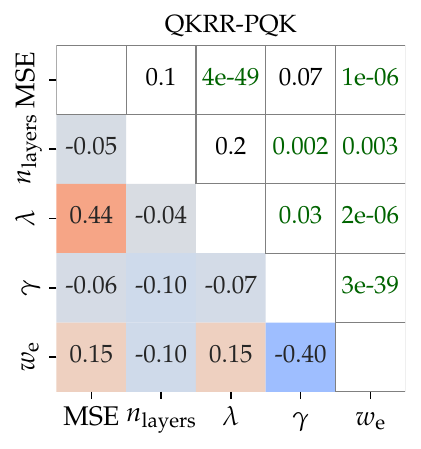"}
    \end{minipage}
    \hfill
    \begin{minipage}[b]{0.23\linewidth}
        \includegraphics[width=\linewidth]{"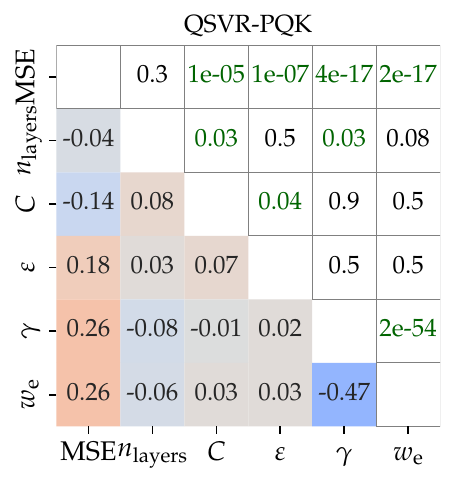"}
    \end{minipage}
    \caption{\label{fig:default-clf-reg-correlations}Overview of the Spearman correlation analyses between QKM hyperparameters and test performance scores, as well as between the hyperparameters themselves, for the QFMNIST dataset family. The results correspond to aggregating across all datasets, encoding circuits, and $n_{\mathrm{layers}}\in [1,8]$. The Spearman correlation coefficients are given on the lower triangles of the respective matrices (with blue for negative to red for positive coefficients), while the upper triangles display corresponding $p$-values. Here, statistically significat correlations ($p\leq 0.05$) are highlighted in green.
    }
\end{figure*}
To obtain even deeper insights into the influence of hyperparameters, Fig.~\ref{fig:default-clf-reg-correlations} provides an exemplary overview on the results of the correlation analyses between QKM hyperparameters and test performance scores, as well as between the hyperparameters themselves for the QFMNIST dataset family. We refer to Fig.~\ref{fig:default-clf-reg-correlations-appendix} in Appendix~\ref{subsec:default-correlation-analysis-appendix} for a summary of all other dataset families investigated in this study. The results are aggregated across datasets within a dataset family and across all nine encoding circuits and $n_{\mathrm{layers}}\in [1,8]$, respectively. We calculate the Spearman coefficient \revision{to measure correlaiton.} Figure~\ref{fig:default-clf-reg-correlations} indicates statistically significant correlations by green-highlighted $p$-values. In line with the aforementioned findings from the analysis of the importance of hyperparameter, we observe moderate to strong statistically significant correlations between test performance scores and the various regularization parameters, i.e., $\lambda$ and $C$ for QKRR and QSVR/QSVC, respectively as well as $\varepsilon$ for QSVR. Additionally, consistent with findings on bandwidth tuning~\citep{Shaydulin2022,canatar2023}, we frequently observe moderate and statistically significant correlations between the feature scaling parameter $w_{\mathrm{e}}$, cf. Eq.~\eqref{eq:embedding-width}, and performance scores (also considering the findings from Appendix~\ref{subsec:default-correlation-analysis-appendix} in Fig.~\ref{fig:default-clf-reg-correlations-appendix}). This parameter often also correlates with the $n_{\mathrm{layers}}$. For QSVR-/QSVC-PQK models, moderate correlations are also observed between the length-scale parameter $\gamma$ of the Gaussian outer kernel and test performance. Finally, we note that the $n_{\mathrm{layers}}$ parameter only shows weak statistically significant correlations to the test performance for the QFMNIST dataset.

As mentioned above, the correlation analysis provides an explanation for the poor performance regimes observed in Fig.~\ref{fig:default-clf-reg-model-performance}. We interpret the strong correlations between regularization parameters and model performance as indicative of ill-conditioned training kernel Gram matrices. Furthermore, it corroborates the observed tendency to overfitting of some models. 

\subsection{\label{subsec:influence-encoding-circs}Influence of Encoding Circuits}

\begin{figure*}[tb]
    \centering
    \begin{minipage}[t]{.49\textwidth}
        \includegraphics[width=\linewidth]{"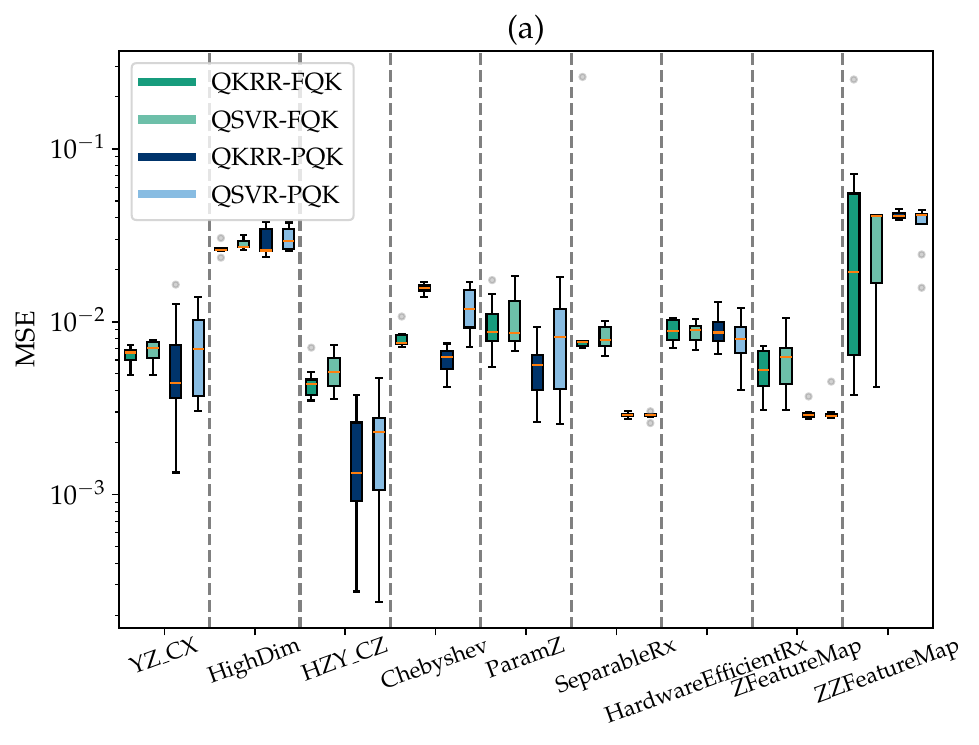"}
    \end{minipage}
    \begin{minipage}[t]{.49\textwidth}
        \includegraphics[width=\linewidth]{"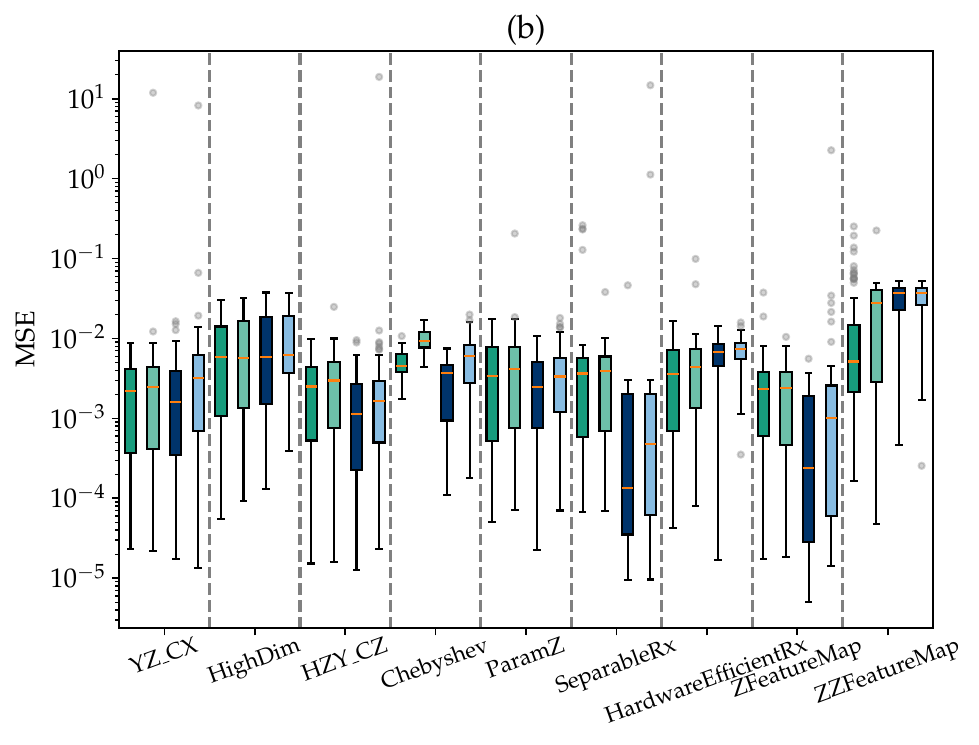"}
    \end{minipage}
    \begin{minipage}[t]{.49\textwidth}
        \includegraphics[width=\linewidth]{"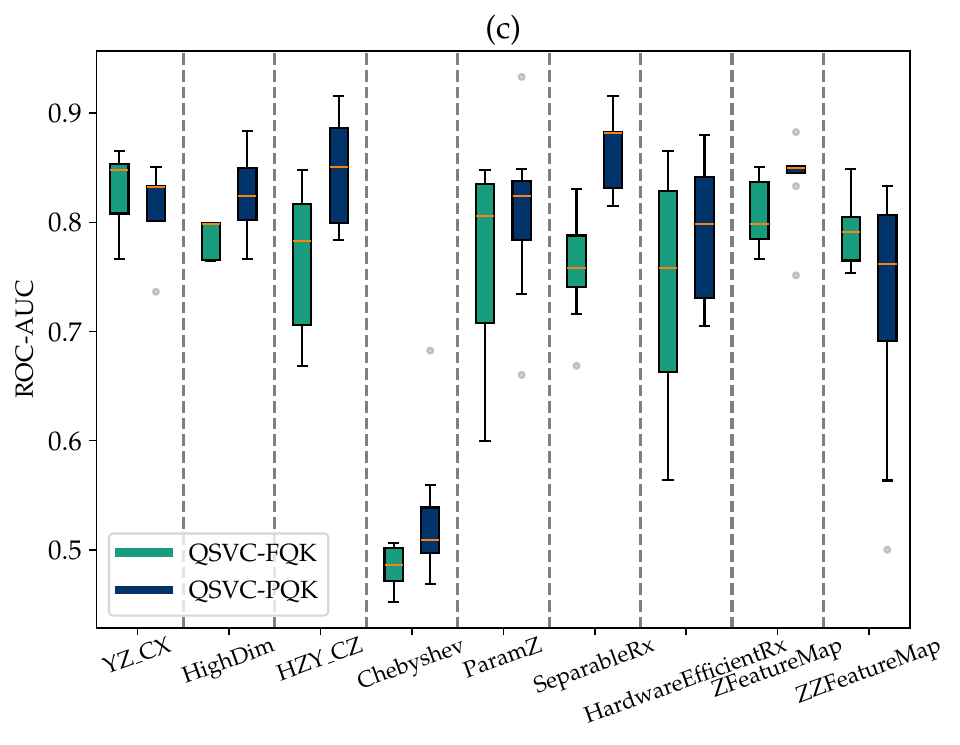"}
    \end{minipage}
    \begin{minipage}[t]{.49\textwidth}
        \includegraphics[width=\linewidth]{"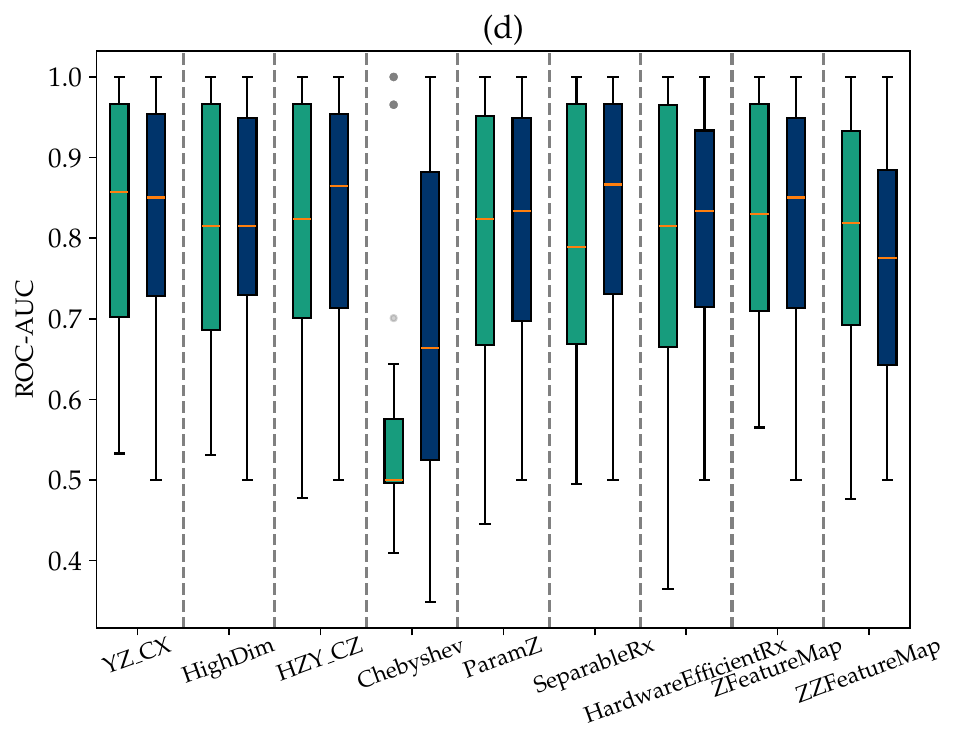"}
    \end{minipage}
    \caption{\label{fig:default-reg-clf-encoding-circs}Impact of data encoding circuits on test performance scores. Results are aggregated across all datasets within the respective family and $n_{\mathrm{layers}}\in [1,8]$. The \textbf{upper panel} exemplifies this for the \textbf{Friedman} regression problem. In~\textbf{(a)} we show results corresponding to the dataset with $d=15$ features. Here, test performance scores, as measured by MSE, are aggregated over $n_{\mathrm{layers}}\in[1,8]$ per regression method (QSVR/QKRR) and quantum kernel (FQK/PQK). The findings in~\textbf{(b)} represent test performance scores aggregated over all datasets in the Friedman family. The \textbf{lower panel} details the \textbf{two curves diff} dataset. In~\textbf{(c)}, we display test performances, as measured by ROC-AUC score, aggregated over $n_{\mathrm{layers}}\in[1,8]$ for the dataset with degree $D=13$. Aggregated results across all datasets are depicted in~\textbf{(d)}.}
\end{figure*}
To investigate the impact of data encoding circuits on the performance of the resulting models, we aggregate results across all datasets within each family and choose $n_{\mathrm{layers}}\in  [1,8]$\,. \revision{The results per encoding circuit are shown} in Fig.~\ref{fig:default-reg-clf-encoding-circs} for the Friedman and the two curves diff datasets. The results of the other dataset families of this study are shown in Fig.~\ref{fig:default-reg-clf-encoding-circs-appendix} of the Appendix~\ref{subsec:appendix-influence-encoding-circs}. When examining the performance scores for a single dataset within a dataset family, significant performance differences among the data encoding circuits emerge. This can be seen in Figs.~\ref{fig:default-reg-clf-encoding-circs}~(a) and~(c). Here, the ``HighDimEncodingCircuit'' and the ``ZZFeatureMap'' perform particularly poor in regression tasks, while the ``ChebyshevPQC encoding circuit'' underperforms in classification tasks, both for FQK and PQK approaches (note that trends for MSE and ROC-AUC performance scores work in opposite directions). However, when considering performance scores aggregated across all datasets within a dataset family, we observe that, except for the ``ChebyshevPQC'' in the QSVC-FQK case and with some trend for the ``ZZFeatureMap'' in regression tasks, all encoding circuits perform comparably in terms of overall performance and variance (cf. Figs.~\ref{fig:default-reg-clf-encoding-circs}~(b) and~(d)). This supports arguments for problem-specific data encoding architecture search~\citep{Altares2021,Rapp2025}. 

In earlier studies~\citep{bowles2024}, it has been found that data encoding circuits without entanglement can perform surprisingly well. In our case, these are the ``SeparableRx'' and ``ZFeatureMap'' encoding circuits. 
With \revision{associated} model performances obtained in Fig.~\ref{fig:default-reg-clf-encoding-circs}, this raises the possibility that driving forces other than ``quantumness'' might be fostering the performance of QKMs. This observation underscores the need for further comprehensive and systematic investigations into the underlying factors contributing to QKM performance.

\revision{In Appendix~\ref{subsec:appendix-data-enc}, we additionally investigate the effect of different data encoding strategies to distribute the features of a data point among the available qubits in a particular encoding circuit. Here, we find that depending on the dataset complexity it can be beneficial to redundantly encode all or merely certain features, respectively. Overall, the observed effects are, however, small compared to other factors such as adequate hyperparameter tuning.}

\subsection{\label{subsec:Indepth-PQK}Analysis of PQK Design Options}

\begin{table*}[tb]
    \centering
    \caption{\label{tab:correlation-indepth-PQK}\emph{Semi-partial} correlation analysis results for exploring the effect of the choice of the outer kernel function $\kappa$ (cf.~\eqref{eq:PQK-general}) and the measurement operator $O_{k}$ (cf. Eqs.~\eqref{eq:1RDM-Ok},~\eqref{eq:2RDM-Ok},~\eqref{eq:1plus2RDM-Ok}) in the definition of a PQK on the resulting test performance scores, while controlling for the effect of all other hyperparameters on the test performance. We investigate PQK-QSVC and PQK-QSVR models for classification and regression tasks, respectively. We compute ROC-AUC scores for evaluating classification accuracy and MSEs for assessing regression results. We employ Spearman $\rho$ rank-order correlation analysis as implemented in the \texttt{pingouin}~\citep{Vallat2018} package. Bold values refer to \textbf{statistically significant correlations}, with $p$-values$\leq 0.05$, while others refer to \emph{not} statistically significant correlation, which are given for the sake of completeness. \emph{We note that while negative correlation coefficients w.r.t. the MSE point at overall better model performance, it is the other way round for correlations with ROC-AUC scores.}} 
    \resizebox{\textwidth}{!}{
        \begin{tabular}{l@{\hskip 10pt}SSSSSSSSSSSS}
            \toprule\toprule
            \multirow{2}{*}{dataset} & \multicolumn{3}{c}{$\rho_{\mathrm{Spearman}}^\kappa$} & \multicolumn{9}{c}{$\rho_{\mathrm{Spearman}}^{O_k}$} \\
            \cmidrule(r){2-4} \cmidrule(r){5-13}
            & \multicolumn{1}{c}{$\kappa^{\text{Gauss}}$} & \multicolumn{1}{c}{$\kappa_{3/2}^{\text{Mat}}$} & \multicolumn{1}{c}{$\kappa^{\text{RQ}}$} & \multicolumn{1}{c}{$X_{k=1}$} & \multicolumn{1}{c}{$Z_{k=1}$} & \multicolumn{1}{c}{$X_{k=1}+Z_{k=1}$} & \multicolumn{1}{c}{$\sum_P P_{k=1}$} & \multicolumn{1}{c}{$X_{k=2}$} & \multicolumn{1}{c}{$Z_{k=2}$} & \multicolumn{1}{c}{$X_{k=2}+Z_{k=2}$} & \multicolumn{1}{c}{$\sum_P P_{k=2}$} & \multicolumn{1}{c}{$P^{1+2}$} \\
            \midrule
            {\emph{correlations w.r.t. MSE}} & ~ & ~ & ~ & ~ & ~ & ~ & ~ & ~ & ~ & ~ & ~ & ~ \\
            {\hspace*{.2mm} Friedman $(d=10)$} & {-0.063} & {0.126} & {-0.068} & {\textbf{0.237}} & {\textbf{0.183}} & {-0.031} & {\textbf{-0.220}} & {\textbf{0.187}} & {0.087} & {0.012} & {-0.102} & {\textbf{-0.352}} \\
            {\hspace*{.2mm} QFMNIST $(d=8)$} & {\textbf{0.129}} & {-0.09} & {0.045} & {\textbf{0.129}} & {0.107} & {-0.086} & {-0.097} & {0.113} & {0.057} & {-0.040} & {-0.051} & {\textbf{-0.139}} \\
            {\hspace*{.2mm} \ce{NH3}-PES} & {\textbf{-0.265}} & {\textbf{0.261}} & {-0.036} & {\textbf{0.245}} & {0.049} & {-0.040} & {-0.084} & {0.106} & {-0.022} & {0.056} & {-0.125} & {\textbf{-0.203}} \\
            {\emph{correlations w.r.t. ROC-AUC}} & ~ & ~ & ~ & ~ & ~ & ~ & ~ & ~ & ~ & ~ & ~ & ~ \\
            {\hspace*{.2mm} two curves diff $(D=13)$} & {\textbf{-0.297}} & {\textbf{0.394}} & {\textbf{-0.115}} & {\textbf{-0.340}} & {-0.125} & {0.038} & {\textbf{0.191}} & {\textbf{-0.265}} & {0.020} & {0.012} & {\textbf{0.210}} & {\textbf{0.256}} \\
            {\hspace*{.2mm} hidden manifold diff $(m=13)$} & {\textbf{0.188}} & {0.038} & {\textbf{-0.215}} & {\textbf{-0.149}} & {-0.020} & {-0.007} & {-0.077} & {-0.086} & {0.036} & {0.025} & {\textbf{0.166}} & {0.113} \\
            \bottomrule\bottomrule
        \end{tabular}
    }
\end{table*}

Projected quantum kernels as defined in Eq.~\eqref{eq:PQK-general} have additional degrees of freedom which can strongly influence the performance of the resulting model. Specifically, any proper outer kernel function $\kappa$ can be applied and the measurement operator $O_{k}$ can be freely selected. To the best of our knowledge, a thorough investigation of the influence of both of these design choices on the model is still lacking.\footnote{Reference~\citep{egginger2024hyperparameter} touches upon similar aspects but has a much narrower focus on the impact of the RDM size on cross-validation accuracy and generalization ability.}

In the following, we systematically examine the impact of Gaussian, Mat\'ern, and RationalQuadratic outer kernel functions $\kappa$, as well as the effects of different measurement operators as defined in Eqs.~\eqref{eq:1RDM-Ok},~\eqref{eq:2RDM-Ok}, and~\eqref{eq:1plus2RDM-Ok} on the model performance. Here we only consider specific instances of the dataset families, i.e., $d=10,8,6$ for the Friedman, QFMNIST and the \ce{NH3}-PES dataset, and $D=m=13$ with $d=4$ features for the two curves diff and hidden manifold diff datasets. We note that although redundant encoding of some or all features may be beneficial, as shown in the previous sections, we restrict ourselves in the following to a number of qubits equal to the number of features in each respective dataset for simplicity. This approach also seems appropriate for answering our research questions.

To quantify the influence of the respective choices on the model performance, while taking into account the corresponding impact of all other hyperparameters, 
we aggregate all simulation results per dataset and conduct (semi-) partial correlation analyses (cf. Appendix.~\ref{sec:appendix-statistis} for details). The results are summarized in Tab.~\ref{tab:correlation-indepth-PQK}. Note that while negative correlation coefficients w.r.t. the MSE point at overall better model performance, the opposite is the case for correlations with ROC-AUC scores. An overview of the model performances used for calculating the values in the table is provided in Appendix~\ref{subsec:appendix-indepth-pqk} in Figs.~\ref{fig:boxplots-indepth-pqk-reg} and~\ref{fig:boxplots-indepth-pqk-clf}. From the table we directly realize that both $\kappa$ \emph{and} the choice of $O_{k}$ for defining the respective projected quantum circuits exhibit statistically significant correlations. For the outer kernel, no universal statement can be made \emph{a priori} regarding which one is generally suitable for effective QKM design---at least not within the scope of our study. The only exception is that for the classification tasks considered here, the RationalQuadratic kernel is not a good choice. Furthermore, it is worth noting that the commonly used Gaussian kernel is not automatically the best choice.

Regarding a proper choice of the measurement operator $O_{k}$, Tab.~\ref{tab:correlation-indepth-PQK} suggests $X_{k=1}$ to be insufficient for all regression and classification tasks considered here. In this case, we frequently observed training kernel Gram matrices entirely filled with ones, thus rendering them impractical for learning. In contrast, utilizing the $P^{1+2}$ measurement operator, cf. Eq.~\eqref{eq:1plus2RDM-Ok}, points towards an overall improvement in model performance (within the scope of this study).

\begin{figure*}[tb]
    \centering
    \includegraphics[width=\textwidth]{"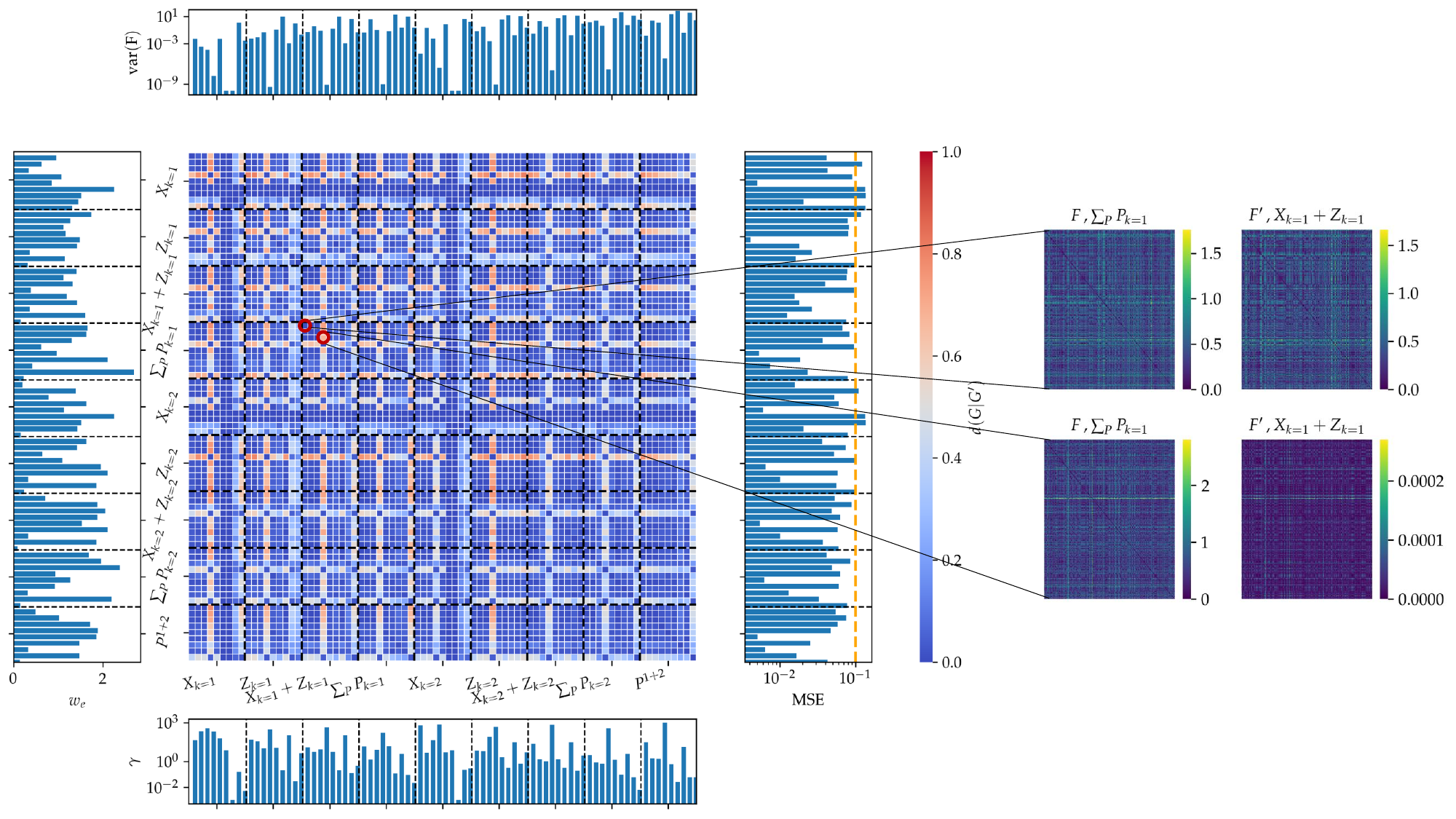"}
    \caption{\label{fig:indepth-pqk-qfmnist}Comprehensive insight into various mechanisms of PQKs exemplified for the QFMNIST dataset with $d=8$ principal components and Gaussian outer kernel function. The central heatmap plot illustrates the $d(G|G^\prime)$ distance measure according to Eq.~\eqref{diff-gram-matrices}, where each section correpsonds to a family of projected quantum circuits as defined by a given measurement operator $O_{k}$. The sections are further subdivided into the different encoding circuits in the order as they appear in Appendix~\ref{sec:appendix-encoding-circuits}. The plot at the top details $\mathrm{Var}(F)$; cf. Eq.~\eqref{eq:F}. The close-up images on the very right correspond to examples of $F$ and $F^\prime$ that lead to $d(G|G^\prime)\to 0$ and $d(G|G^\prime)\neq 0$, respectively. The heatmap visualization of $F$ can be viewed as illustrations of embeddings into the corresponding Hilbert space. The histogram of MSEs right to the central heatmap helps to identify Gram matrices that lead to good/bad performances (with bad defined as MSE $>0.1$, indicated by orange dashed line), respectively. The $w_{\mathrm{e}}$- and $\gamma$-plots on the left and on the bottom, respectively helps to understand the machanisms which render these cases into the weak-performing regime. Here, $w_{\mathrm{e}}$ becomes comparatively small, while $\gamma$ becomes relatively large.}
\end{figure*}

In Ref.~\citep{bowles2024} the authors have raised the question whether the projected quantum circuit is not so much responsible for learning, but rather the subsequent outer (Gaussian) kernel applied to the features computed by Eq.~\eqref{QNN}. Referring to Tab.~\ref{tab:correlation-indepth-PQK} per dataset, it becomes evident that both the choice of $O_{k}$ (and thus the projected quantum circuit itself), and the outer kernel $\kappa$ are crucial for achieving sufficiently good model performance. Therefore, the question of which factor is more important cannot be definitively answered based on this analysis alone. Consequently, we further list all (statistically significant) correlations of the various hyperparameters of the PQK QKMs for a given outer kernel, both among themselves and with model performance in Appendix~\ref{subsec:appendix-indepth-pqk} in Tab.~\ref{tab:full-correlation-indepth-PQK}. Thereof, it again becomes clear that depending on the dataset and outer kernel function, both hyperparameters that contribute to the definition of the projected quantum circuits (i.e., $n_{\mathrm{layers}}$ and $w_{\mathrm{e}}$) and those solely associated with the outer kernel exhibit statistically significant correlations.

To better understand these findings we quantify the distance between Gram matrices $G$ with normalized entries $G_{ij}=k(\mathbf{x},\mathbf{x}^\prime)$~\citep{bowles2024}
\begin{align}
    \label{diff-gram-matrices}
    d(G|G^\prime) &= \frac{\sum_{ij}\left(G_{ij} - G_{ij}^\prime\right)^2}{|G|}\,,
\end{align}
where $|G|$ refers to the number of entries in $G$. For Gaussian outer kernels it directly follows from Eq.~\eqref{eq:PQK-std} that the projected quantum circuit contribution $F_{\boldsymbol{\theta}}(\mathbf{x},\mathbf{x}^\prime)$ becomes
\begin{align}
    \label{eq:F}
    F_{\boldsymbol{\theta}}(\mathbf{x},\mathbf{x}^\prime) &= -\frac{\log[k_{\mathbf{\theta}}^{\text{PQK}}(\mathbf{x},\mathbf{x}^\prime)]}{\gamma}\,.
\end{align}
Thus, calculating $d(G|G^\prime)$ also defines a difference measure between projected quantum circuit contributions $F$ and $F^\prime$, which moreover also holds for the Mat\'ern and RationalQuadratic outer kernels, cf. Eqs.~\eqref{eq:Matern} and~\eqref{eq:rationalq}, respectively. Figure~\ref{fig:indepth-pqk-qfmnist} exemplifies this for the QFMNIST ($d=8$) dataset for Gaussian outer kernel function. Each section in the central $d(G|G^\prime)$ heatmap plot corresponds to a family of projected quantum circuits defined by a given measurement operator $O_{k}$ and is further subdivided into the different encoding circuits in the order as they appear in Appendix~\ref{sec:appendix-encoding-circuits}. There are \revision{both} regions with finite distances between two Gram matrices and regions with vanishing differences. This can be explained by the $\mathrm{Var}(F)$ plot at the top as well as the zoomed-in examples of the $F$ heatmaps on the very right. Vanishing $d(G|G^\prime)$ distances correspond to Gram- and $F$-matrices that show favourable variance behavior or, in other words, lead to similar embeddings $F$ and $F^\prime$ in the corresponding Hilbert space (upper zoom-in plot). In contrast, finite distances (lower zoom-in plot) result from one Gram-/$F$-matrix being ill-conditioned, i.e., $\mathrm{var}(G)\to 0$ and thus $\mathrm{var}(F)\to 0$ or vice versa. Consequently, these matrices correspond to weakly performing  models, while the others show better MSE scores as can be observed from the respective histogram plot on the right. Considering the $w_{\mathrm{e}}$- and $\gamma$-histograms on the left and on the bottom helps to understand the mechanisms which render these cases into the weak-performing regime. Here, $w_{\mathrm{e}}$ becomes comparatively small, while $\gamma$ becomes relatively large. With this mechanism we can \emph{a posteriori} explain the origin of bad performing PQK models in Figs.~\ref{fig:default-clf-reg-model-performance} and~\ref{fig:default-reg-clf-encoding-circs}. The analogous plots for the QFMNIST($d=8$) dataset for the remaining outer kernel functions are detailed in the Appendix~\ref{sec:appendix-detailed-results}.

This analysis confirms our hypothesis that there is a synergy and interdependence between projected quantum circuits and outer kernels as well as QKM hyperparameters. As such, we conclude that PQK-based QKMs require careful examination of outer kernels \emph{and} suitable measurement operators to perform well on a given dataset. 

\section{\label{sec:discussion}Discussion}
In this benchmarking study, we systematically analyzed various design criteria of QKMs that are based on FQKs and PQKs for several regression and classification tasks. In order to derive universal insights, we conducted hyperparameter optimizations and training of more than $20,000$ models.

\revision{Our findings emphasize the importance of hyperparameters such as regularization and bandwidth-tuning as well as outer kernel length scale and the choice of measurement operators for PQKs, in effective training of QKMs. Interestingly, the number of qubits and layers in data encoding circuits appear less crucial. Performance comparisons reveal no substantial differences between FQKs and PQKs. For less complex datasets, the particular choice of the kernel method is not really important as, on average, they are performing equally well. While PQKs slightly outperform classical baselines for simpler datasets, both quantum and classical models struggle as the dataset complexity increases, indicating no clear quantum advantage.}

\revision{Another remarkable observation significantly extends the findings of Ref.~\citep{bowles2024} and generlly reveals that circuits without entanglement perform on par with or better than those with entangling gates. We show this across various encoding circuits, using both FQKs and PQKs within QSVM and QKRR approaches and addressing both regression and classification problems. Thus, we highlight the need to unravel the driving forces of QKM performance if ``quantumness'' turns out as debatable mechanism.}

We note that although we encompass multiple different dataset families, the effective number of distinct families, the size of each dataset, and the specifics of splitting into training and test sets represents a restriction, which limits the generalizability of our results. Moreover, we randomly initialized the trainable parameters of data encoding circuits. While this resulted in comparatively good performance scores in all cases, we found in some cases that training them can improve the QKM performance, although there is no guaranty, cf. Appendix~\ref{sec:appendix-kta}. Hence, we recommend to carefully scrutinize this in future studies.

\revision{Our 15-qubit simulations find no clear PQK performance advantage despite expected FQK trainability limits as a consquence of exponential concentration~\cite{thanasilp2024exponential}, likely due to bandwidth tuning improving generalization~\citep{Shaydulin2022,canatar2023} but, as shown in Refs.~\citep{Slattery2023,egginger2024hyperparameter}, at the expense of becoming classically tractable. To properly understand the connections between bandwidth tuning, exponential concentration and classical tractability in terms of the geometric difference metric~\citep{Huang2021} will be part of a future work.}

\revision{In the context of our results, we believe that future research should focus on developing and applying datasets with larger complexity to gain a deeper understanding of quantum kernel methods. One idea could be to start form the Fourier representation of quantum kernels~\citep{Schuld2021,schuld2021supervised} and investigate how this can be leveraged for the QKM design as well as for the identification of promising dataset properties. If such a dataset is found, then thoroughly investigating corresponding implications on generalization properties and exponential concentration~\cite{thanasilp2024exponential} as well as classical tractability~\cite{Huang2021} of those quantum kernels would be of great interest. Another crucial aspect is to systematically explore problem-specific data encoding strategies.}

\section{\label{sec:conclusion}\revision{Conclusion}}
\revision{This study provides valuable insights into the design of effective QKMs, offering essential guidlines for achieving good model performance. Key factors include data preprocessing, bandwidth-tuning, and careful optimization of classical hyperparameters for both FQKs and PQKs. While the number of qubits and layers of underlying data encoding circuits becomes only critical as the dataset complexity increases, the choice of outer kernel functions and measurement operators is vital for PQKs. Here, we revealed that classical and quantum parts drive learning synergistically.}

\revision{Despite identifying these universal patterns, generally unraveling the true quantum contribution of QKMs remains elusive. Although we showed that more complex datasets require more expressive encodings, our findings challenge how QKMs leverage quantum-specific advantages, as encoding circuits without entanglement perform equally well or even better. With this, our work underscores the need for a dual approach to QML research: identifying datasets with potential quantum advantage and refining the corresponding model designs to fully exploit quantum capabilities.}

\bmhead{Acknowledgments}
This work was supported by the German Federal Ministry of Economic Affairs and 
Climate Action through the project AQUAS (grant no. 01MQ22003D), as well as the German
Federal Ministry of Education and Research through the project H2Giga Degrad-EL\textsuperscript{3}
(grant no. 03HY110D). Furthermore, the authors acknowledge Fraunhofer IAO for providing access to the GPU simulation cluster (CLARSA), which was established as part of the Competence Center Quantum Computing Baden-Württemberg. Here, J.S. is grateful to Vamshi Mohan Katukuri for his support with and excellent administration of the computing clusters. Furthermore, we would like to thank Moritz Link for his excellent contributions in an early stage of this work. Moreover, we are particularly grateful for insightful discussions with Roberto Fl\'orez Ablan.

\bmhead{Availability of data and material} The code developed for this study is available via GitLab~\citep{schnabel24gitlab}. Data used in this study is publicly available and can be accessed via the respective references provided in the manuscript. A guidance how to use the code and potentially reproduce the results is provided in the repository.

\section*{Declarations}

\bmhead{Conflict of interest} The authors declare that they have no conflict of interest.



\appendix
\setcounter{page}{1}
\renewcommand{\thepage}{A\arabic{page}}
\setcounter{figure}{0}
\renewcommand{\thefigure}{A\arabic{figure}}

\begin{appendices}


\section{\label{appendix:detailed-theory}\revision{Detailed Theoretical Background}}

\subsection{\label{appendix:conventional-kernel-theory}\revision{Details on Conventional Kernel Theory}}
\revision{The key idea behind the conventional kernelized approach to (supervised) machine learning is to find and analyze patterns by transforming the respective learning problem from the original input data domain $\mathcal{X}$ to a higher-dimensional (potentially infinite-dimensional) \emph{feature space} $\mathcal{F}$, where the learning tasks \revision{can often be expressed in a linear form.} This mapping is accomplished by a \emph{feature map} $\phi: \mathcal{X}\to\mathcal{F}; \xdata\mapsto\phi(\xdata)$. Kernels, are real- or complex-valued symmetric and positive semi-definite functions of two input data points, $k: \mathcal{X}\times\mathcal{X}\to\mathbb{C}$. In this regard, another central concept is that of \emph{reproducing kernel Hilbert space} (RKHS) \revision{which} uniquely determines the kernel and vice versa~\citep{Aronszajn1950} and in addition, for every kernel there exists at least one feature map such that~\citep{schuld2021supervised}
\begin{align}
    \label{eq:kernel-appendix}
    k(\xdata, \xdata') = \langle\phi(\xdata), \phi(\xdata')\rangle_{\mathcal{F}}\,.
\end{align}
Moreover, every feature map gives rise to a kernel. Less formally one can think of the RKHS as a space whose elementary functions, the kernels, assign a similarity measure between two data points $\xdata$ and $\xdata^\prime$.}

\revision{Instead of explicitly computing the transformation $\phi(x)$ to the high-dimensional feature space, the \emph{kernel trick} accomplishes the same result in the original input space through the kernel function. Besides, the Gaussian (RBF) kernel, we employ further common kernel functions in this work (specifically, for $\kappa$ in Eq.~\eqref{eq:PQK-general}, i.e. the PQK definition). First, the Mat\'{e}rn kernel~\citep{GPRBook} with $\nu=\nicefrac{3}{2}$, i.e.
\begin{align}
    \label{eq:Matern}
    \kappa^{\text{Mat}}_{3/2}(\mathbf{x},\mathbf{x}^\prime) &= \left(1+\frac{\sqrt{3}\lVert\mathbf{x}-\mathbf{x}^\prime\rVert}{\ell}\right) \notag \\ &\times\exp\left(-\frac{\sqrt{3}\lVert\mathbf{x}-\mathbf{x}^\prime\rVert}{\ell}\right)\,,
\end{align}
and the RationalQuadratic kernel~\citep{KernelCookbook}, i.e., 
\begin{align}
    \label{eq:rationalq}
    \kappa^{\text{RQ}}(\mathbf{x},\mathbf{x}^\prime) &= \left(1 + \frac{\lVert\mathbf{x}-\mathbf{x}^\prime\rVert^2}{2\alpha\ell^2}\right)^{-\alpha}\,.
\end{align}
}

\revision{A key result in kernel theory is the \emph{representer theorem}~\citep{Schoellkopf2001representer, schölkopf2002learning}, which states that the function $h$ that minimizes a regularized empirical risk loss, can always be represented as a finite (length of the training sample $N$) weighted linear combination of the kernel between some $\xdata\in\mathcal{X}$ and the training data $\xdata_i$, i.e.,
\begin{align}
    \label{eq:representer}
    h(\xdata) &= \sum_{i=1}^N c_i k(\xdata,\xdata_i)\,.
\end{align}
Note that the number of terms in the sum is independent of the dimension of $\mathcal{H}_k$. The determination of the coefficients $\lbrace c_i\rbrace$ is a convex optimization problem~\citep{DiMarcantonio2023}. This result is central to kernel methods used for supervised learning problems, e.g., SVMs~\citep{schölkopf2002learning}, Gaussian Processes~\citep{GPRBook} or KRR~\citep{Murphy}. All these algorithms are based on the kernel \emph{Gram matrix} $K_{ij} = [k(\xdata_i, \xdata_j)]$, as detailed below for KRR and SVR.}

\subsection{\label{appendix:KRR-vs-SVR}\revision{KRR versus SVR}}

\revision{Kernel Ridge Regression (KRR) and Support Vector Regression (SVR) are both kernel-based regression techniques, but they differ in their formulation, optimization objectives, and practical applications. 
Given a dataset $\lbrace (x_i, y_i)\rbrace_{i=1}^N$, KRR minimizes the following objective function~\citep{schölkopf2002learning}
\begin{equation}
    \min_{\mathbf{w}}\sum_{i=1}^N (y_i - \mathbf{w}^T\phi(x_i))^2 + \lambda\lVert\mathbf{w}\rVert^2\,,
\end{equation}
where $\phi(x)$ is a feature map as introduced in Sec.~\ref{sec:background}, $\mathbf{w}$ is the weight vector in the transformed space, and $\lambda$ is a regularization parameter. Using the representer theorem, the solution takes the form Eq.~\eqref{eq:representer}, where the linear coefficient vecotr $\mathbf{c}$ is obtained by solving
\begin{align}
    \mathbf{c} &= (\mathbf{K} + \lambda\mathbf{I})^{-1}\mathbf{y}\,.
\end{align}
Therefore, the key characteristics of KRR are
\begin{enumerate}
    \item Matrix inversion makes it computationally expensive, i.e. $\mathcal{O}(N^3)$ for large datasets
    \item The quadratic loss function penalizes large error more heavily than small ones
    \item The Tikhonov regularization $\lambda$ balances smoothness and fit to data  
\end{enumerate}
The SVR aims to find a function $f(x)$ that predicts the targets $y$ with a margin of tolerance $\varepsilon$. For this, it uses the $\varepsilon$-intesive loss function and solves the following constrained optimization problem~\citep{schölkopf2002learning,scikit-learn}
\begin{equation}
    \min_{\mathbf{w},b,\xi,\xi^*}\frac{1}{2}\lVert\mathbf{w}\rVert^2 + C\sum_{i=1}^N(\xi_i + \xi_i^*)\,,
\end{equation}
subject to the constraints
\begin{align}
    y_i - \mathbf{w}^T\phi(x_i) - b &\leq \varepsilon + \xi_i\,,\\
    \mathbf{w}^T\phi(x_i) + b - y_i &\leq \varepsilon + \xi_i^*\,,\\
    \xi_i,\xi_i^* &\geq 0\,.
\end{align}
Here, $\varepsilon$ is a threshold margin for error tolerance, $\xi_i, \xi_i^*$ are slack variables that allow deviations beyond $\varepsilon$, and $C>0$ is the regularization parameter controlling the tradeoff between margin and error. The dual formulation leads to the prediction function
\begin{align}
    f(x) &= \sum_{i=1}^N(\alpha_i - \alpha_i^*)K(x,x_i) + b\,,
\end{align}
where $\alpha_i, \alpha_i^*$ are the dual coefficients. The key characteristics of SVR are
\begin{enumerate}
    \item The $\varepsilon$-intensive loss encourages sparsity and robustness to outliers
    \item Solving the quadratic problem can be also computationally demanding, i.e. $\mathcal{O}(N^2)$ to $\mathcal{O}(N^3)$, but is often more scalable than KRR for large datasets	
\end{enumerate}
To this end, the choice between KRR and SVR depends on the dataset size, the desired model complexity, and the tradeoff between computational efficiency and model performance. KRR is well-suited for small datasets with Gaussian noise, while SVR employs a sparse optimization and is more robust to outliers as well as better suited for large datasets.}

\revision{We refer to Refs.~\citep{schölkopf2002learning,GPRBook,Murphy} for a thorough introduction into  kernel theory and kernel methods.}

\section{\label{sec:appendix-statistis}Some Background on Statistical Methods}
Since we subject our results a statistical analysis to detect statistically significant correlations between hyperparameters and hyperparameters and model performance, we briefly describe the underlying concepts in the following. 

Correlation analysis is a statistical technique used to measure and describe the strength and direction of the relationship between two variables. The Pearson and Spearman correlation coefficients are two primary types of quantifying correlation. The \emph{Pearson} correlation coefficient measures the linear relationship between two datasets $X$ and $Y$. Strictly speaking, Pearson's correlation requires each dataset to be normally distributed as well as assumes homoscedasticity, i.e., the spread of the data points is consistent across the range of values. It is~\citep{Vallat2018}
\begin{equation}
    \label{eq:pearson-corr-coeff}
    r = \frac{\sum_i (x_i-\bar{x})(y_i-\bar{y})}{\sqrt{\sum_i(x_i-\bar{x})^2}\sqrt{\sum_i(y_i-\bar{y})^2}}\,,
\end{equation}
where $x_i\in X$ and $y_i\in Y$ are the individual sample points, while $\bar{x}$ and $\bar{y}$ are the means for the $x_i\in X$ and $y_i\in Y$ samples, respectively. Correlations of $\pm 1$ imply a perfect positive and negative linear relationship, respectively, with $0$ indicating the absence of association. The \emph{Spearman} correlation coefficient is a non-parametric measure of the monotonicity of the relationship between two datasets. Unlike the Pearson correlation it does not assume that both datasets are normally distributed. Correlations $\pm 1$ imply an exact positive and negative monotonic relationship, respectively. Mathematically, the Spearman correlation coefficient is defined as the Pearson correlation coefficient between the rank variables. The Spearman's rank correlation is moreover less sensitive to outliers.

To obtain reliable correlation estimates it is crucial to ensure a sufficiently large sample size. Additionally, one should ideally ensure that inferred correlation coefficients are statistically significant. This can be done by computing p-values. The p-value is the probability of obtaining a correlation coefficient as extreme as, or more extreme than the observed value under the null hypothesis $H_0$, which assumes that there is no correlation between the two variables. For interpreting p-values in correlation analysis one defines a significance level, which is commonly set to $0.001$, $0.01$ or $0.05$. Then it holds
\begin{itemize}
    \item $p\leq 0.05$: The correlation is statistically significant and one can reject $H_0$
    \item $p > 0.05$: The correlation is \emph{not} statistically significant and one fails to reject $H_0$.
\end{itemize}
Typically one performs two-sided statistical tests to compute the $p$-value, i.e., it checks for the $H_0$ and for an alternative hypothesis $H_1$, assuming the correlation is different from zero. It is important to adjust the $p$-values accordingly to account for increased risk of false positives. There are several methods available, for more details we refer to the documentation of the python package \texttt{pingoin}~\citep{Vallat2018}. 

Another essential concept of correlation analysis are \emph{partial correlations}~\citep{Kim2015} to measure the degree of association between two variables $x$ and $y$ while controlling for the effect of one or more additional variables $z$. This helps to isolate the direct relationship between two primary variables of interest, removing the influence of the control variables. Practically, this is achieved by calculating the correlation coefficient between the residuals of two linear regressions~\citep{Vallat2018}
\begin{equation}
    x \sim z\,,\, y\sim z\,.
\end{equation}
Like the correlation coefficient, the partial correlation coefficient takes on a value in $[-1,1]$. The semi-partial correlation works analogously, with the exception that the set of controlling variables in only removes from either $x$ or $y$.

\section{\label{sec:appendix-encoding-circuits}Detailed Overview of Encoding Circuits}

In the following we provide plots of each encoding circuits used in this study and briefly describe (cf. sQUlearn documentation) the corresponding features. Features are denoted with a feature vector \texttt{x[i]}, while variationally trainable parameters ($\boldsymbol{\theta}$) are labeled \texttt{p[i]}.
\subsection{YZ\_CX\_EncodingCircuit}
\begin{figure}[h]
    \includegraphics[width=\columnwidth]{"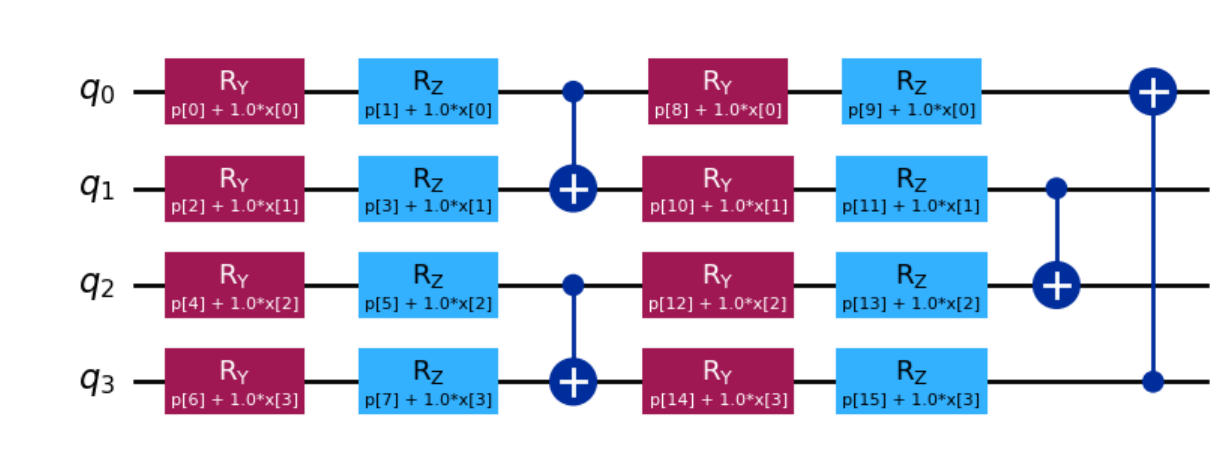"}
    \caption{Exemplaric illustration of the \texttt{YZ\_CX\_EncodingCircuit} from Ref.~\citep{Haug2023}.}
    \phantomsection
    \label{fig:yz_cx}
\end{figure}

This encoding circuit was introduced in Ref.~\citep{Haug2023} and is originally designed for encoding high-dimensional features. An example with four qubits and features and two layers is shown in Fig.~\ref{fig:yz_cx}.

\subsection{HighDimEncodingCircuit}
\begin{figure}[h]
    \includegraphics[width=\columnwidth]{"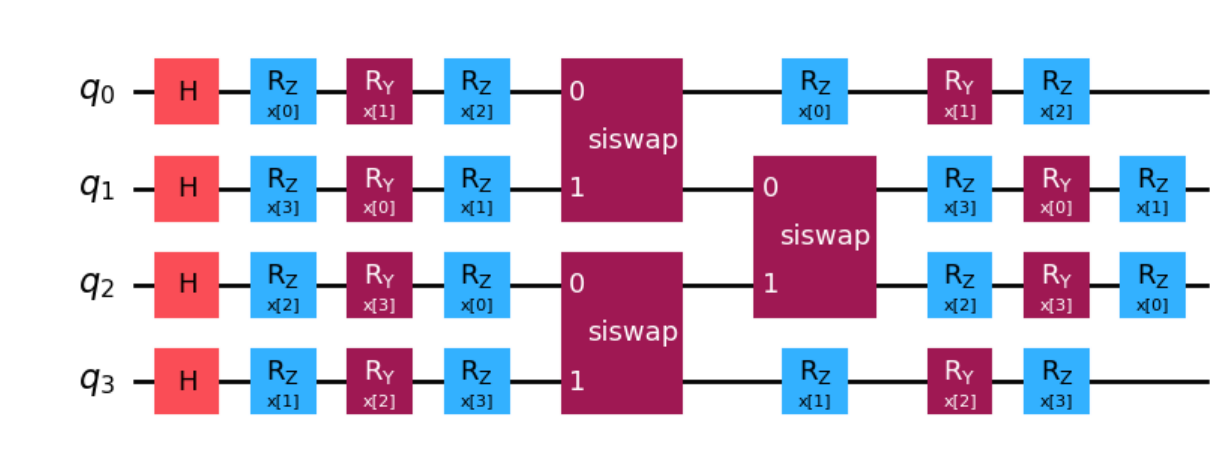"}
    \caption{Exemplaric representation of the \texttt{HighDimEncodingCircuit} from Ref.~\citep{Peters2021}.}
    \phantomsection
    \label{fig:highdim}
\end{figure}

The \texttt{HighDimEncodingCircuit} from Ref.~\citep{Peters2021}, was introduced to deal with high-dimensional data from the domain of cosmology and is constructed such that it preserves the magnitude of the entries of the quantum kernel matrix that otherwise typically vanish due to the exponentially growing Hilbert space. An example with four qubits and four features as well as two layers is depicted in Fig.~\ref{fig:highdim}.

\subsection{HZY\_CZ\_EncodingCircuit}
\begin{figure}[h]
    \includegraphics[width=\columnwidth]{"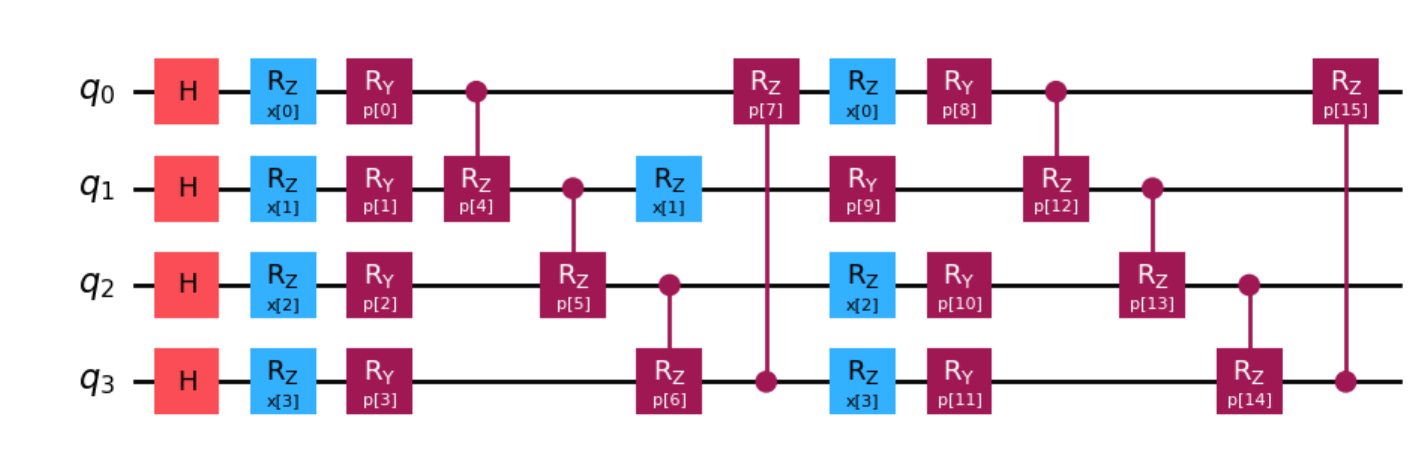"}
    \caption{Exemplaric representation of the \texttt{HZY\_CZ\_EncodingCircuit} from Ref.~\citep{Hubregtsen2021}.}
    \phantomsection
    \label{fig:hubregtsen}
\end{figure}

This encoding circuit was introduced in Ref.~\citep{Hubregtsen2022} and is shown exemplarily in Fig.~\ref{fig:hubregtsen} for four qubits, four features and two layers.
\subsection{ChebyshevPQC}
\begin{figure}[h]
    \includegraphics[width=\columnwidth]{"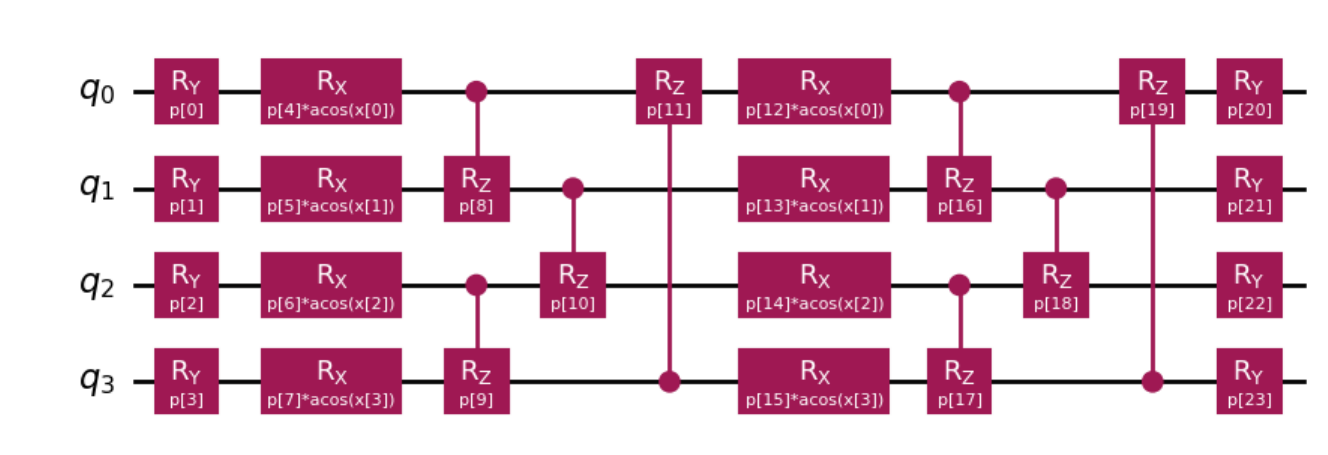"}
    \caption{Exemplaric representation of the \texttt{ChebyshevPQC} as introduced in Ref.~\citep{Kreplin2024reductionoffinite}.}
    \phantomsection
    \label{fig:cheby}
\end{figure}

The \texttt{ChebyshevPQC} encoding circuit was introduced in Ref.~\citep{Kreplin2024reductionoffinite} to provide a basis of Chebyshev polynomials. This is realized by the non-linear feature encodings via $\arccos(x)$ mappings. In the mentioned paper it was shown that this data encoding works well within a QNN setting. An example illustration with four qubits, four features and two layers in given in Fig.~\ref{fig:cheby}.

\subsection{ParamZFeatureMap}
\begin{figure}[h]
    \includegraphics[width=\columnwidth]{"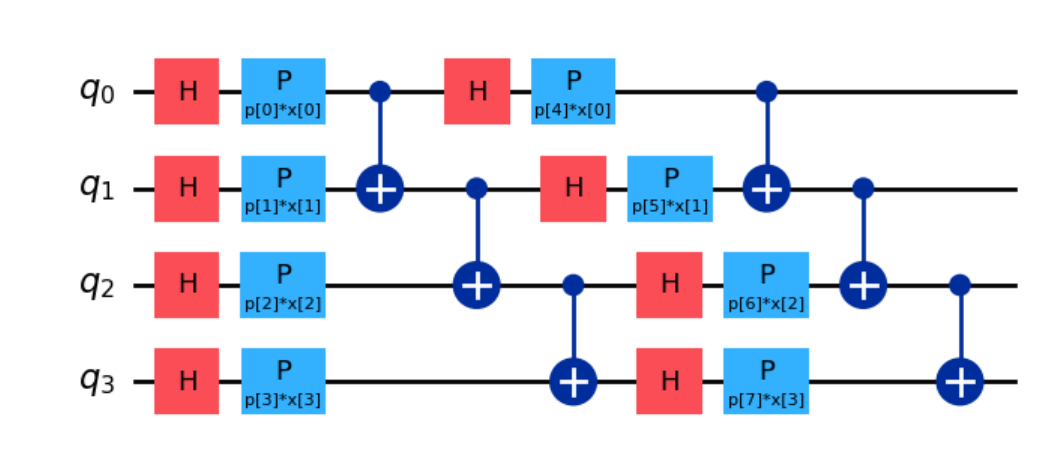"}
    \caption{Exemplaric representation of the \texttt{ParamZFeatureMap}, which implements Qiskit's \texttt{ZFeatureMap} with additional CNOT gates between the rotation layers.}
    \phantomsection
    \label{fig:par_z_ent}
\end{figure}

This encoding circuit is inspired by Qiskit's \texttt{ZFeatureMap} and allows for rescaling the input data with variationally trainable parameters and introduces additional CNOT gates between the default rotation gates. An example is given in Fig.~\ref{fig:par_z_ent}.

\subsection{SeparabaleEncodingRx}
\begin{figure}[h]
    \includegraphics[width=.6\columnwidth]{"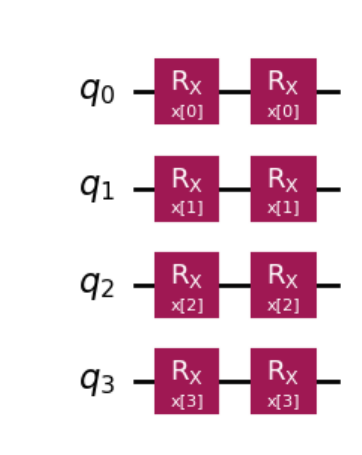"}
    \caption{Exemplaric representation of the \texttt{SeparableRxEncoding} from Ref.~\citep{canatar2023}.}
    \phantomsection
    \label{fig:se_rx}
\end{figure}

This paradigamtic encoding circuit was used in Ref.~\citep{canatar2023} to analytically study the effect of bandwidth tuning in quantum kernels. In this work, we include this data encoding to reveal learning capabilities of QKMs in absence of entanglement. An example is shown in Fig.~\ref{fig:se_rx}.

\subsection{HardwareEfficientEmbeddingRx}
\begin{figure}[h]
    \includegraphics[width=\columnwidth]{"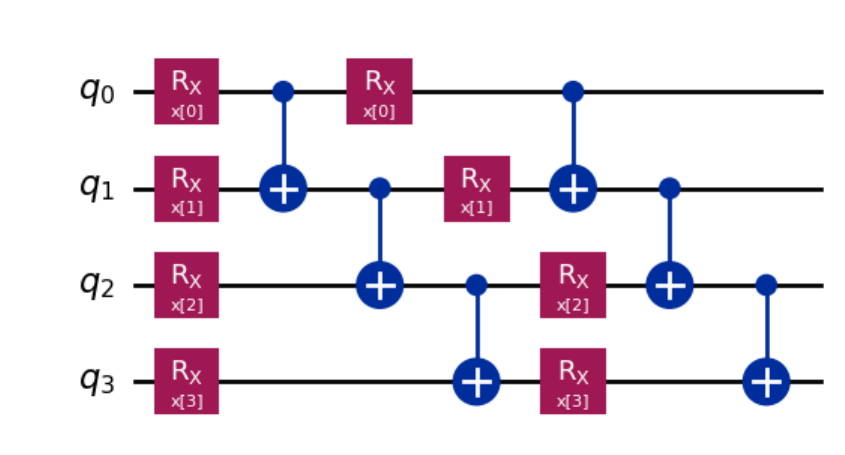"}
    \caption{Exemplaric representation of the \texttt{HardwareEfficientEmbeddingRx} from Ref.~\citep{thanasilp2024exponential}.}
    \phantomsection
    \label{fig:hee_rx}
\end{figure}

The \texttt{HardwareEfficientEmbeddingRx} circuit is taken from Ref.~\citep{thanasilp2024exponential}, where it served as basis to study exponential concentration in QKMs. An example is shown in Fig.~\ref{fig:hee_rx}.

\subsection{ZFeatureMap}
\begin{figure}[h]
    \includegraphics[width=.8\columnwidth]{"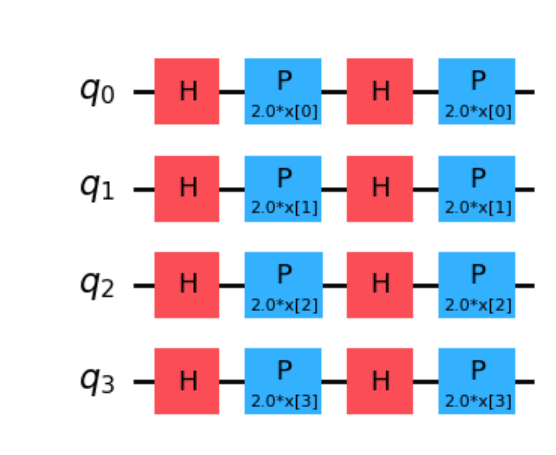"}
    \caption{Exemplaric representation of Qiskit's \texttt{ZFeatureMap}, cf. Ref.~\citep{qiskit2024}.}
    \phantomsection
    \label{fig:zfmap}
\end{figure}

This encoding circuit is taken from Qiskit~\citep{qiskit2024}. It defines a first-order Pauli-Z-evolution circuit. An example is given in Fig.~\ref{fig:zfmap}.

\subsection{ZZFeatureMap}
\begin{figure}[h]
    \includegraphics[width=\columnwidth]{"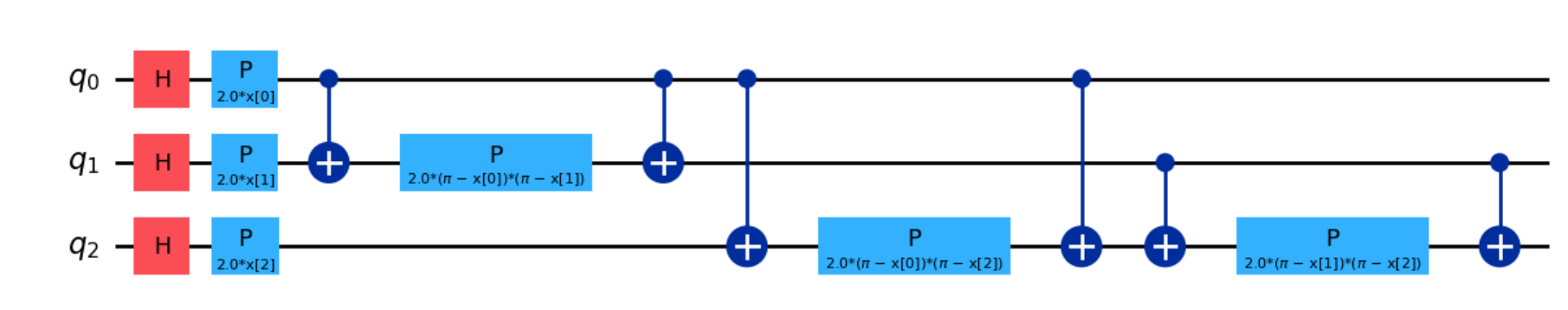"}
    \caption{Exemplaric representation of Qiskit's \texttt{ZZFeatureMap}, cf. Ref.~\citep{qiskit2024}.}
    \phantomsection
    \label{fig:zzfmap}
\end{figure}

This encoding circuit is taken from Qiskit~\citep{qiskit2024}. It defines a second-order Pauli-Z-evolution circuit. An example is given in Fig.~\ref{fig:zzfmap}.

\section{\label{sec:appendix-details-dataset}\revision{Details on Datasets}}

\revision{\bmhead{Friedman} The Friedman \#1 regression problem is described in Refs.~\citep{Friedman1991,Breiman1996}. The dataset consists of $d \geq 5$ independent features distributed uniformly on the interval $[0,1]$. The labels $y$ satisfy
\begin{equation}
    \label{friedman1}
    \begin{split}
        y(\xdata) &= 10\sin(\pi x_1 x_2) + 20(x_3 - 0.5)^2 \\ &+ 10 x_4 + 5x_5 + \sigma\mathcal{N}(0,1)\,,
    \end{split}
\end{equation}
where $x_i$ with $i\in\lbrace 1,2,3,4,5\rbrace$ is the $i$-th component of the data point $\xdata$ and $\sigma$ is the standard deviation of Gaussian noise applied to the output. We set $\sigma=0.01$ for all simulations. Only $d=5$ features contribute to the computation of the labels, while the remaining features are independent. Due to its non-linearity as well as its interaction between features and noise characteristics, this dataset mimics real-world regression scenarios. For this study, we generate datasets for $d=5,\ldots 15$ features.}

\section{\label{sec:appendix-qkmtuner}\revision{Details on QKMTuner}}
\revision{As indicated in Sec.~\ref{subsec:setup-and-implementation}, we developed the tool \texttt{QKMTuner} to facilitate the extensive hyperparameter search of this study. Its core functionalities are sketched in Fig.~\ref{fig:sketch-implementation} consisting of two main routines: a hyperparameter optimization within a grid-search and a hyperparameter importance analysis. The grid-search is given a list of data encoding circuits for each of which it builds a user-defined $n_{\mathrm{qubits}}\times n_{\mathrm{layers}}$-grid on which a hyperparameter search is performed for each grid point and each data encoding circuit for a given dataset.
For evaluating hyperparameter importances, the respective method only takes a list of data encoding circuits and automatically searches for the optimal hyperparameters including $n_{\mathrm{qubits}}$ and $n_{\mathrm{layers}}$. 
In both cases, the hyperparameters are determined by maximizing the minimum between the mean and median of five-fold cross-validation scores. This choice of the objective function prevents the scores from becoming too optimistic and thus helps to prevent overfitting towards an easy fold. The respective scoring method can be set manually, whereas in this work we use the area under the receiver operating characteristic curve (ROC-AUC) from prediction scores for classification and the (negative) mean squared error (MSE) for regression tasks, respectively. For hyperparameter sampling we use the tree-structured parzen estimator~\citep{watanabe2023treestructuredparzenestimatorunderstanding}, albeit different algorithms as provided by Optuna are supported.}

\section{\label{sec:reproducibility}Note on Reproducibility of Results}
To ensure reproducible results across executions, we set the \texttt{random\_state} parameter of \emph{scikit-learn's} estimators and splitters. Here, we follow the recommendations in the scikit-learn documentation on robustness of cross-validation results; i.e., we pass a \texttt{RandomState} instance to the estimator (here QSVC is the only one with a \texttt{random\_state} parameter) while we pass an integer to the cross-validation and the train-test splitters. Regarding the latter, we note that comprehensively validating the stability of results would require to test different interger seeds. In our case, this is infeasible due to the enormous study size. However, since we  aim at unravling general patterns and trends in QKMs, we do not require to pinpoint individual model performances and hence the necessity of stabilizing results across differents seed is not that important.

\section{\label{sec:appendix-detailed-results}Detailed Results}
Here we support and complement the key findings of this study as presented in Sec.~\ref{sec:results} of the main text. 

\begin{figure*}[tb]
    \centering
    \begin{minipage}[t]{.49\textwidth}
        \includegraphics[width=\linewidth]{"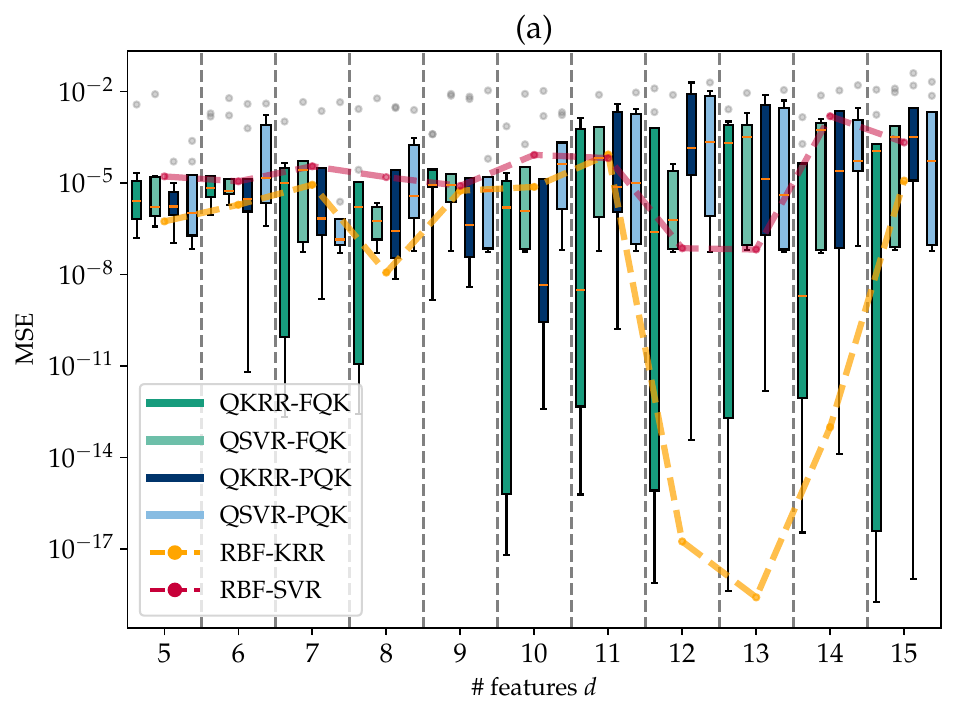"}
    \end{minipage}
    \begin{minipage}[t]{.49\textwidth}
        \includegraphics[width=\linewidth]{"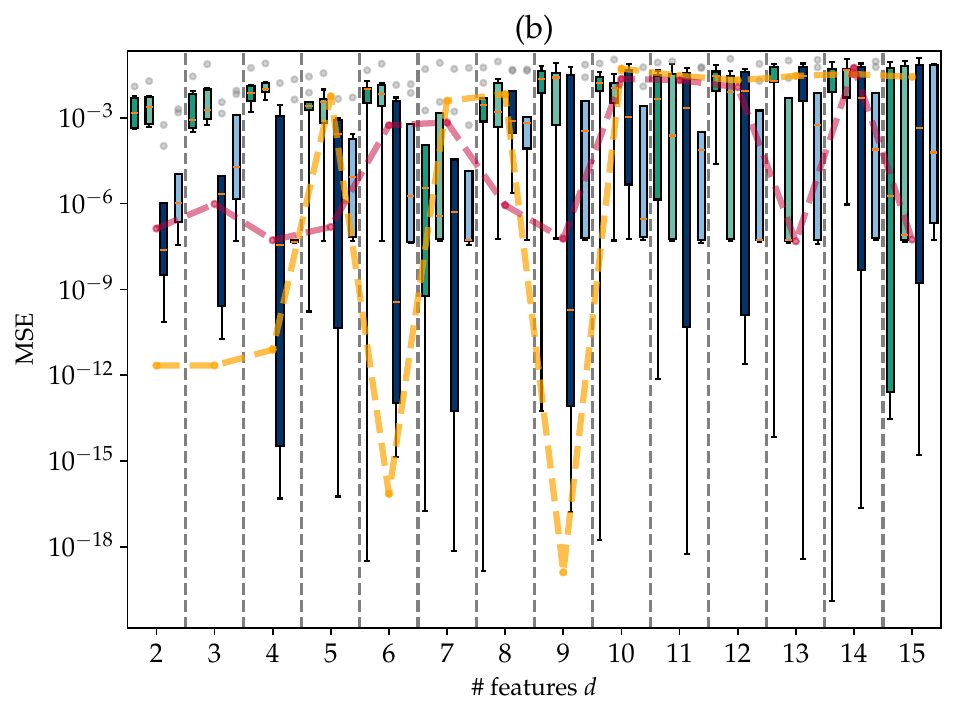"}
    \end{minipage}
    \begin{minipage}[t]{.49\textwidth}
        \includegraphics[width=\linewidth]{"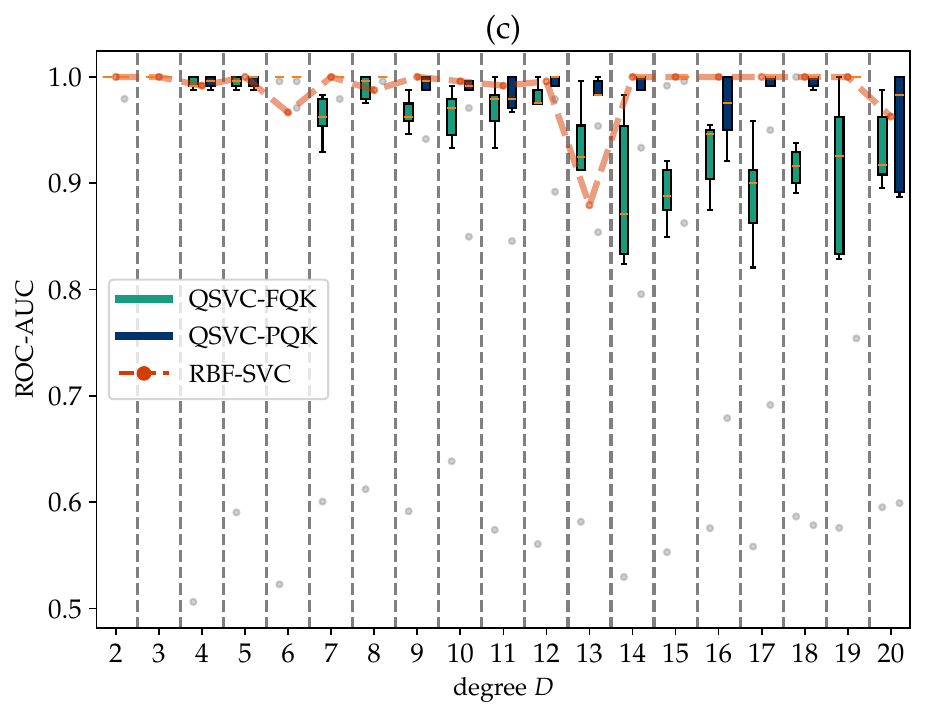"}
    \end{minipage}
    \begin{minipage}[t]{.49\textwidth}
        \includegraphics[width=\linewidth]{"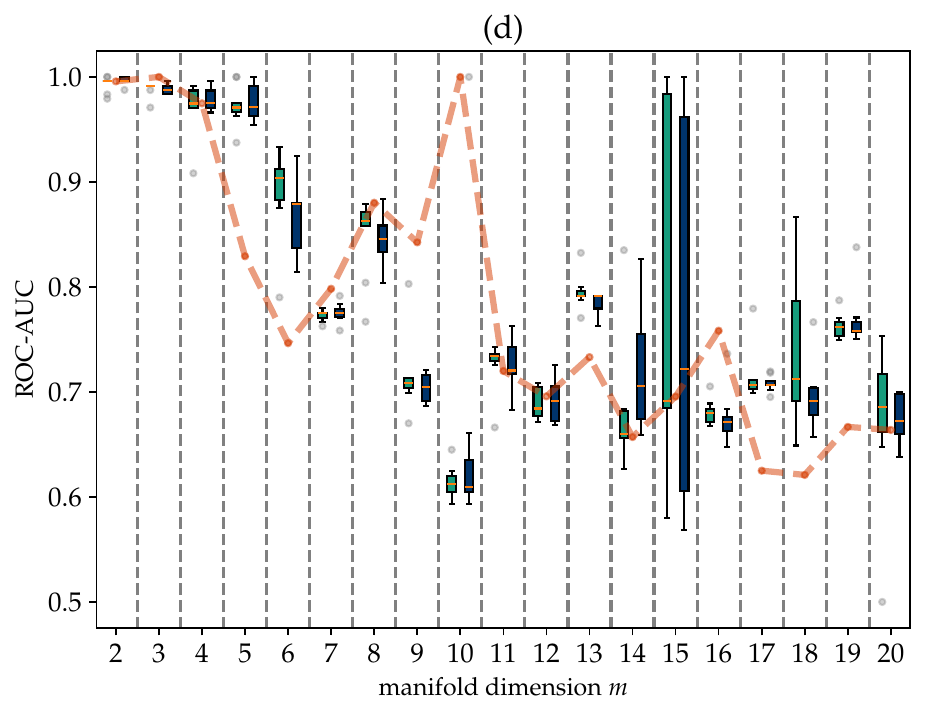"}
    \end{minipage}
    \caption{Analogous illustration to Fig.~\ref{fig:default-clf-reg-model-performance} with associated \emph{training} performance scores. Results for each QKM and dataset are aggregated across data encoding circuits with corresponding optimal $n_{\mathrm{layers}}^*$, yielding minimum/maximum test performance scores for regression/classification, respectively. \revision{For comparison, we provide classical KRR/SVR and SVC results each based on a RBF kernel. Corresponding hyperparameters are also optimized using Optuna with parameter ranges chosen as given in Sec.~\ref{subsec:setup-and-implementation}.} The \textbf{upper panel} displays corresponding findings for \textbf{regression} datasets. Here,~\textbf{(a)} shows the \textbf{Friedman} dataset family with datsets for $\#\mathrm{features} d=[5,15]$. In~\textbf{(b)} we give results corresponding to the \textbf{QFMNIST} dataset family with datasets having number of principal components $d=[2,15]$. The \textbf{lower panel} shows the \textbf{classification} tasks. The results in~\textbf{(c)} correspond to the \textbf{two curves diff} family with $d=4$ and degree $D=[2,20]$. In~\textbf{(d)} we present the \textbf{hidden manifold diff} family with $d=4$ and manifold dimension $m=[2,20]$.}
    \phantomsection
    \label{fig:default-clf-reg-train-performance}
\end{figure*}

\subsection{\label{subsec:model-performance-correlation-appendix}Details on Model Performance}
In Sec.~\ref{subsec:model-performance} we mention in connection with outliers in Fig.~\ref{fig:default-clf-reg-model-performance} that both regression and classification datasets exhibit a propensity for overfitting. The Friedman and QFMNIST regression datasets in Figs.~\ref{fig:default-clf-reg-train-performance}~(a) and~(b), respectively, reveal that each dataset shows lower whiskers with $\mathrm{MSE}\to 0$. Comparing to the corresponding test performance scores in Figs.~\ref{fig:default-clf-reg-model-performance}~(a) and~(b), this may also indicate signatures of overfitting especially for datasets with larger $d$. The corresponding training performance scores for the two curves diff and hidden manifold diff datasets are displayed in Figs.~\ref{fig:default-clf-reg-train-performance}~(c) and~(d), respectively. As discussed in the main text, since we investigate a large variety of combinations, it is unavoidable that some of them result in models that are too expressive for some tasks. Moreover, we point out that one should carefully investigate different training- and test set sizes in future studies. 

\subsection{\label{subsec:default-correlation-analysis-appendix}Additions to Hyperparameter Correlation Analysis}
\begin{figure*}[tb]
    \centering
    \begin{minipage}[b]{0.23\linewidth}
        \includegraphics[width=\linewidth]{"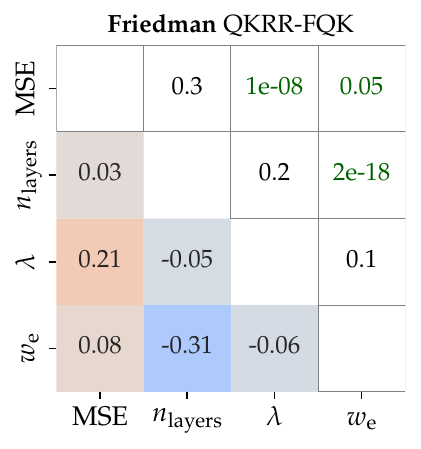"}
    \end{minipage}
    \hfill
    \begin{minipage}[b]{0.23\linewidth}
        \includegraphics[width=\linewidth]{"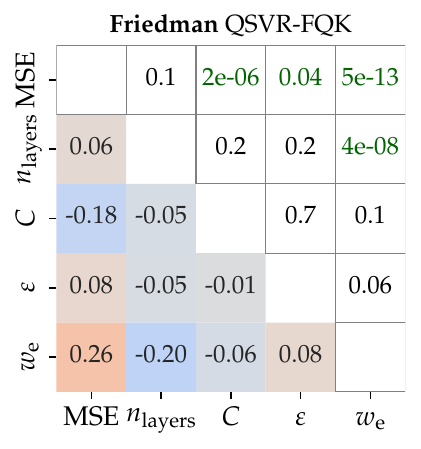"}
    \end{minipage}
    \hfill
    \begin{minipage}[b]{0.23\linewidth}
        \includegraphics[width=\linewidth]{"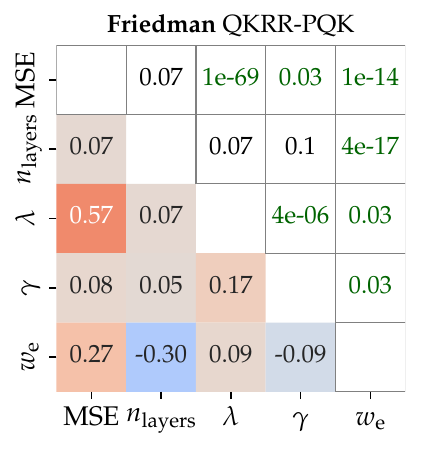"}
    \end{minipage}
    \hfill
    \begin{minipage}[b]{0.23\linewidth}
        \includegraphics[width=\linewidth]{"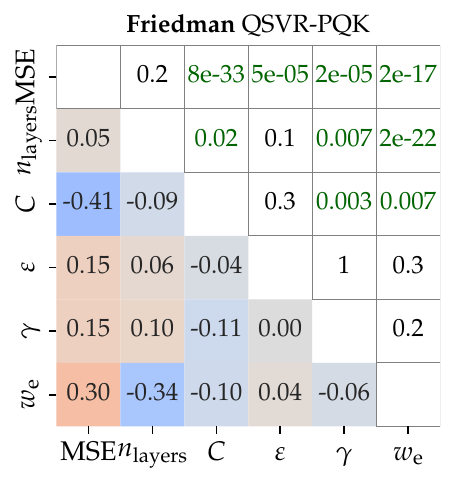"}
    \end{minipage}
    \vfill
    \begin{minipage}[b]{0.23\linewidth}
        \includegraphics[width=\linewidth]{"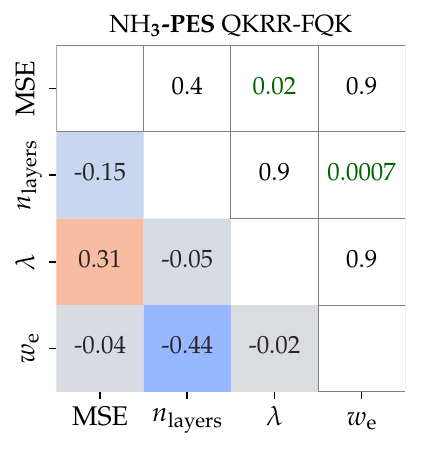"}
    \end{minipage}
    \hfill
    \begin{minipage}[b]{0.23\linewidth}
        \includegraphics[width=\linewidth]{"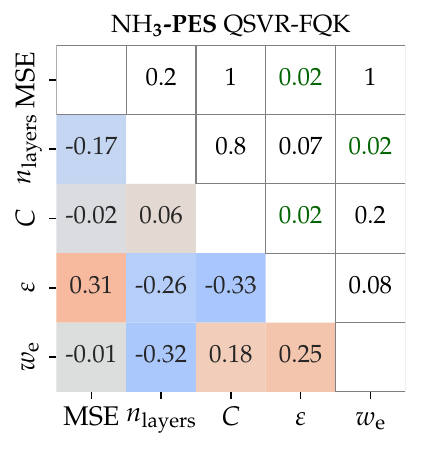"}
    \end{minipage}
    \hfill
    \begin{minipage}[b]{0.23\linewidth}
        \includegraphics[width=\linewidth]{"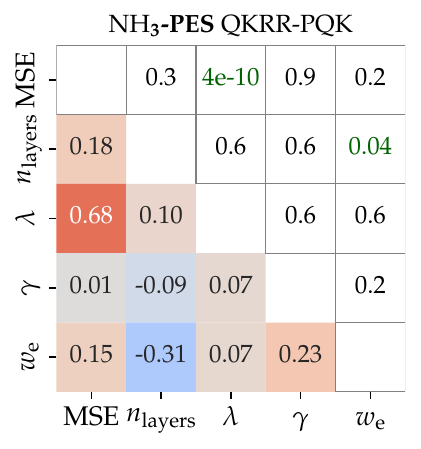"}
    \end{minipage}
    \hfill
    \begin{minipage}[b]{0.23\linewidth}
        \includegraphics[width=\linewidth]{"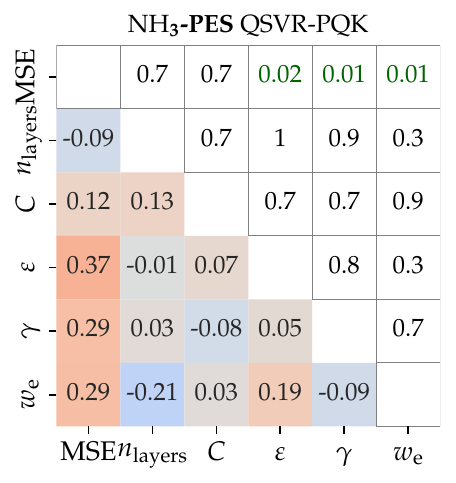"}
    \end{minipage}
    \vfill
    \begin{minipage}[t]{.23\textwidth}
        \includegraphics[width=\linewidth]{"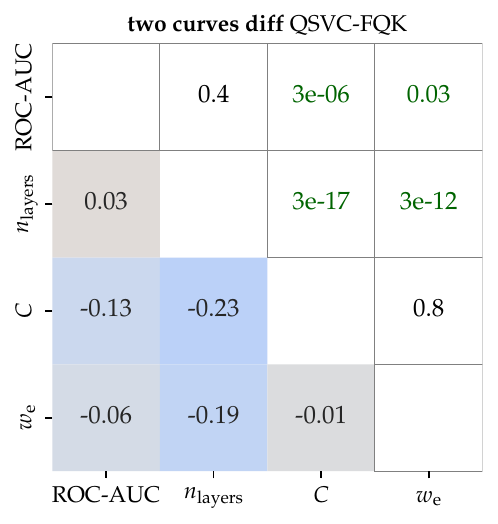"}
    \end{minipage}
    \hfill
    \begin{minipage}[t]{.23\textwidth}
        \includegraphics[width=\linewidth]{"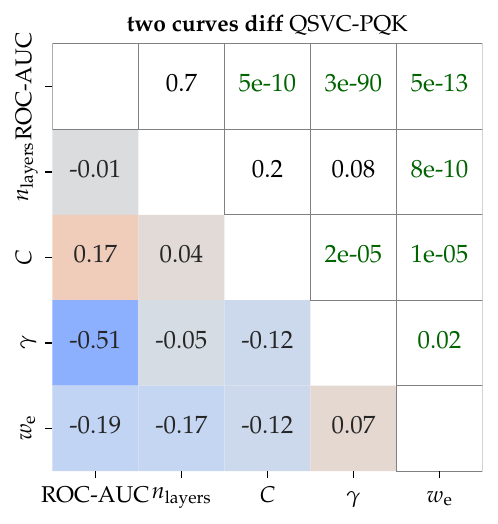"}
    \end{minipage}
    \hfill
    \begin{minipage}[t]{.23\textwidth}
        \includegraphics[width=\linewidth]{"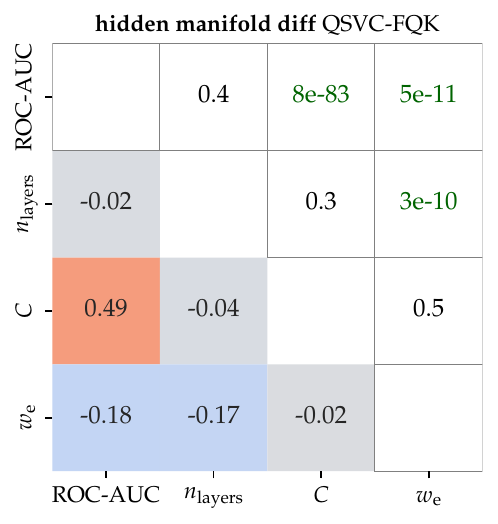"}
    \end{minipage}
    \hfill
    \begin{minipage}[t]{.23\textwidth}
        \includegraphics[width=\linewidth]{"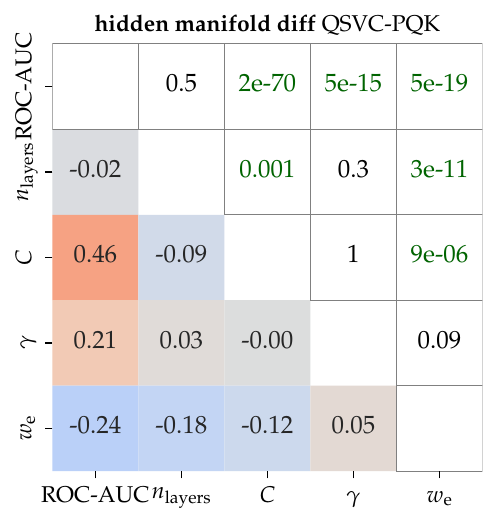"}
    \end{minipage}
    \caption{Summary of the Spearman correlation analyses between QKM hyperparameters and test performance scores, as well as between the hyperparameters themselves as discussed in Sec.~\ref{subsec:influence-hyperparameters}. This figure complements the respective discussion in the main text with the findings for the other datasets. The results correspond to aggregating data across all datasets within the associated family, encoding circuits, and $n_{\mathrm{layers}}\in [1,8]$. The Spearman correlation coefficients are given on the lower triangles of the respective matrices (with blue for negative to red for positive coefficients), while the upper triangles display corresponding $p$-values. Here, statistically significat correlations ($p\leq 0.05$) are highlighted in green. Since we use the MSE to evaluate regression performance and the ROC-AUC score for classification, we note that the signs of respective correlation coefficient work in opposite direction; i.e., e.g., negative correlation with MSE means better regression performance, while it is the other way round for classification. The \textbf{first row} corresponds to \textbf{Friedman} and the \textbf{second row} to the \textbf{\ce{NH3}-PES} regression tasks. The \textbf{third row} illustrates \textbf{two curves diff} on the left and \textbf{hidden manifold diff} on the right.}
    \phantomsection
    \label{fig:default-clf-reg-correlations-appendix}
\end{figure*}

To complement the report of correlation analyses between QKM hyperparameters and test performance scores, as well as between the hyperparameters themselves in Sec.~\ref{subsec:influence-hyperparameters}, we provide a summary of all other dataset families not shown in the main text in Fig.~\ref{fig:default-clf-reg-correlations-appendix}.

\subsection{\label{subsec:appendix-influence-encoding-circs}Details on the Influence of Encoding Circuits}
\begin{figure*}[t]
    \centering
    \begin{minipage}[t]{.49\textwidth}
        \includegraphics[width=.92\linewidth]{"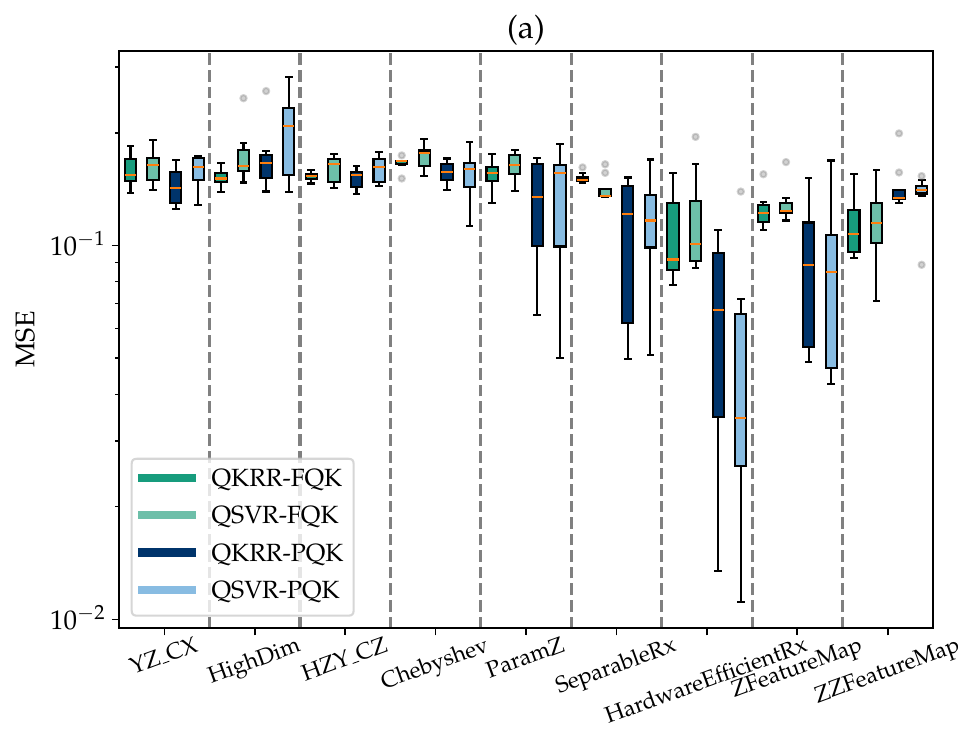"}
    \end{minipage}
    \begin{minipage}[t]{.49\textwidth}
        \includegraphics[width=.92\linewidth]{"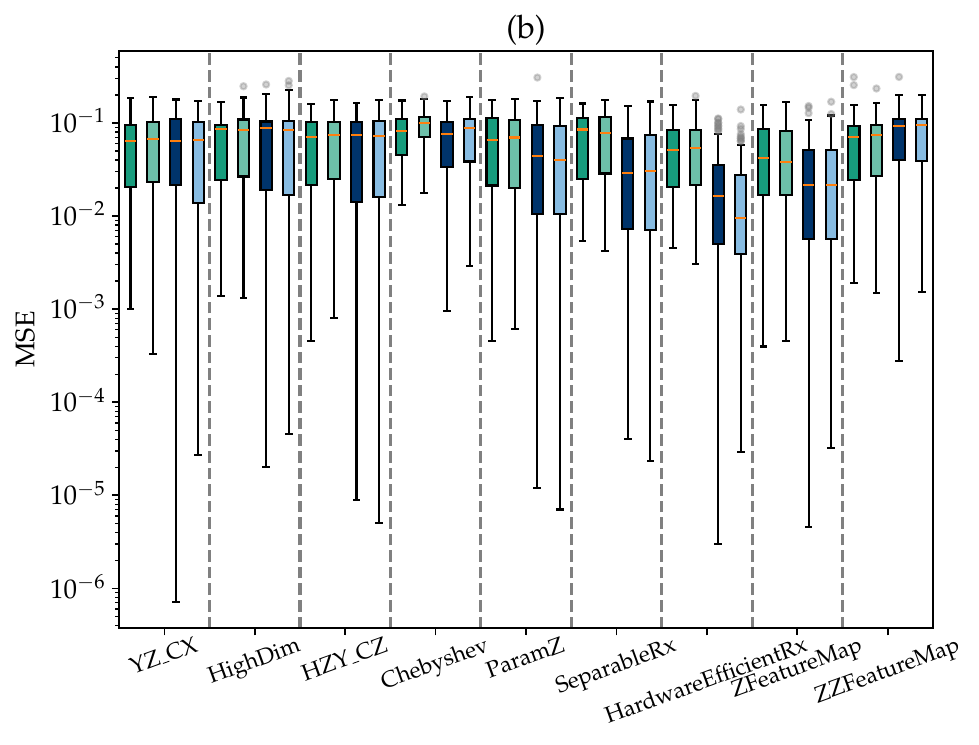"}
    \end{minipage}
    \begin{minipage}[t]{.49\textwidth}
        \includegraphics[width=.92\linewidth]{"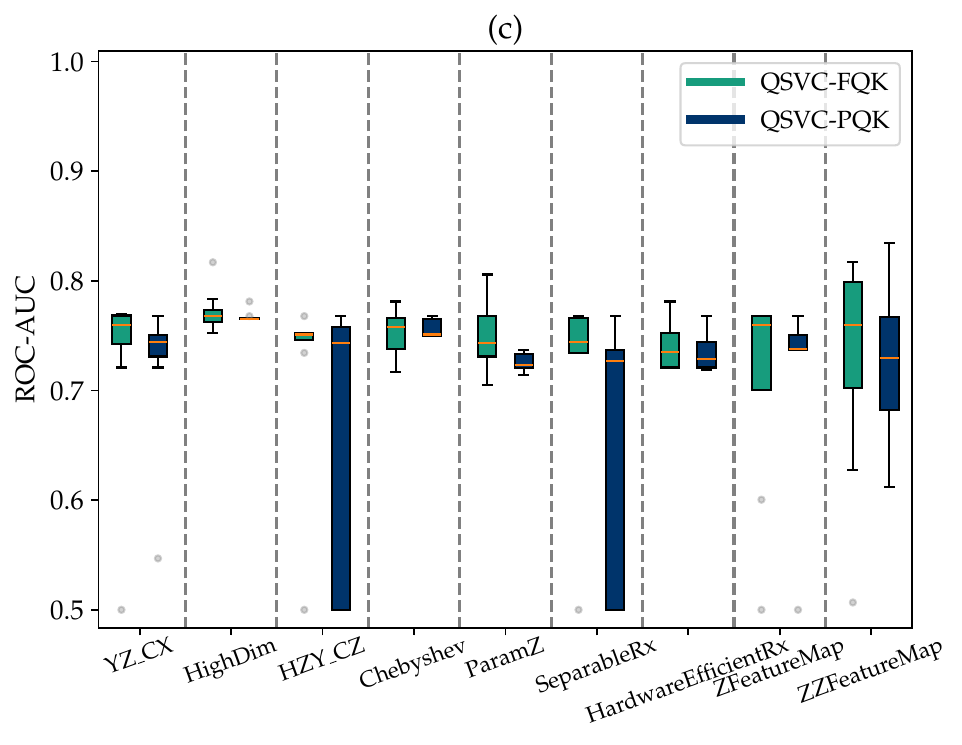"}
    \end{minipage}
    \begin{minipage}[t]{.49\textwidth}
        \includegraphics[width=.92\linewidth]{"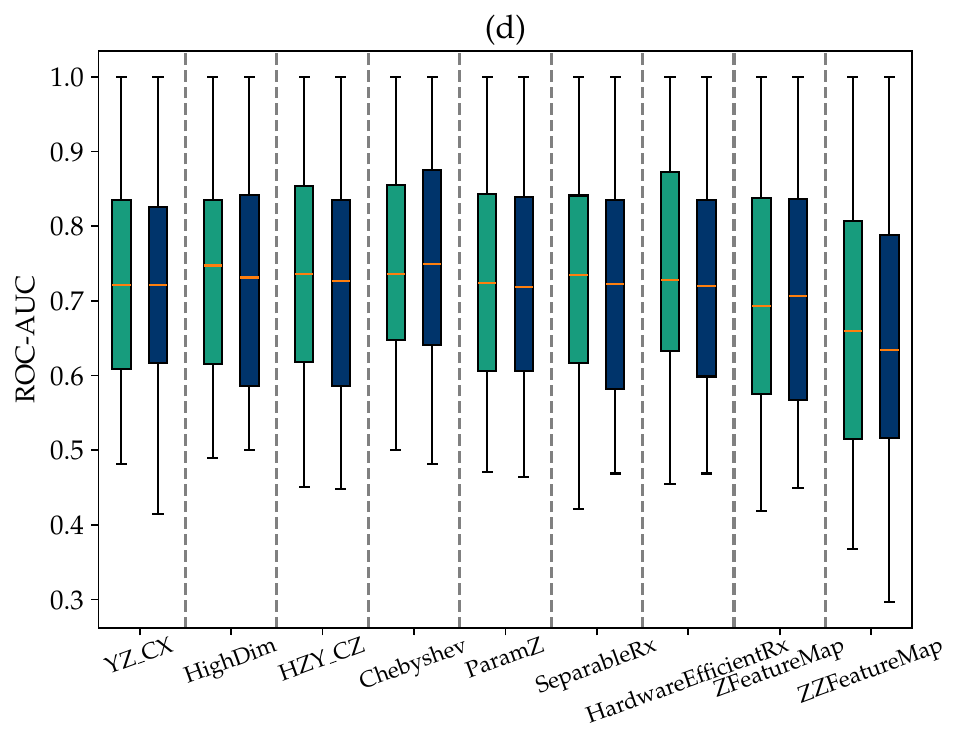"}
    \end{minipage}
    \caption{Complementing illustration to Fig.~\ref{fig:default-reg-clf-encoding-circs} of the main text with the datasets not shown therein. \textbf{Upper panel:} Regression \textbf{QFMNIST}. With~\textbf{(a)} Test performance aggregated over $n_{\mathrm{layers}}\in[1,8]$ per regression method (QSVR/QKRR) and quantum kernel for the dataset with number of features $d=15$ and~\textbf{(b)} test performance scores aggregated over all datasets in the QFMNIST family and over respective $n_{\mathrm{layers}}$. \textbf{Lower panel:} Classification \textbf{hidden manifold diff}. With~\textbf{(c)} Test performance aggregated over $n_{\mathrm{layers}}\in[1,8]$ for the dataset with manifold dimension $m=13$ and~\textbf{(d)} test performance scores aggregated over all datasets in the hidden manifold diff family and over respective $n_{\mathrm{layers}}$.}
    \phantomsection
    \label{fig:default-reg-clf-encoding-circs-appendix}
\end{figure*}

When investigating the impact of data encoding circuits on QKM model performance in Sec.~\ref{subsec:influence-encoding-circs}, we merely presented results of the Friedman (regression) and the two curves diff (classification) datasets in Fig.~\ref{fig:default-reg-clf-encoding-circs}. Here, we provide analogue findings in Fig.~\ref{fig:default-reg-clf-encoding-circs-appendix} for the QFMNIST and the hidden manifold diff datasets.

\subsection{\label{subsec:appendix-data-enc}Influence of Data Encoding Strategies}

\begin{figure}[tb]
\includegraphics[width=.75\columnwidth]{"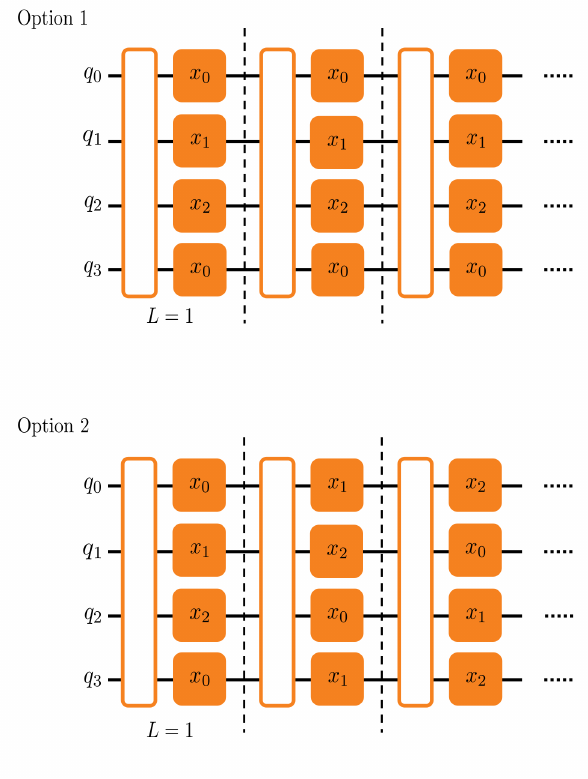"}
\caption{\label{fig:encoding-strategies}Schematic illustration of two different data encoding options considered in the investigations of Figs.~\ref{fig:encoding-strategy-qfmnist} and ~\ref{fig:encoding-strategy-two-curves-diff}.}
\end{figure}

\revision{As already implicitly indicated in Sec.~\ref{subsec:influence-hyperparameters}, it is known that the number of qubits and layers of data encoding circuits determine the Fourier frequency spectrum of the resulting quantum model and that encoding features redundantly can help in getting more accurate results~\citep{schuld2021supervised,Schuld2021}. However, from a practical point of view it is not clear how to best distribute the features of a data point among the available qubits in a particular encoding circuit. To shed some light on this aspect, we compare QKMs built from selected encoding circuits following the two encoding schemes illustrated in Fig.~\ref{fig:encoding-strategies}. In both embedding schemes, features are encoded in the qubits from top to bottom. Differences in the two encoding strategies occur when the number of qubits is larger than the number of features. In ``option 1'', after all features have been encoded once in a layer, the encoding is repeated and then cut off once the final qubit is reached. In ``option 2'', the enumeration of features is not reset at the next layer. As a consequence, each feature in ``option 1'' is always assigned to the same qubit, whereas in ``option 2'' the features can be shuffled across qubits.}

\begin{figure}[tb]
    \centering
    \includegraphics[width=\columnwidth]{"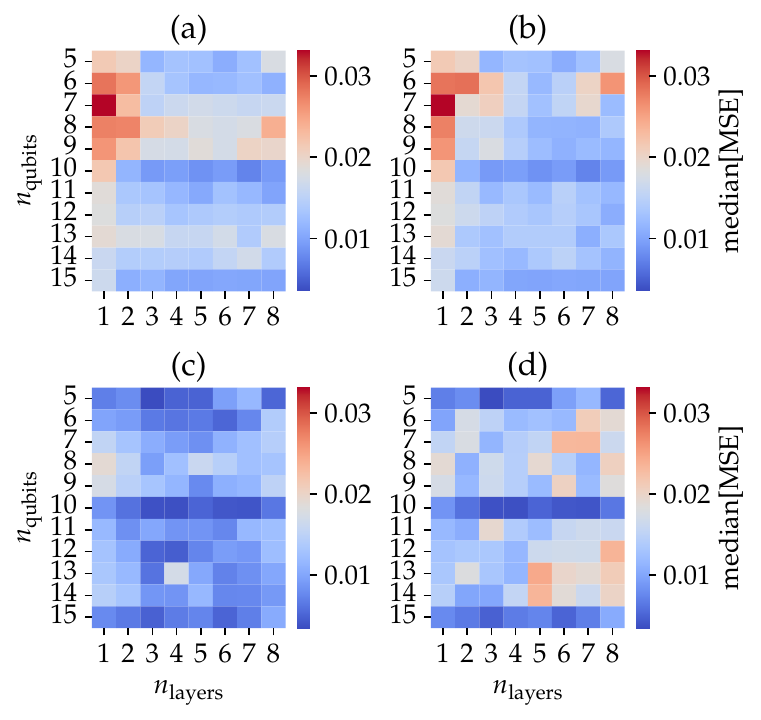"}
    \caption{\label{fig:encoding-strategy-qfmnist}Investigation of different encoding strategies as illustrated in Fig.~\ref{fig:encoding-strategies} for FQK and PQK approaches for the \textbf{QFMNIST} regression dataset with $d=5$ principal components. The results are aggregated over QKRR and QSVR approaches as well as over the corresponding encoding circuits, and then the median of the MSE test scores was calculated in each case. The upper panel corresponds to FQKs, while the lower panel shows PQKs, whereas the left column represents ``option 1'' and the right column ``option 2'', respectively. As such, we have: \textbf{(a)} FQK, option 1, \textbf{(b)} FQK, option 2, \textbf{(c)} PQK, option 1 and \textbf{(d)} PQK, option 2.}
\end{figure}

\revision{Using the QFMNIST dataset with $d=5$ as an example, Fig.~\eqref{fig:encoding-strategy-qfmnist} shows the findings for this regression tasks. The results for each quantum kernel are aggregated across the regression methods and the respective encoding circuits. Then, the median of the resulting MSE test score is calculated. For FQK, both embedding options show that largest scores are achieved 
for integer multiples of the number of features \revision{($n_{\mathrm{qubits}}=10$)}. It can be seen that higher scores occur more frequently with embedding ``option 2''. For PQK, it appears that no feature redundancies are necessary in either case. Here, however, embedding ``option 1'' often delivers larger scores.} 

\begin{figure}[tb]
    \centering
    \includegraphics[width=\columnwidth]{"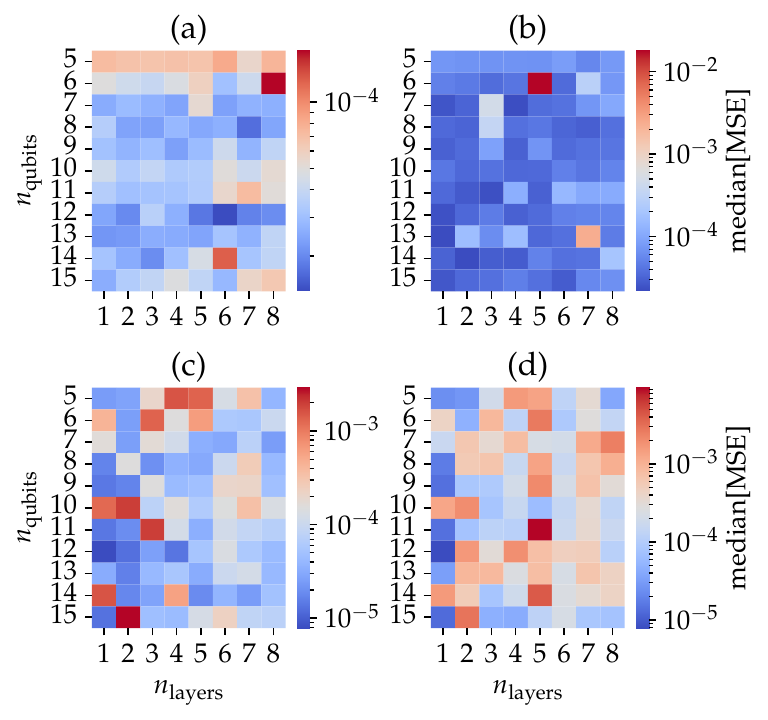"}
    \caption{Investigation of different encoding strategies as illustrated in Fig.~\ref{fig:encoding-strategies} for FQK and PQK approaches for the \textbf{Friedman} regression dataset $d=5$ features. The results were aggregated over QKRR and QSVR approaches as well as over the corresponding encoding circuits, and then the median of the MSE test scores was calculated in each case. The upper panel corresponds to FQKs, while the lower panel shows PQKs, whereas the left column represents ``option 1'' and the right column ``option 2'', respectively. As such, we have: \textbf{(a)} FQK, option 1, \textbf{(b)} FQK, option 2, \textbf{(c)} PQK, option 1 and \textbf{(d)} PQK, option 2.}
    \phantomsection
    \label{fig:encoding-strategy-friedman}
\end{figure}

The results for testing the two different encoding strategies for the Friedman dataset for $d=5$ features 
are displayed in Fig.~\ref{fig:encoding-strategy-friedman}. Contrary to the QFMNIST results in Fig.~\ref{fig:encoding-strategy-qfmnist}, in all cases, except for FQK, option 2, most accurate results are obtained for $n_{\mathrm{qubits}}=12$. Here, FQK and PQK approaches show similar behavior. The second encoding option for FQK appears to be less sensitive for the respective feature redundancy. Notably, that the redundancy does not have to be an integer multiple of the original number of features.

\begin{figure}[tb]
    \centering
    \includegraphics[width=\columnwidth]{"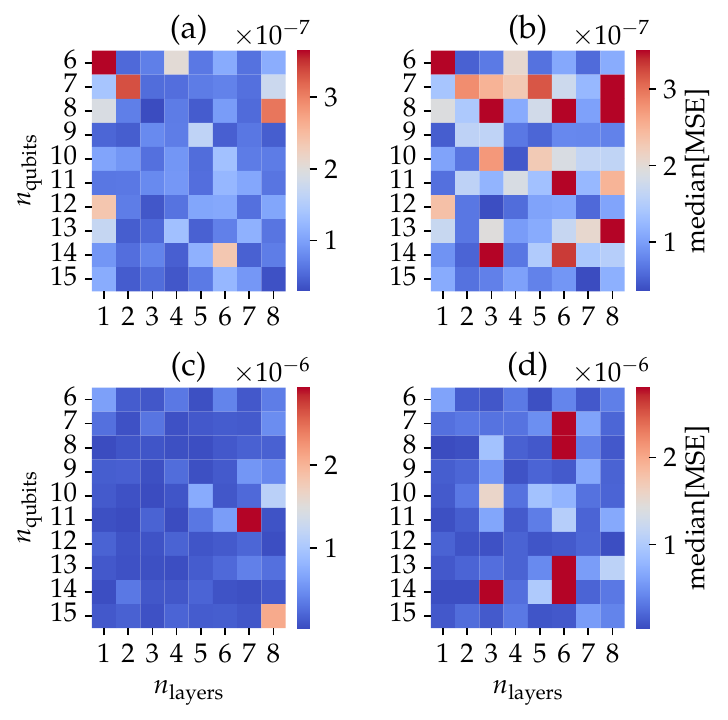"}
    \caption{Investigation of different encoding strategies as illustrated in Fig.~\ref{fig:encoding-strategies} for FQK and PQK approaches for the \textbf{\ce{NH3}} regression dataset (number of features $=6$). The results were aggregated over QKRR and QSVR approaches as well as over the corresponding encoding circuits, and then the median of the MSE test scores was calculated in each case. The upper panel corresponds to FQKs, while the lower panel shows PQKs, whereas the left column represents ``option 1'' and the right column ``option 2'', respectively. As such, we have: \textbf{(a)} FQK, option 1, \textbf{(b)} FQK, option 2, \textbf{(c)} PQK, option 1 and \textbf{(d)} PQK, option 2.}
    \phantomsection
    \label{fig:encoding-strategy-nh3}
\end{figure}

Figure~\ref{fig:encoding-strategy-nh3} shows the result for testing the two different encoding strategies for the \ce{NH3} dataset. Most accurate results are obtained for integer multiples of the number of features.

\begin{figure}[tb]
    \centering
    \includegraphics[width=\columnwidth]{"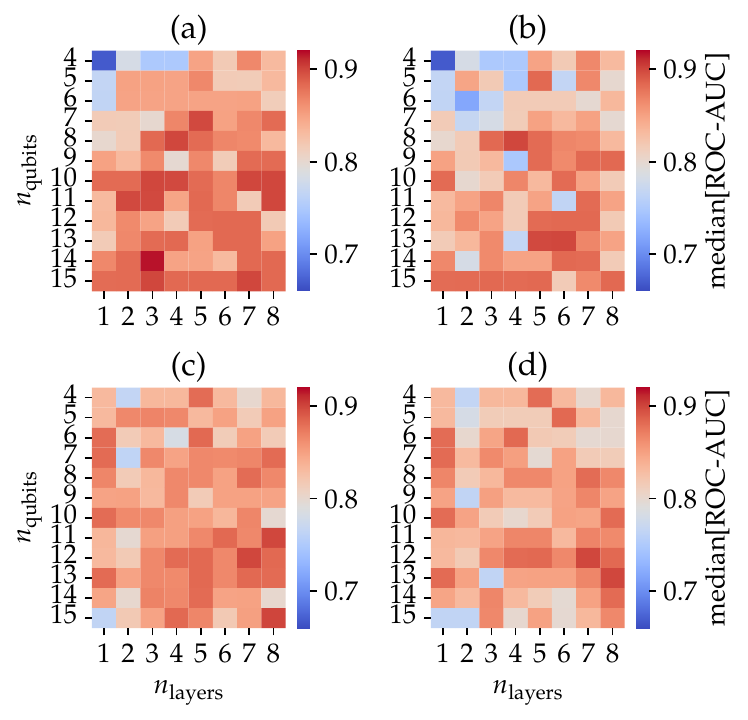"}
    \caption{\label{fig:encoding-strategy-two-curves-diff}Investigation of different encoding strategies as illustrated in Fig.~\ref{fig:encoding-strategies} for FQK- and PQK-QSVC approaches, respectively for the \textbf{two curves diff} classification dataset with $D=13$. Note, the feature dimension is kept constant at $d=4$. The results are aggregated over the respective encoding circuits and then the median of the ROC-AUC test scores is calculated in each case. The upper panel corresponds to FQKs, while the lower panel shows PQKs, whereas the left column represents ``option 1'' and the right column ``option 2'', respectively. As such, we have: \textbf{(a)} FQK, option 1, \textbf{(b)} FQK, option 2, \textbf{(c)} PQK, option 1 and \textbf{(d)} PQK, option 2.}
\end{figure}

\revision{Figure~\ref{fig:encoding-strategy-two-curves-diff} displays the results for the two curves diff classification dataset for $D=13$ (with $d=4$). Again, results are aggregated over the respective encoding circuits and then the median of the ROC-AUC test scores is evaluated. It can be seen that in contrast to the previous example, both embedding options deliver the best classification scores for non-integer multiples of the number of features (for PQK and FQK). Here, ``option 1'' performs better in each case. In addition, larger values of $n_{\text{layers}}$ and $n_{\text{qubits}}$ lead to increasing classification scores. The difference to the previous example (Fig.~\ref{fig:encoding-strategy-qfmnist}) might be due to significantly larger dataset complexity. This also explains the overall larger difference in performance scores for the two curves diff.}

\begin{figure}[tb]
    \centering
    \includegraphics[width=\columnwidth]{"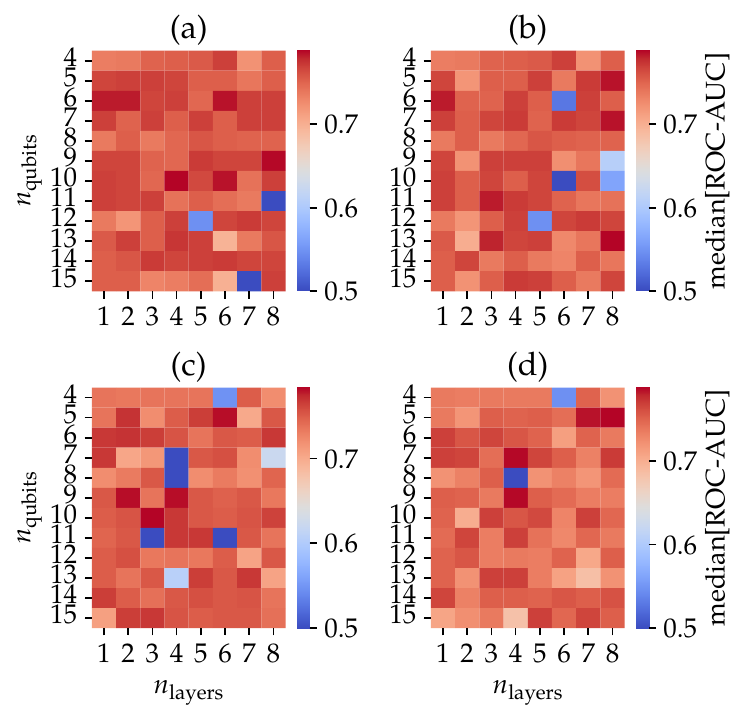"}
    \caption{Investigation of different encoding strategies as illustrated in Fig.~\ref{fig:encoding-strategies} for FQK- and PQK-QSVC approaches, respectively for the \textbf{hidden manifold diff} classification dataset with manifold dimension $m=13$. Note, the feature dimension is kept constant at $d=4$ The results were aggregated over the respective encoding circuits and then the median of the ROC-AUC test scores was calculated in each case. The upper panel corresponds to FQKs, while the lower panel shows PQKs, whereas the left column represents ``option 1'' and the right column ``option 2'', respectively. As such, we have: \textbf{(a)} FQK, option 1, \textbf{(b)} FQK, option 2, \textbf{(c)} PQK, option 1 and \textbf{(d)} PQK, option 2.}
    \phantomsection
    \label{fig:encoding-strategy-hidden-manifold-diff}
\end{figure}

Figure~\ref{fig:encoding-strategy-hidden-manifold-diff} provides insights into the effect of the two different encoding strategies for the hidden manifold diff dataset with $m=13$. The findings discussed previously for the two curves diff dataset still apply.

\clearpage
\subsection{\label{subsec:appendix-indepth-pqk}Details on the Analysis of PQK Design Options}
\begin{figure}[tb]
    \centering
    \includegraphics[width=\columnwidth]{"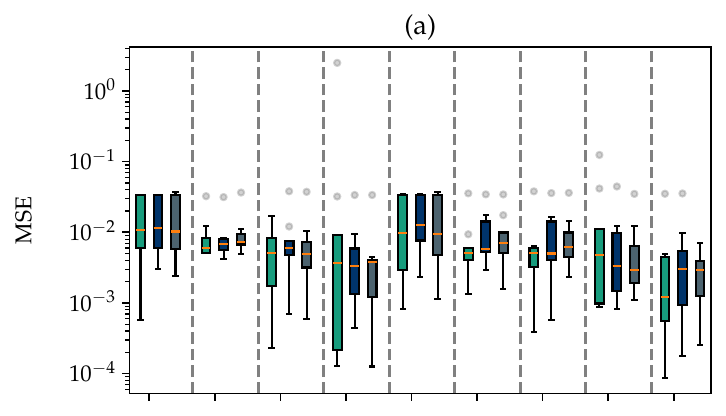"}
    \includegraphics[width=\columnwidth]{"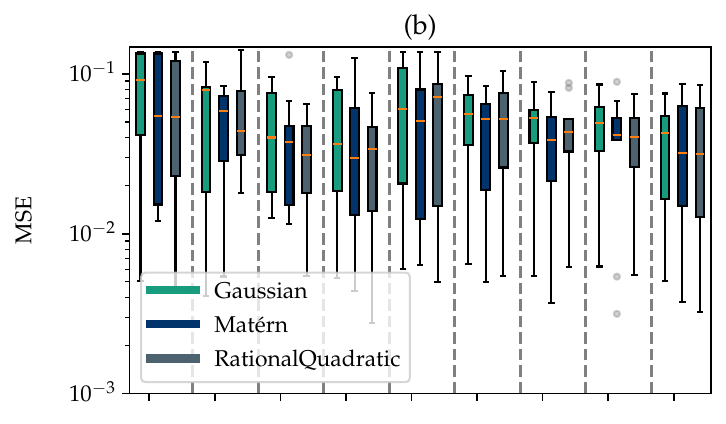"}
    \includegraphics[width=\columnwidth]{"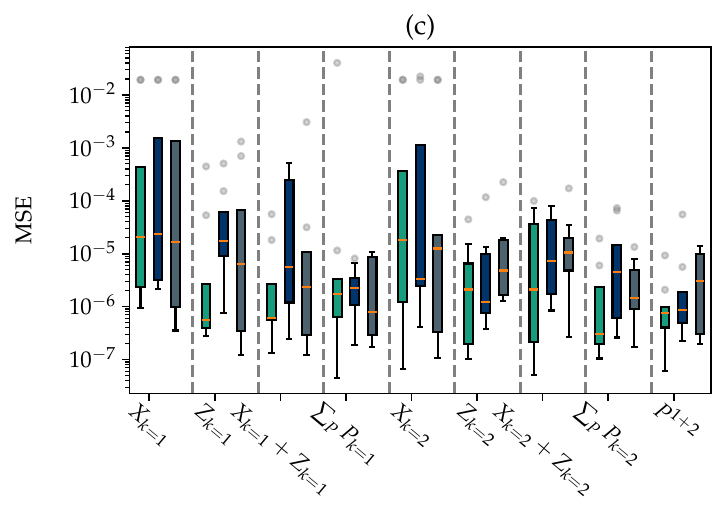"}
    \caption{MSE test performances for \textbf{QSVR-PQK models} as a function of the measurement operators given in Eqs.~\eqref{eq:1RDM-Ok}, ~\eqref{eq:2RDM-Ok} and~\eqref{eq:1plus2RDM-Ok} for Gaussian, Mat\'er and RationalQuadratic outer kernel functions $\kappa$. This supports the discussion around Tab.~\ref{tab:correlation-indepth-PQK} in Sec.~\ref{subsec:Indepth-PQK} of the main text. The simulation results corresponding to the following \textbf{regression} datasets are shown: \textbf{(a)} Friedman dataset for $d=10$ features. \textbf{(b)} QFMNIST for $d=8$ PCA components. \textbf{(c)} The \ce{NH3}-PES. The data for each $O_{k}$ are aggregated across the encoding circuits considered in this work.}
    \phantomsection
    \label{fig:boxplots-indepth-pqk-reg}
\end{figure}
\subsubsection{Overview on Test Performances}
\begin{figure*}[tb]
    \centering
    \begin{minipage}[t]{.49\textwidth}
        \includegraphics[width=\linewidth]{"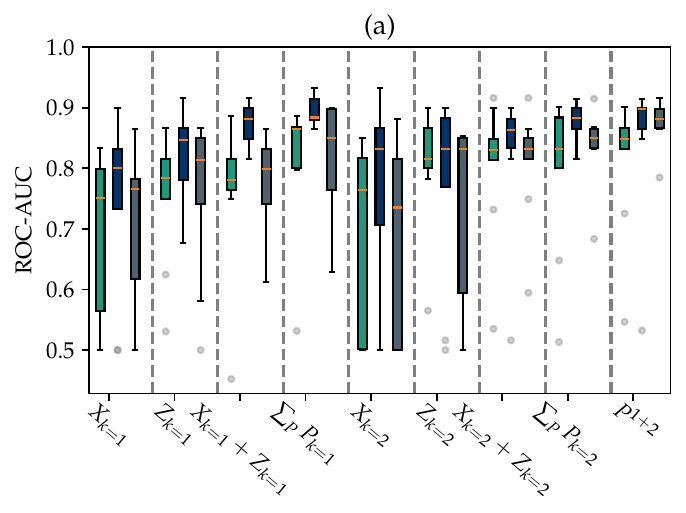"}
    \end{minipage}
    \begin{minipage}[t]{.49\textwidth}
        \includegraphics[width=\linewidth]{"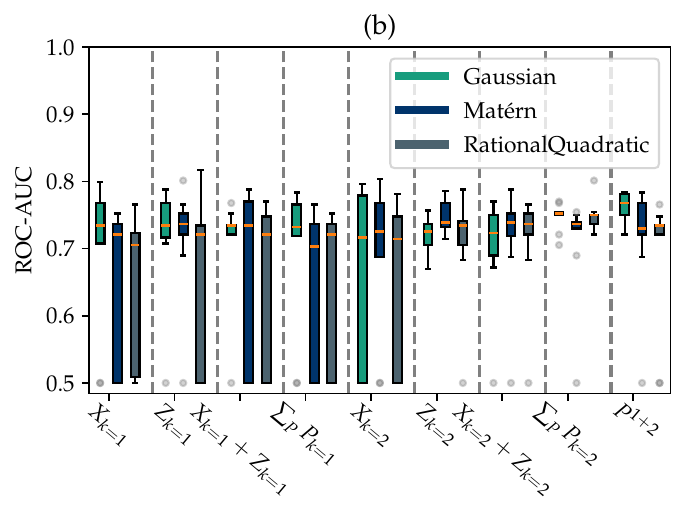"}
    \end{minipage}
    \caption{ROC-AUC test score for \textbf{QSVC-PQK models} as a function of the measurement operators given in Eqs.~\eqref{eq:1RDM-Ok}, ~\eqref{eq:2RDM-Ok} and~\eqref{eq:1plus2RDM-Ok} for Gaussian, Mat\'er and RationalQuadratic outer kernel functions $\kappa$. This supports the discussion around Tab.~\ref{tab:correlation-indepth-PQK} in Sec.~\ref{subsec:Indepth-PQK} of the main text. The results corresponding to the following \textbf{classification} datasets are shown: \textbf{(a)} two curves diff dataset for $d=4$ and degree $D=13$. \textbf{(b)} hidden manifold diff dataset for $d=4$ and manifold dimension $m=13$. The data for each $O_{k}$ are aggregated across the encoding circuits considered in this work.}
    \phantomsection
    \label{fig:boxplots-indepth-pqk-clf}
\end{figure*}

In the main text we investigate (semi-) partial correlations in Tabs.~\ref{tab:correlation-indepth-PQK} and~\ref{tab:full-correlation-indepth-PQK} between the choice of the outer kernel function $\kappa$ and corresponding hyperparameters as well as the measurement operator $O_{k}$ and resulting model test performances. In the following, we give a detailed insights into the test scores for each outer kernel (Gaussian, Mat\'ern, RationalQuadratic) as a function of the measurement operator in Figs.~\ref{fig:boxplots-indepth-pqk-reg} and~\ref{fig:boxplots-indepth-pqk-clf} for the regression and classification tasks, respectively considered in this study. The data for each $O_{k}$ are aggregated across the encoding circuits considered here. Comparing these plots with the numbers listed in Tab.~\ref{tab:correlation-indepth-PQK} helps to understand the reported statistically significant correlation. For instance, Fig.~\ref{fig:boxplots-indepth-pqk-reg}~(b) indicates that for each $O_{k}$ the Gaussian outer kernel leads to weaker test performance scores compared to the rest. This in turn reflects in the positive Spearman correlation coefficient in Tab.~\ref{tab:correlation-indepth-PQK}. Analogous conclusions can be drawn for the observed trends with respect to the different measurement operators considered in this work.

The same applies to, e.g., the two curves diff dataset in Fig.~\ref{fig:boxplots-indepth-pqk-clf}~(a). 
While the Gaussian kernel tends to yield relatively low ROC-AUC scores, the Mat\'ern kernel generally shows a trend towards higher test scores. Again, this reflects in the negative and positive, respectively Spearman correlation coefficients in Tab.~\ref{tab:correlation-indepth-PQK}. Once more, we note that analogous implications hold for findings regarding the different measurement operators.

\subsubsection{Extended Correlation Analysis}
In order to support the efforts in Sec.~\ref{subsec:Indepth-PQK} to gain a deeper understanding of what are basic mechanisms in PQKs responsible for learning, we provide all (statistically significant) semi-partial correlations of the various hyerparameters of the QSVR/QSVC-PQK model for a given outer kernel $\kappa$, both among themselves and with model test performance in Tab.~\ref{tab:full-correlation-indepth-PQK}. Here, we note that while negative correlation coefficients w.r.t. the MSE point at overall better model performance, it is the other way round for correlations with ROC-AUC scores.
\begin{table}[tb]
    \centering
    \caption{Statistically significant semi-partial correlations of the various hyperparameter of the QSVR/QSVC-PQK model for a given outer kernel function $\kappa$, both among themnselves and with test performance scores. This table supports the corresponding discussion in Ref.~\ref{subsec:Indepth-PQK}. \emph{We note that while negative correlation coefficients w.r.t. the MSE point at overall better model performance, it is the other way round for correlations with ROC-AUC scores.}}
    \phantomsection
    \label{tab:full-correlation-indepth-PQK}
        \begin{tabular}{lllS}
            \toprule\toprule
            {$\kappa(\text{Dataset})$} & {$X$} & {$Y$} & {$\rho_{\mathrm{Spearman}}(X,Y)$} \\
            \midrule
            \multirow{3}{*}{$\kappa^{\text{Gauss}}(\text{Friedman})$} & $n_{\mathrm{layers}}$ & $\gamma$ & {0.344} \\
            ~ & $C$ & MSE & {-0.331} \\
            ~ & $w_{\mathrm{e}}$ & MSE & {0.267} \\ \cdashline{1-4}
            \multirow{2}{*}{$\kappa^{\text{Gauss}}(\text{QFMNIST})$} & $C$ & MSE & {0.353} \\
            ~ & $\gamma$ & $w_{\mathrm{e}}$ & {-0.262} \\ \cdashline{1-4} 
            \multirow{5}{*}{$\kappa^{\text{Gauss}}(\text{\ce{NH3}-PES})$} & $n_{\mathrm{layers}}$ & $\gamma$ & {0.293} \\
            ~ & $C$ & $\varepsilon$ & {0.319} \\
            ~ & $C$ & MSE & {0.233} \\  
            ~ & $\varepsilon$ & MSE & {0.390} \\
            ~ & $w_{\mathrm{e}}$ & MSE & {0.322} \\\cdashline{1-4} 
            \multirow{2}{*}{$\kappa^{\text{Gauss}}(\text{two curves diff})$} & $n_{\mathrm{layers}}$ & $\gamma$ & {-0.270} \\
            ~ & $\gamma$ & {ROC-AUC} & {-0.306} \\\cdashline{1-4} 
            \multirow{2}{*}{$\kappa^{\text{Gauss}}(\text{hidden manifold diff})$} & $n_{\mathrm{layers}}$ & $w_{\mathrm{e}}$ & {-0.264} \\
            ~ & $w_{\mathrm{e}}$ & {ROC-AUC} & {-0.355} \\
            \midrule
            \multirow{6}{*}{$\kappa_{3/2}^{\text{Mat}}(\text{Friedman})$} & $n_{\mathrm{layers}}$ & $C$ & {-0.260} \\
            ~ & $n_{\mathrm{layers}}$ & $w_{\mathrm{e}}$ & {-0.275} \\
            ~ & $n_{\mathrm{layers}}$ & MSE & 0.465 \\
            ~ & $C$ & MSE & {-0.402} \\
            ~ & $\varepsilon$ & MSE & {0.242} \\
            ~ & $\ell$ & MSE & {-0.246} \\\cdashline{1-4} 
            \multirow{2}{*}{$\kappa_{3/2}^{\text{Mat}}(\text{QFMNIST})$} & $n_{\mathrm{layers}}$ & MSE & {0.244} \\
            ~ & $\ell$ & $w_{\mathrm{e}}$ & {0.337} \\\cdashline{1-4} 
            \multirow{4}{*}{$\kappa_{3/2}^{\text{Mat}}(\text{\ce{NH3}-PES})$} & $n_{\mathrm{layers}}$ & $C$ & {0.272} \\
            ~ & $\varepsilon$ & MSE & {0.250} \\
            ~ & $\ell$ & $w_{\mathrm{e}}$ & {0.403} \\
            ~ & $w_{\mathrm{e}}$ & MSE & {0.228} \\\cdashline{1-4} 
            \multirow{4}{*}{$\kappa_{3/2}^{\text{Mat}}(\text{two curves diff})$} & $C$ & $\ell$ & {0.490} \\
            ~ & $C$ & {ROC-AUC} & {0.323} \\
            ~ & $\ell$ & $w_{\mathrm{e}}$ & {0.305} \\
            ~ & $w_{\mathrm{e}}$ & {ROC-AUC} & {-0.318} \\\cdashline{1-4}
            \multirow{4}{*}{$\kappa_{3/2}^{\text{Mat}}(\text{hidden manifold diff})$} & $n_{\mathrm{layers}}$ & $\ell$ & {-0.370} \\
            ~ & $n_{\mathrm{layers}}$ & $w_{\mathrm{e}}$ & {-0.345} \\
            ~ & $C$ & $\ell$ & {0.238} \\
            ~ & $w_{\mathrm{e}}$ & {ROC-AUC} & {-0.299} \\
            \midrule
            \multirow{5}{*}{$\kappa^{\text{RQ}}(\text{Friedman})$} & $n_{\mathrm{layers}}$ & $\ell$ & {-0.290} \\
            ~ & $C$ & $\alpha$ & {-0.264} \\
            ~ & $\alpha$ & $\ell$ & {-0.382} \\
            ~ & $\alpha$ & MSE & {-0.441} \\
            ~ & $w_{\mathrm{e}}$ & MSE & {0.254} \\\cdashline{1-4}
            \multirow{3}{*}{$\kappa^{\text{RQ}}(\text{QFMNIST})$} & $C$ & $\alpha$ & {-0.431} \\
            ~ & $\alpha$ & $\ell$ & {-0.263} \\
            ~ & $\ell$ & $w_{\mathrm{e}}$ & {0.562} \\\cdashline{1-4}
            \multirow{4}{*}{$\kappa^{\text{RQ}}(\text{\ce{NH3}-PES})$} & $C$ & $\ell$ & {0.365} \\
            ~ & $\varepsilon$ & MSE & {0.281} \\
            ~ & $\alpha$ & $w_{\mathrm{e}}$ & {0.275} \\
            ~ & $\alpha$ & MSE & {-0.301} \\\cdashline{1-4}
            \multirow{5}{*}{$\kappa^{\text{RQ}}(\text{two curves diff})$} & $n_{\mathrm{layers}}$ & $C$ & {-0.236} \\
            ~ & $C$ & $\alpha$ & {-0.304} \\
            ~ & $\alpha$ & $\ell$ & {-0.251} \\
            ~ & $\alpha$ & $w_{\mathrm{e}}$ & {0.312} \\
            ~ & $\ell$ & {ROC-AUC} & {0.473} \\\cdashline{1-4}
            \multirow{7}{*}{$\kappa^{\text{RQ}}(\text{hidden manifold diff})$} & $n_{\mathrm{layers}}$ & $\ell$ & {-0.341} \\
            ~ & $n_{\mathrm{layers}}$ & $w_{\mathrm{e}}$ & {-0.477} \\
            ~ & $C$ & {ROC-AUC} & {0.456} \\
            ~ & $\alpha$ & $\ell$ & {0.535} \\
            ~ & $\alpha$ & $w_{\mathrm{e}}$ & {0.256} \\
            ~ & $\ell$ & {ROC-AUC} & {-0-358} \\
            ~ & $w_{\mathrm{e}}$ & {ROC-AUC} & {-0.250} \\
            \bottomrule\bottomrule
        \end{tabular}
\end{table}

\subsubsection{Complementary Plots to Figure~\ref{fig:indepth-pqk-qfmnist}}

\begin{figure*}[tb]
    \centering
    \includegraphics[width=.9\linewidth]{"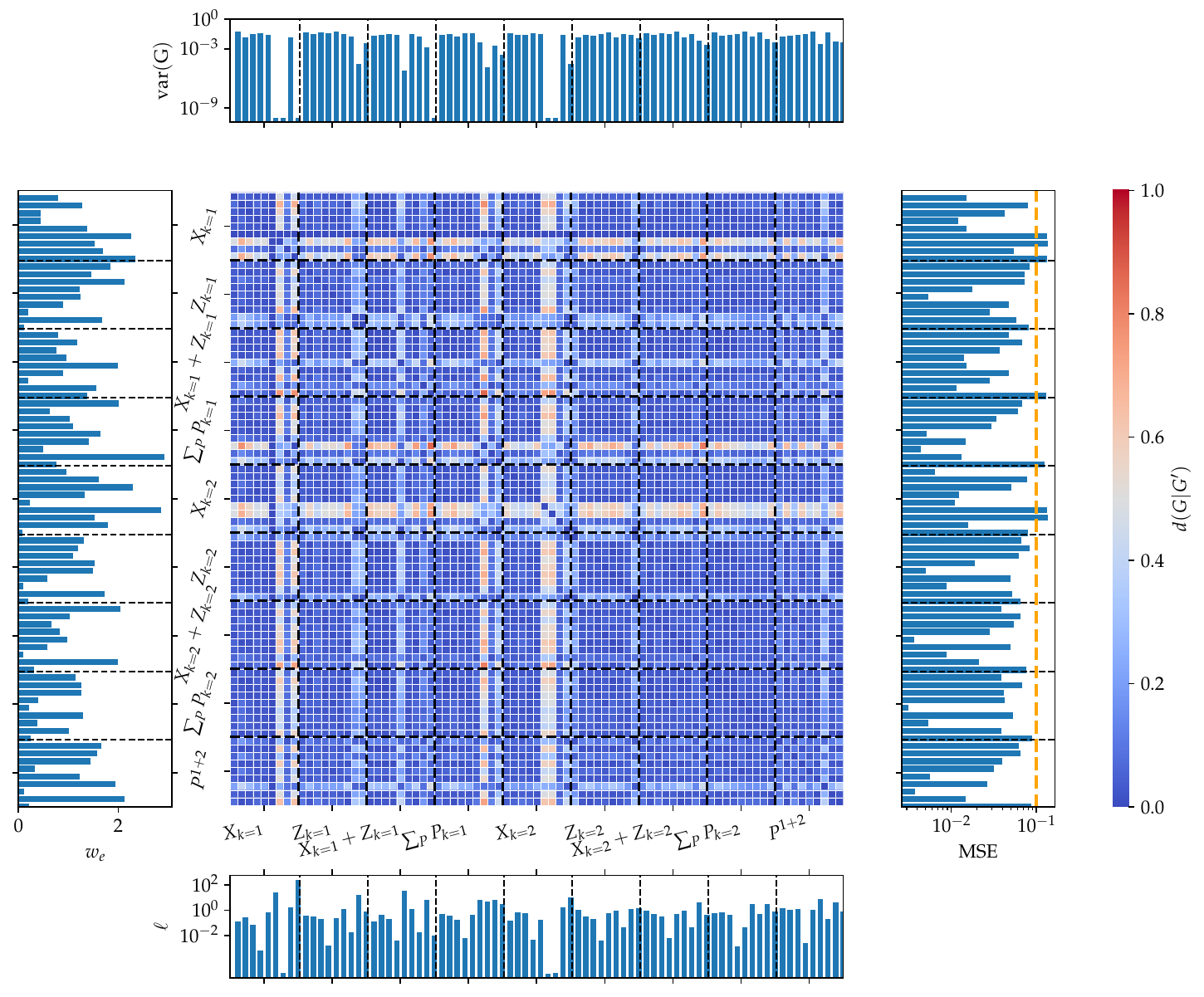"}
    \caption{Analogous illustration to Fig.~\ref{fig:indepth-pqk-qfmnist} as discussed in Sec.~\ref{subsec:Indepth-PQK} for the QFMNIST dataset with $d=8$ components and the Mat\'ern outer kernel.}
    \phantomsection
    \label{fig:qfmnist_indepth-pqk_magnum-opus_matern}
\end{figure*}

\begin{figure*}[tb]
    \centering
    \includegraphics[width=.9\linewidth]{"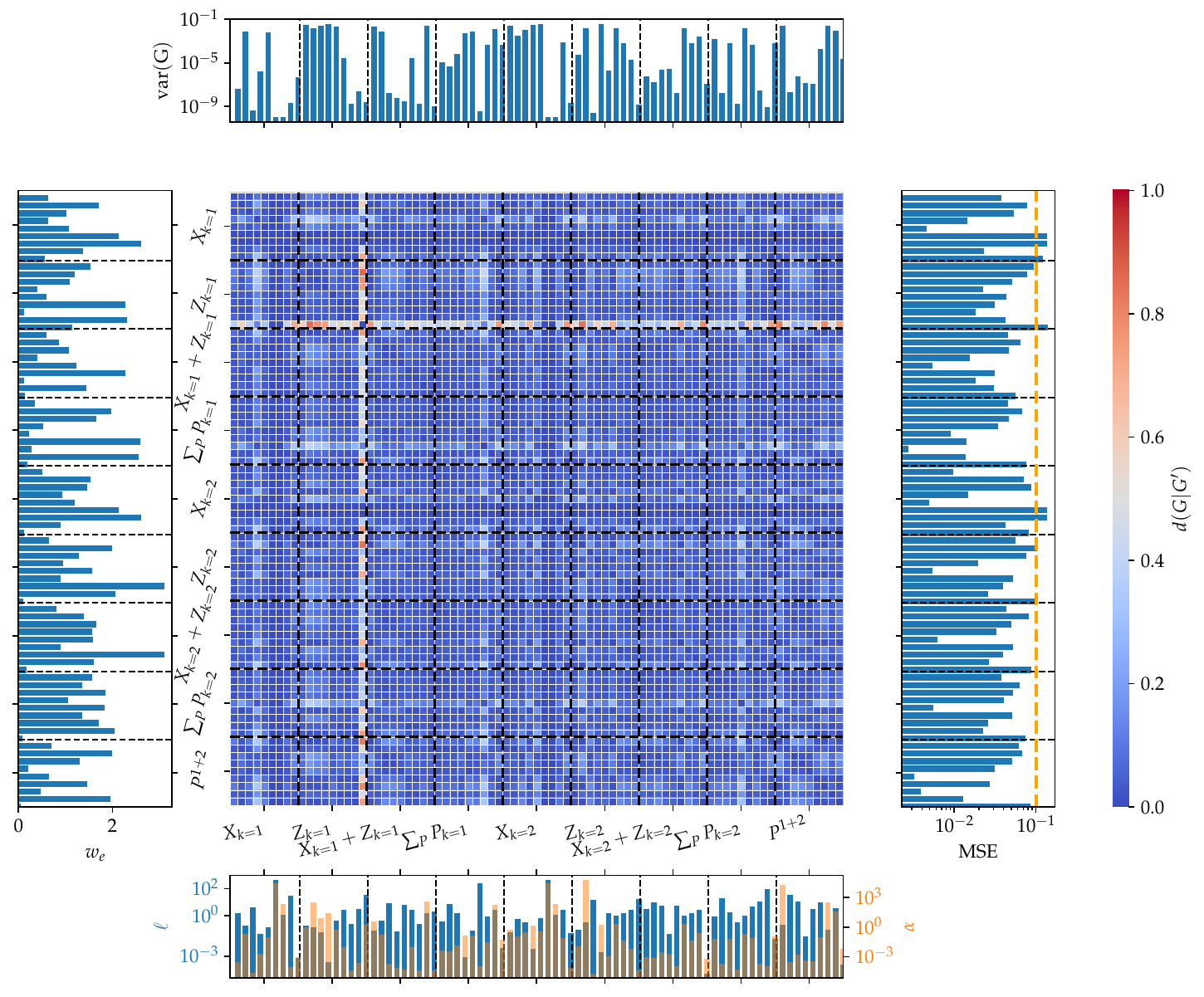"}
    \caption{Analogous illustration to Fig.~\ref{fig:indepth-pqk-qfmnist} as discussed in Sec.~\ref{subsec:Indepth-PQK} for the QFMNIST dataset with $d=8$ components and the RationalQuadratic outer kernel.}
    \phantomsection
    \label{fig:qfmnist_indepth-pqk_magnum-opus_rq}
\end{figure*}

In Sec.~\ref{subsec:Indepth-PQK} we thoroughly investigate the underlying mechanisms that are eventually responsible for learning in PQK approaches. While comprehensive correlation analyses in Tabs.~\ref{tab:correlation-indepth-PQK} and~\ref{tab:full-correlation-indepth-PQK} provide profound insights, a deeper understanding of these findings follows from studying the distance measure as defined in Eq.~\eqref{diff-gram-matrices}. For Gaussian outer kernels we can additionally define the projected quantum circuit contributions $F_{\boldsymbol{\theta}}$ as given in Eq.~\eqref{eq:F}. In the main text, we detail the corresponding discussion for the QFMNIST ($d=8$) dataset in case of the Gaussian outer kernel. The central illustration for this is provided in Fig.~\ref{fig:indepth-pqk-qfmnist}. In addition to these explanations we display the corresponding findings for the Mat\'ern and RationalQuadratic outer kernels in Figs.~\ref{fig:qfmnist_indepth-pqk_magnum-opus_matern} and~\ref{fig:qfmnist_indepth-pqk_magnum-opus_rq}, respectively. 
In contrast to studying $\mathrm{var}(F)$ in case of Gaussian outer kernels, we can leverage the analogous notion of $\mathrm{var}(G)$ to explain (non-)vanishing $d(G|G')$ distances. Finite distances result from one Gram matrix being ill-conditioned, i.e., $\mathrm{var}(G)\to 0$, while vanishing distances correspond to favourable variance behaviors. By considering the feature scaling $w_{\mathrm{e}}$ and length-scale parameters ($\ell$ and $\alpha$) on the left and on the bottom, respectively helps to understand the mechanisms that render these cases into the weak-performing regime.

Similarly, we can provide corresponding discussions and illustrations for all remaining datasets of this study upon reasonable request.

\clearpage
\section{\label{sec:appendix-kta}Optimization of Trainable Parameters in Data Encoding Circuits}
\begin{table}[t]
    \centering
    \caption{\label{tab:KTA-res}KTA results for two curves diff ($D=13$) and hidden manifold diff ($m=13$) datasets for both QSVC-FQK and QSVC-PQK approaches. All results correspond to the respective best performing ``ChebyshevPQC'' encoding circuits.}
    \begin{tabular}{llSS}
        \toprule\toprule
        {Dataset} & {QKM} & {ROC-AUC before} & {ROC-AUC after} \\
        \midrule
        \multirow{2}{*}{two curves diff} & {FQK} & {0.506} & {0.801} \\
        ~ & {PQK} & {0.682} & {0.834} \\
        \multirow{2}{*}{hidden manifold diff} & {FQK} & {0.781} & {0.683} \\
        ~ & {PQK} & {0.768} & {0.641} \\
        \bottomrule\bottomrule
    \end{tabular}
\end{table}
Here, we provide a brief insight into the optimization of the variational parameters $\boldsymbol{\theta}$, which are generally incorporated for defining data encoding circuits as given in Eq.~\eqref{eq:data-encoding}. As mentioned in Sec.~\ref{subsec:research-design-models}, these trainable parameters, if present, were randomly initialized with a fixed seed. To study the effectiveness of parameter optimization, we optimize the model corresponding to the best ``ChebyshevPQC'' configuration for both the two curves diff ($D=13$), cf. Fig.~\ref{fig:default-reg-clf-encoding-circs}~(c) and the hidden manifold diff ($m=13$) datasets. For this purpose, the kernel-target-alignment (KTA) metric is maximized over the training set~\citep{Hubregtsen2021,alvarezestevez2024}. We employ the respective sQUlearn implementation with the corresponding Adam optimizer.

The results listed in Tab.~\ref{tab:KTA-res} indicate that optimization can lead to a significant improvement in the case of the two curves diff dataset, but, this improvement is not guaranteed, as demonstrated by the counterexample of the hidden manifold diff dataset. This discrepancy may be attributed to a poorly tuned optimizer. A detailed analysis of this issue, as well as the investigation of a broader range of datasets and data encoding circuits, however, is beyond the scope of this work. Nevertheless, we highlight the potential of KTA optimization for datasets with high complexity with correlating sub-optimal QKMs.

\end{appendices}

\end{document}